%% file: prb.tex
\documentclass[aps,prb,twocolumn,superscriptaddress,footinbib,notitlepage,showpacs]{revtex4-1}
\usepackage[utf8]{inputenc}
\usepackage{longtable}
\usepackage{morefloats}
\usepackage[dvips]{graphicx,color}
\usepackage[dvipsnames]{xcolor}
\usepackage{epsfig,graphicx,amsfonts,amsbsy}
\usepackage{amsmath,amsfonts,amsthm,amssymb}
\usepackage{url}
\usepackage{verbatim}
\usepackage[colorlinks=true,allcolors=blue]{hyperref}
\usepackage{array}
\usepackage{multirow}
\usepackage{tabularx}
\usepackage{multirow}
\usepackage{braket} 
\usepackage{dsfont}
\usepackage{bm}
\usepackage{natbib}
\usepackage{placeins}
\usepackage{import}

\graphicspath{{final_fig/}}

\begin{document}

\title{Path-integral Monte Carlo study of electronic states in quantum dots in an external magnetic field}

\author{Csaba T\H oke}
\affiliation{BME-MTA Exotic Quantum Phases ``Lend\"ulet" Research Group, Budapest University of Technology and Economics, Institute of Physics, Budafoki \'ut 8, H-1111 Budapest, Hungary}
\affiliation{Department of Theoretical Physics,Budapest University of Technology and Economics, Institute of Physics, Budafoki \'ut 8, H-1111 Budapest, Hungary}
\author{Tam\'as Haidekker Galambos}
\affiliation{BME-MTA Exotic Quantum Phases ``Lend\"ulet" Research Group, Budapest University of Technology and Economics, Institute of Physics, Budafoki \'ut 8, H-1111 Budapest, Hungary}
\affiliation{Department of Theoretical Physics,Budapest University of Technology and Economics, Institute of Physics, Budafoki \'ut 8, H-1111 Budapest, Hungary}
\affiliation{Department of Physics, University of Basel, Klingelbergstrasse 82, CH-4056 Basel, Switzerland}

\date{\today}

\begin{abstract}
  We explore the correlated electron states in harmonically confined few-electron quantum dots in an external
  magnetic field by the path-integral Monte Carlo method for a wide range of the field and the Coulomb interaction strength.
  Using the phase structure of a preceding unrestricted Hartree-Fock calculation for phase fixing, we find a rich variety
  of correlated states, often completely different from the prediction of mean-field theory.
  These are finite temperature results, but sometimes the correlations saturate with decreasing temperature,
  providing insight into the ground-state properties.
\end{abstract}

\maketitle


\section{Introduction}
\label{sec:intro}

Quantum dots established in gated submicron regions of a two-dimensional electron gas, contacted either vertically or laterally,
have been studied intensively in the last decades \cite{Chakraborty92,Kouwenhoven01,Reimann02,Alhassid00}.
Such structures trap a few to a few hundred electrons, and are often referred to as artificial atoms.
Unlike in real atoms, where the electron-electron interaction plays a quantitatively important but
qualitatively minor role,
in weakly confined quantum dots the electron density can be tuned to small values,
and the Coulomb repulsion among electrons can play a prominent role.
Many experiments are adequately described in terms of weak-coupling theories (Hartree-Fock, density functional theory,
or the constant interaction model).
This is no doubt due to the strong screening of the Coulomb interaction, most importantly, by the leads in vertical quantum dots.
With lateral gates, however, the intermediate coupling regime is accessible,
while the strongly coupled regime requires weaker confinement in cleaner samples, or the use of semiconductors
that have an effective mass larger than GaAs.

Electrons in quantum dots show an especially intriguing behavior in the presence of an external magnetic field, as
the relevant length scales---the confinement length, the magnetic length, and the effective Bohr radius of the
interaction---can be tuned to comparable values, letting confinement, interaction and magnetic effects compete.
In the presence of an external magnetic field, the ground-state properties have been studied by exact diagonalization (in truncated Hilbert spaces) \cite{Maksym90,Wagner92,Mikhailov02,Cha03,Tavernier03,Tavernier06},
Hartree-Fock mean-field theory \cite{Yannouleas99,Yannouleas00,Reusch01,Reusch03,Yannouleas03b},
post-Hartree-Fock projection techniques \cite{Yannouleas01,Yannouleas02,Yannouleas02b,Yannouleas03c,DeGiovannini07,DeGiovannini08},
and the diffusion Monte Carlo method with phase fixing \cite{Bolton96,Guclu05,Guclu05b}.
Moreover, at finite temperature, the energy scale $k_\text{B}T$ of thermal excitations brings a fourth player into the game.
Finite temperature theoretical studies rely upon either the finite temperature Hartree-Fock theory \cite{Dean01} or the exact diagonalization of the complete truncated Hamiltonian \cite{Maksym90} (as distinct from finding the ground state and the lowest excitations by the L\'anczos method).

The path-integral Monte Carlo (PIMC) method has been applied to quantum dots only for zero external magnetic fields
\cite{Egger99,Egger99b,Harting00,Filinov01,Lozovik03,Reusch03b,Weiss05}.
No doubt this is due to the sign problem, which arises in quantum Monte Carlo simulations of fermionic systems,
and the typically more severe phase problem, which occurs whenever the presence of an external magnetic field or spin-orbit coupling
prevents us from using real-valued wave functions and density matrices, irrespective of the statistics of the simulated particles.
For zero magnetic field, the multilevel blocking method has been shown to mitigate the sign problem \cite{Mak98,Egger99,Egger00},
but the generalization of this method to magnetic problems does not seem to be straightforward.
It is also possible to use phase fixing in path-integral Monte Carlo; this requires the use of a guiding many-body density matrix.
Recently we have shown \cite{Haidekker18} that a relatively crude way of fixing the phase may nevertheless yield
valuable results in PIMC.
Encouraged by this finding, here we apply PIMC for characterizing the electronic structures that arise in harmonically confined dots.
We fix the phase by using the outcome of the unrestricted Hartree-Fock (UHF) approximation to the many-body problem.
For tiny systems we could use the phase of the density matrix of the Hartree-Fock effective Hamiltonian.
But we can study much larger systems by using the density matrix built up of the Hartree-Fock ground state only;
this can be regarded as the zero-temperature limit of the former approximation.
As we will see, a phase-fixing ansatz still leaves sufficient freedom in the PIMC method to let the simulated system discover strongly
correlated phases that qualitatively differ from the guide density matrix that is applied for fixing the phase.
This situation is quite different from the use of phase fixing in diffusion Monte Carlo \cite{Ortiz93,Bolton96,Guclu05b,Ghosal06},
a method that becomes very inefficient unless the trial wave function is a good approximation to the true ground state,
as the trial wave function is also used for importance sampling during the random walk.

We emphasize that we are interested both in the performance of the simulation method and the correlated behavior of the electron system.
We are particularly interested in methods that make the small magnetic field range accessible.
This range is especially important in experiments \cite{Zhitenev97}.
But most efficient correlated methods work better in the large magnetic field limit,
e.g., exact diagonalization (configuration interaction) as restricted to the lowest Landau level \cite{Maksym90},
or composite fermion diagonalization \cite{Jeon04b,Jeon07}.
We will see that PIMC is still efficient when the cyclotron energy is about 20\% of the confinement energy,
which for typical experimental parameters corresponds to less than 0.4 T \cite{Zhitenev97},
and the lowest Landau level approximation is definitely invalid.
However, below this range PIMC also becomes increasingly difficult.

Preliminary information on the ordering within the quantum dot is obtained by inspecting the spatial
structure of the UHF ground state.
Within PIMC, on the other hand, only the correlation functions yield structural information, as the spatial profiles are
subject to a random reorientation during the random walk process involved in the simulation,
apparently not hindered by phase fixing.
We find that the predictions of the PIMC method are often at odds with the UHF theory, even at small temperatures,
where essentially ground-state properties are expected to dominate.
For example, at the coupling strength where spin polarization becomes complete,
UHF predicts the reentrance of featureless rotationally invariant states,
but PIMC discerns the incipient structure of the Wigner molecule in the correlations.
In some cases, PIMC completely revises the spatial structure suggested by UHF.
For example, for $N=6$ electrons PIMC detects a transition from a hexagonal Wigner molecule to a pentagonal one at strong coupling
at some magnetic fields.
PIMC also finds correlated structures that never occur in UHF, as we will see below.
These findings also affect the validity of the post-Hartree-Fock wave functions derived for the Wigner molecules
as obtained by projecting the symmetry-breaking UHF state to a fixed angular momentum subspace
\cite{Yannouleas01,Yannouleas02,Yannouleas02b,Yannouleas03c,DeGiovannini07,DeGiovannini08},
as these wave functions inherit the symmetry of the UHF ground state.

In spite of the valuable structural information, however, the PIMC method with phase fixing is unable to yield a
complete revised phase diagram at present.
The reason is that PIMC is performed in subspaces of fixed $z$ component of the total spin, and in the absence of a reliable
estimator of the (conditional) free energy in these subspaces, the information on the spin remains limited.

The paper is structured as follows.
In Sec.~\ref{sec:model}, we introduce the quantum dot model we study, and define the dimensionless parameters that are relevant in the
analysis.
In Sec.~\ref{sec:methods}, we review the path-integral Monte Carlo method, the unrestricted Hartree-Fock method,
and the use of phase fixing in PIMC.
In Sec.~\ref{sec:results}, we present the predictions of these methods in detail for small dots, containing $N=3$ (quantum
dot lithium) to $N=8$ electrons (quantum dot oxygen).
We summarize our findings and conclude on the phase-fixed PIMC method in Sec.~\ref{sec:conclusion}.
The cumulant approximation to the action in the presence of an external magnetic field,
which is the main technical invention of this paper,
is derived in the Appendix both for the Coulomb interaction and the harmonic confinement potential.


\section{Model}
\label{sec:model}

We study a two-dimensional quantum dot of rotational symmetry; we assume that the confinement potential is parabolic.
The Hamiltonian is
\begin{multline}
  \mathcal H=\sum_{i=1}^N\frac{(\mathbf p_i - e\mathbf A(\mathbf r_i))^2}{2m^\ast} + \frac{1}{2}m^\ast\omega_0^2\sum_{i=1}^N r_i^2\\
  + \frac{e^2}{4\pi\epsilon_r\epsilon_0}\sum_{i<j}\frac{1}{|\mathbf r_i-\mathbf r_j|},
\end{multline}
where we use the symmetric gauge $\mathbf A(\mathbf r)=\frac{B}{2}(-y,x,0)$; $\hbar\omega_0$ is the confinement energy,
and $\epsilon_r$ and $m^\ast$ are material-specific parameters (e.g., 12.7 and 0.067 m$_e$ in GaAs, respectively).
The noninteracting part of the Hamiltonian can also be written, for a single particle for simplicity, as
\begin{equation}
  \mathcal H_0=-\frac{\hbar^2}{2m^\ast}\nabla^2+\frac{1}{2}m^\ast\omega^2 r^2-\frac{\omega_c}{2} L_z,
\end{equation}
where $\omega_c=\frac{eB}{m^\ast}$ is the cyclotron frequency, $\omega=\sqrt{\omega_0^2 + \omega_c^2/4}$,
and $L_z$ is the angular momentum perpendicular to the plane of the electron system.
We will use $\ell=\sqrt\frac{\hbar}{m^\ast\omega}$ as the unit of length throughout this paper.
The eigenstates of $\mathcal H_0$ are the well-known Fock-Darwin orbitals,
\begin{equation}
  \label{eq:fockdarwinstate}
\eta_{nl}(\mathbf r)=\frac{e^{il\phi}}{\sqrt{2\pi}}\sqrt\frac{n!}{(n+|l|)!}e^{-r^2/4}
\left(\frac{r}{\sqrt2}\right)^{|l|}L^{|l|}_n\left(\frac{r^2}{2}\right),
\end{equation}
where $l$ is the eigenvalue of $L_z$.
The corresponding energy is
\begin{equation}
  \label{eq:fockdarwinenergy}
\epsilon_{nl} = \hbar\omega(2n+|l|+1)-\frac{\hbar\omega_c}{2}l.
\end{equation}
In the limit $\omega_0\to0$, $\epsilon_{nl}$ tends to the $l$-independent quantized energies of Landau levels,
while in the $B\to0$ limit it becomes identical to the energy of the two-dimensional harmonic oscillator.

The interacting electron system in the quantum dot is characterized by two dimensionless parameters:
the ratio of the Coulomb energy scale to the confinement,
\begin{equation}
\lambda=\frac{e^2/(4\pi\epsilon_r\epsilon_0\ell)}{\hbar\omega}=\frac{\ell}{a^\ast_\text{B}},
\end{equation}
where $a^\ast_\text{B}=\frac{\hbar^24\pi\epsilon_r\epsilon_0}{m^\ast e^2}$ is the effective Bohr radius of the host semiconductor,
and the dimensionless measure of the magnetic field,
\begin{equation}
\gamma=\frac{\omega_c}{\omega}.
\end{equation}
[Sometimes the ratios $\lambda_0=\frac{e^2/(4\pi\epsilon_r\epsilon_0\ell_0)}{\hbar\omega_0}$ and
$\gamma_0=\frac{\omega_c}{\omega_0}$ are used, where $\ell_0=\sqrt\frac{\hbar}{m^\ast\omega_0}$
is the oscillator length of the confinement potential.
Experimentally, neither $\omega$ nor $\omega_0$ is constant, as the latter depends on the gate voltage
and the former also depends on the magnetic field.]


\section{Methods}
\label{sec:methods}

The path-integral Monte Carlo method \cite{Ceperley95} evaluates physical quantities as
derivatives of the partition function; the latter is expressed as an imaginary-time path integral,
and path integration is performed by standard Monte Carlo methods.
We consider the thermal density matrix of the system,
\begin{equation} 
\label{eq:dm}
\rho(R,R';\beta) = \sum_{n}e^{-\beta\epsilon_n}\Psi_n(R)\Psi^\ast_n(R'),
\end{equation} 
where
$\beta=\frac{1}{k_\text{B}T}$ is the inverse temperature,
$R\equiv(\mathbf r_1,\mathbf r_2,\dots,\mathbf r_N)$ collects $dN$ particle coordinates,
$d$ is the dimensionality of the system,
$N$ is the number of particles,
$\{\Psi_n\}$ is a complete set of many-body eigenstates,
and $\{\epsilon_n\}$ are the corresponding energies.
We express $\rho(R,R';\beta)$ in terms of density matrices that correspond to a higher temperature:
\begin{multline}
\rho(R,R';\beta) = \int dR_1\cdots\int dR_{M-1} \rho(R,R_1;\tau)\\
\times\rho(R_1,R_2;\tau)\dots\rho(R_{M-1},R';\tau).
\end{multline}
Here, $\tau\equiv\beta/M$ is the imaginary time step between adjacent slices.
There are several plausible schemes for approximating $\rho(R_{m-1},R_m;\tau)$ \cite{Ceperley95}.
The simplest of these, the so-called primitive action, is based on the Suzuki-Trotter formula,
and it results in the product of a purely kinetic and an interaction part.
Unfortunately, the primitive approximation to the action is well-known to be inadequate
for the Coulomb interaction because of the sharp divergence of the latter at short range \cite{Ceperley95}.
Here, we adopt the cumulant approximation to the action \cite{Ceperley83}.
Its standard formulation, however, has to be changed in the presence of an external magnetic field, as sketched below
and discussed in detail in Appendix.

In the presence of an external magnetic field, the density matrix unavoidably becomes complex-valued.
The estimators of physical quantities are then obtained as sums of complex values.
This results in an extremely low signal-to-noise ratio in the Monte Carlo integration, even for small systems.
We avoid this problem by phase fixing \cite{Ortiz93,Bolton96,Haidekker18}.
In the case of PIMC, we have to fix the phase of the many-body density matrix in Eq.~(\ref{eq:dm}), not of a wave function.
The paths are sampled by the probability density function
$\prod_{m=1}^{M}|\rho(R_{m-1},R_m;\tau)|$, where we assume that the diagonal density matrix is the integration kernel of the estimator,
and we let $R=R'\equiv R_0=R_M$.
Using a trial (or guide) many-body density matrix $\rho_T(R,R';\beta) =|\rho_T(R,R';\beta)|e^{i\varphi_T(R,R';\beta)}$ for fixing the phase,
this means that an additional effective interaction
\begin{equation}
V_\text{eff}(R) = \frac{\hbar^2}{m^\ast}\left(\nabla\varphi_T(R,R';\beta) - \frac{e}{\hbar}A(R)\right)^2
\end{equation}
appears in the action, besides the physical potential terms.
Notice that $|\rho_T(R,R';\beta)|$ plays no role in this procedure.
While for the effective interaction a semiclassical approximation is used \cite{Haidekker18},
i.e., $V_\text{eff}$ is evaluated on a straight line that connects $R$ and $R'$,
for the harmonic confinement and for the Coulomb repulsion among electrons, we use the cumulant approximation to the action.
Single-particle propagation in the presence of an external magnetic field is described by the free density matrix
\begin{multline}
\label{eq:openbc}
\rho_0(\mathbf r,\mathbf r';\beta)=\frac{1}{2\pi\ell_c^2}\frac{\sqrt u}{1-u}\\
\times\exp\left(-\frac{1+u}{1-u}\frac{\left|\mathbf r-\mathbf r'\right|^2}{4\ell_c^2}+
\frac{i(x'y-xy')}{2\ell_c^2}
\right),
\end{multline}
where $\ell_c=\sqrt{\frac{\hbar}{eB}}$ is the magnetic length and $u=e^{-\beta\hbar\omega_c}$.
The cumulant approximation is
\begin{equation}
\label{eq:cumulant1}
U_\text{C}(R_0,R_1,\tau)=\tau\int dR \mu_\text{C}(R|R_0,R_1,\tau)V(R),
\end{equation}
where the complex amplitude of visiting configuration $R$ sometime during the propagation of the electron configuration
from $R_0$ to $R_1$ is
\begin{equation}
\label{eq:cumulant2}
\mu_\text{C}(R|R_0,R_1,\tau)=\frac{1}{\tau}\int_0^\tau dt \mu(R,t|R_0,R_1,\tau),
\end{equation}
with
\begin{multline}
\label{eq:cumulant3}
\mu(R,t|R_0,R_1,\tau) =\\
\frac{\left(\prod_{i=1}^N\rho_0(\mathbf r_{0,i},\mathbf r_i;t)\right)
    \left(\prod_{i=1}^N\rho_0(\mathbf r_{i},\mathbf r_{1,i};\tau-t)\right)}
     {\prod_{i=1}^N\rho_0(\mathbf r_{0,i},\mathbf r_{1,i};\tau)}.
\end{multline}
It is easy to check that $\mu$ is gauge-independent, which transfers to $\mu_\text{C}$ and to the cumulant action $U_\text{C}$.
The evaluation of Eqs.~(\ref{eq:cumulant1})-(\ref{eq:cumulant3}) is described in detail in Appendix.

We will use the outcome of a preliminary unrestricted Hartree-Fock calculation for phase fixing.
The UHF method is quite standard and it has been applied to quantum dot problems many times
\cite{Pfannkuche93,Palacios94,Fujito96,Yannouleas99,Yannouleas00b,Yannouleas03b,Reusch01,Reusch03}.
We just note that we use UHF in the basis of Fock-Darwin states, Eq.~(\ref{eq:fockdarwinstate}),
calculate the Coulomb matrix elements following Ref.~\onlinecite{Anisimovas98}.
The number of basis states was $N_\text{basis}=55$ for $N\le6$ and  $N_\text{basis}=66$ for $N=7,8$.
As the result of the self-consistent solution of the Pople-Nesbet-Berthier equations in each $S_z=(N_\uparrow-N_\downarrow)/2$ sector,
where $N_\uparrow+N_\downarrow=N$, two sets of orbitals
\[
\{\eta^{\text{HF},\uparrow}_i(\mathbf r)\}_{i=1,\dots,N_\text{basis}}\quad\text{and}\quad
\{\eta^{\text{HF},\downarrow}_i(\mathbf r)\}_{i=1,\dots,N_\text{basis}}
\]
are obtained for each spin.
We assume these orbitals are ordered in increasing order of the corresponding eigenvalues
$\{\epsilon^\uparrow_i\}$ and $\{\epsilon^\downarrow_i\}$ for $i=1,\dots,N_\text{basis}$.
The UHF wave function for the choice $\Omega_\uparrow$, $\Omega_\downarrow$ of orbitals is
\begin{multline}
  \Psi^\text{HF}_{\Omega_\uparrow,\Omega_\downarrow}(R)=
  \text{Det}\left[\eta^{\text{HF},\uparrow}_i(\mathbf r_j)_{i\in\Omega_\uparrow,1\le j\le N_\uparrow}\right]\\
  \times\text{Det}\left[\eta^{\text{HF},\downarrow}_i(\mathbf r_{N_\uparrow+j})_{i\in\Omega_\downarrow,1\le j\le N_\downarrow}\right].
\end{multline}
In particular, the UHF ground state $\Psi^\text{HF}_0(R)$ is specified by $\Omega_\uparrow=\{1,2,\dots,N_\uparrow\}$
and $\Omega_\downarrow=\{1,2,\dots,N_\downarrow\}$.
These mean-field wave functions are unrestricted in the sense that they are not necessarily eigenstates of either
the magnitude of the total spin $S^2$ or the total angular momentum $L_z=\sum_{i=1}^N L_{z,i}$.
At the same time, the UHF ground state typically has lower energy than the restricted Hartree-Fock ground-state,
which has well-defined total spin and angular momentum quantum numbers.
UHF correctly predicts the ``magic numbers'' $N=2,6,12,20,\dots$ in the addition energies,
and it justifies Hund's rules for deciding the ground-state spin at zero magnetic field.
A UHF calculation often informs us about the internal structure of the symmetry breaking state even at strong interactions,
where the validity of mean-field theories may be questionable.
Notably, the post-Hartree-Fock projection techniques
\cite{Yannouleas01,Yannouleas02,Yannouleas02b,Yannouleas03c,DeGiovannini07,DeGiovannini08}
that restore the correct symmetry of the wave function take an unrestricted Hartree-Fock ground state as their starting point.

We stress that a UHF calculation is just a preliminary step to our quantum Monte Carlo approach.
Even though a large number of UHF studies of quantum dots exist in the literature,
a thorough exploration of the charge and spin structures in terms of the magnetic field ($\gamma$ or $\gamma_0$)
and the coupling strength of the Coulomb interaction ($\lambda$ or $\lambda_0$) is apparently not available, even for small dots.
This is understandable, as UHF has important limitations and its predictions should always be handled with care.
Nevertheless, as the output of UHF is an input to our correlated PIMC calculations, we need a
more or less complete UHF phase diagram, even if it has known shortcomings.
For example,
UHF overestimates the energy of spin unpolarized states more than that of fully spin polarized ones \cite{Pfannkuche93},
hence it locates the spin transitions at too small $\lambda$'s.
At high values of $\lambda$, the lowest UHF energy in distinct $S_z$ sectors differ only slightly,
hence the prediction of the spin by UHF therefore becomes somewhat ambiguous.
The spin-singlet--spin-triplet oscillations for $N=2$ are not reproduced by UHF, nor are the
more complicated spin recurrence patterns for higher electron numbers \cite{Bolton96,Mikhailov02,Tavernier03}.
And we will see cases where the spin $S_z$ of the UHF ground state in the strong coupling limit
decreases from the highest value $S_z=N/2$, which occurs when the spatial structure of the ground state also shows an unusual symmetry.

One easy choice for the trial density matrix is in terms of the UHF ground state,
\begin{equation}
  \label{eq:uhftrial0}
  \rho_\text{T,0}(R,R';\beta) = \Psi^\text{HF}_0(R)\Psi^\text{HF}_0(R')^\ast.
\end{equation}
For small systems, one can try to build a trial density matrix from all UHF wave functions:
\begin{equation}
  \label{eq:uhftrial}
  \rho_\text{T}(R,R';\beta) = \sum_{\Omega_\uparrow,\Omega_\downarrow}e^{-\beta E_{\Omega_\uparrow,\Omega_\downarrow}}
  \Psi^\text{HF}_{\Omega_\uparrow,\Omega_\downarrow}(R)\Psi^\text{HF}_{\Omega_\uparrow,\Omega_\downarrow}(R')^\ast,
\end{equation}
where $E_{\Omega_\uparrow,\Omega_\downarrow}$ is the energy of state $\Psi^\text{HF}_{\Omega_\uparrow,\Omega_\downarrow}$.

Note that both the UHF and the PIMC calculations are performed in fixed $S_z$ subspaces.
As UHF is a ground-state method, the true UHF ground state is found by the comparison of the lowest
energy states in each subspace.
Thus jumps in $S_z$ yield partial information on the total spin $S\ge S_z$ of the ground state.

Regarding PIMC, we note that permutations are sampled for up and down-spins separately.


\section{Results}
\label{sec:results}

We will focus on the effect of the electron-electron interaction and neglect the Zeeman energy completely.
The latter is known to be small in GaAs structures ($g=-0.44$), but it is significant in InSb quantum dots ($g\approx-50$).
The Zeeman energy can be reintroduced trivially if necessary; it always favors states of greater total spin, but leaves
the spatial structure intact within the spin-related phase domains.

\begin{figure}[htbp]
\begin{center}
 \def\svgwidth{\columnwidth}
 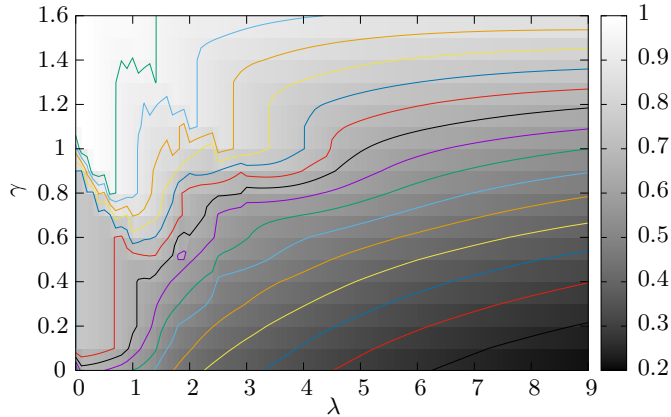
\end{center}
\caption{\label{fig:uhflll}
  The relative weight of the UHF ground state in the lowest Landau level, i.e.,
  $\left\langle\Psi_0^\text{HF} | \mathcal P_\text{LLL} |\Psi_0^\text{HF}\right\rangle/N$ for $N=8$ electrons (quantum dot oxygen),
  where $\mathcal P_\text{LLL}$ is the projection to the lowest Landau level.
  The contours start at 0.25 by the arc around the lower right corner; the increment is 0.05.
}
\end{figure}

Unrestricted Hartree-Fock calculations are performed at fixed $\gamma$ as $\lambda$ is increased from 0 to 9 in steps of 0.1.
We use the converged orbitals of the previous $\lambda$ value as the starting point of the iterative solution
of the UHF problem at $\lambda+0.1$.
When the symmetry of the ground state changes during such an upward sweep, we perform a downward sweep
starting with the orbitals of the new symmetry.
The goal is to avoid metastable solutions of Pople-Nesbet-Berthier equations.
The UHF ground state is found eventually by energy comparison for the two sweeps.
The $B=0$ limit is included in the phase diagrams for comparison with the literature,
but our primary focus is on $B>0$.
As Fig.~\ref{fig:uhflll} shows for the case of $N=8$, the ground state is not predominantly in the lowest Landau
level, at least in the UHF approximation, for moderate values of $\gamma$.

In PIMC simulations, we use a time step $\tau\hbar\omega=0.01$,
hence the number of slices is between 80 and 320 for $\beta^\ast=\beta\hbar\omega=0.8$ to 3.2.
We have checked that the predictions stabilize at this time step.
We update 7, 15, or 31 slices in a multi-slice move, as explained in Ref.~\onlinecite{Haidekker18}.

While ordering is manifest in the charge and spin densities by UHF, we cannot expect this to be the case by PIMC.
We gain insight into the ordering through the angular correlation function (ACF)
\begin{equation}
  g(\theta)=\left\langle\sum_{i<j}\delta(\theta - \theta_{i,j})\right\rangle,
\end{equation}
where $\theta_{i,j}=\cos^{-1}\left(\frac{\mathbf r_i\cdot\mathbf r_j}{r_i r_j}\right)$ is the angle between
two electron positions, taken as nonnegative.
We will also use the spin-up ACF $g_{\uparrow\uparrow}(\theta)$,
the spin-down ACF $g_{\downarrow\downarrow}(\theta)$,
the different-spin ACF $g_{\uparrow\downarrow}(\theta)$, and
the same-spin ACF $g_\text{same}(\theta)=g_{\uparrow\uparrow}(\theta)+g_{\downarrow\downarrow}(\theta)$.
We also often plot the angle-averaged radial density,
\begin{equation}
\rho(r) = \frac{1}{2\pi}\int_0^{2\pi} \rho(r,\phi) d\phi,
\end{equation}
and its spin-specific versions $\rho_\uparrow(r)$ and $\rho_\downarrow(r)$.

Notice that the quantum dot problem can also be discussed classically \cite{Bedanov94}.
For some particle numbers, e.g., $N=3,7,10,12,19$, we get especially stable configurations, which can be
interpreted as finite pieces of a triangular lattice; while for others like $N=5,6,8$, there are
several classical configurations that are very close in energy \cite{Reusch03}.
The quantum mechanical picture is definitely more complicated, primarily because of zero point motion and the presence of the spin.
Nevertheless we will see below that the correlated behavior is more complex in the latter cases.


\subsection{Quantum dot lithium, $N=3$}

\begin{figure}[htbp]
\begin{center}
\includegraphics[width=0.48\columnwidth, keepaspectratio]{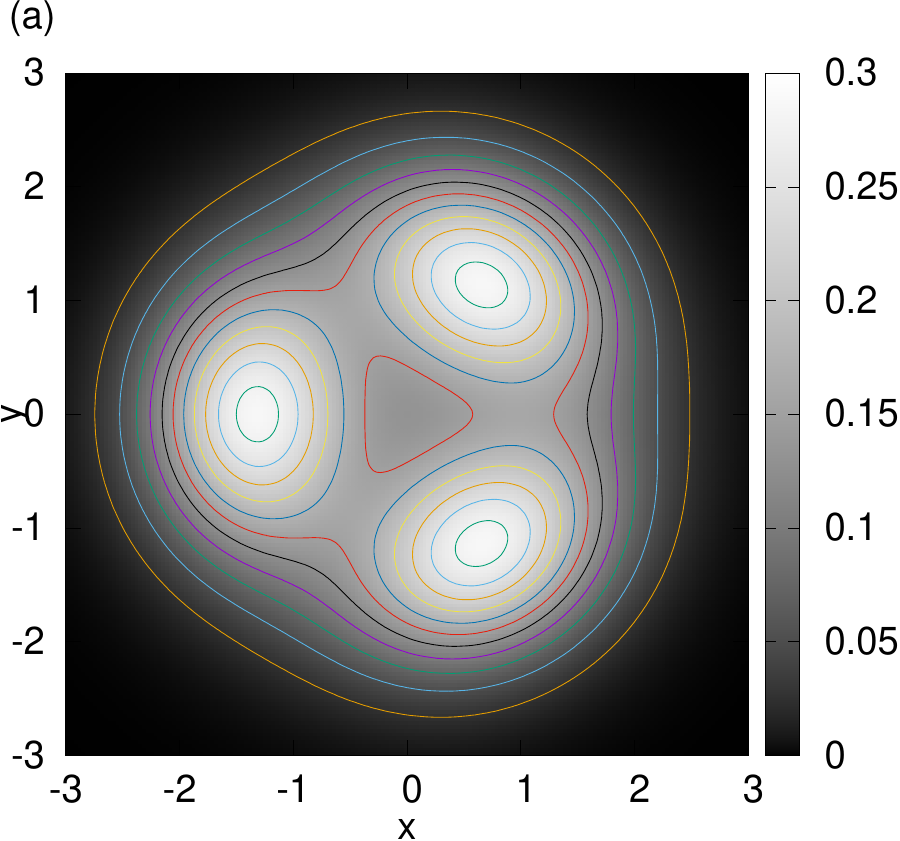}
\includegraphics[width=0.48\columnwidth, keepaspectratio]{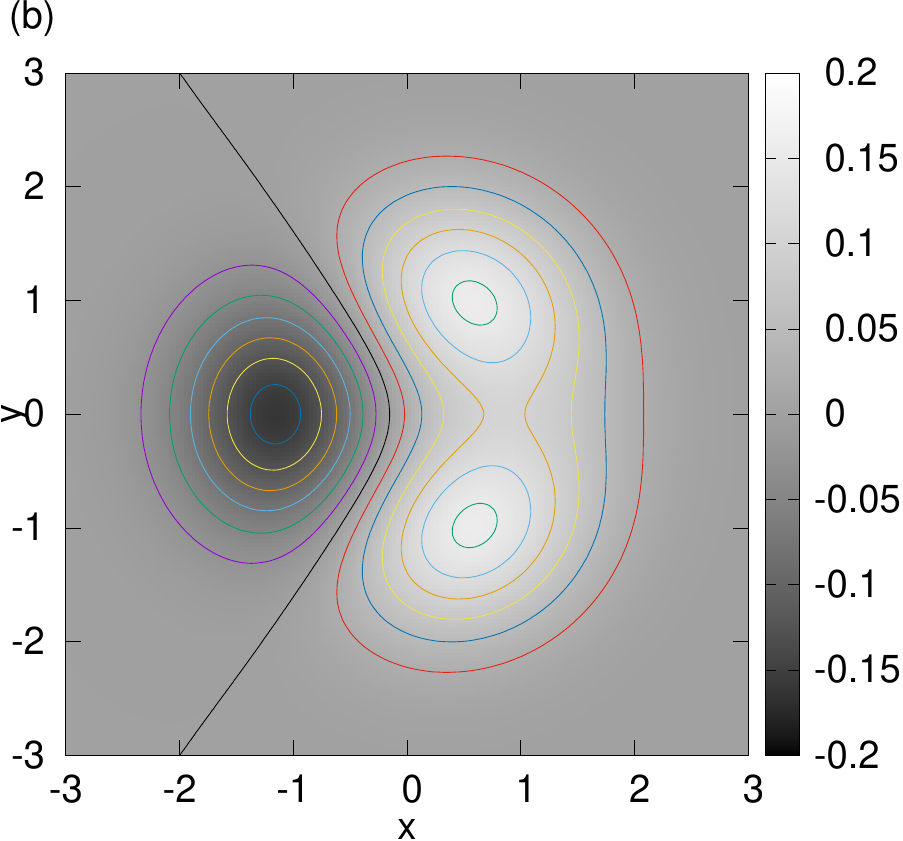}
\end{center}
\caption{\label{fig:lithium}
  Examples of UHF ground-state structures in quantum dot lithium.
  (a) The particle density at $\gamma=0.6$ and $\lambda=3$ in state $(\frac{3}{2},C_{3v})$.
  (b) The spin density at $\gamma=0.6$ and $\lambda=1.8$ in state $(\frac{1}{2},C_{s})$.
}
\end{figure}

For three particles there are four relevant ground-state structures.
We denote them by the combination of the $z$-component of the spin and the symmetry group of the spin density.
(Note that this may not be the same as the symmetry of the wave function.)
(i) The fully polarized, rotationally invariant state is called $(\frac{3}{2},C_\infty)$.
(ii) The fully polarized, broken symmetry state is called $(\frac{3}{2},C_{3v})$.
(iii) The partially polarized, rotationally invariant state is called $(\frac{1}{2},C_\infty)$.
(iv) The partially polarized, broken symmetry state is called $(\frac{1}{2},C_{s})$.
Figure \ref{fig:lithium} shows examples for the symmetry-breaking states $(\frac{3}{2},C_{3v})$ and $(\frac{1}{2},C_{s})$.

\begin{figure}[htbp]
\begin{center}
\includegraphics[width=\columnwidth,keepaspectratio]{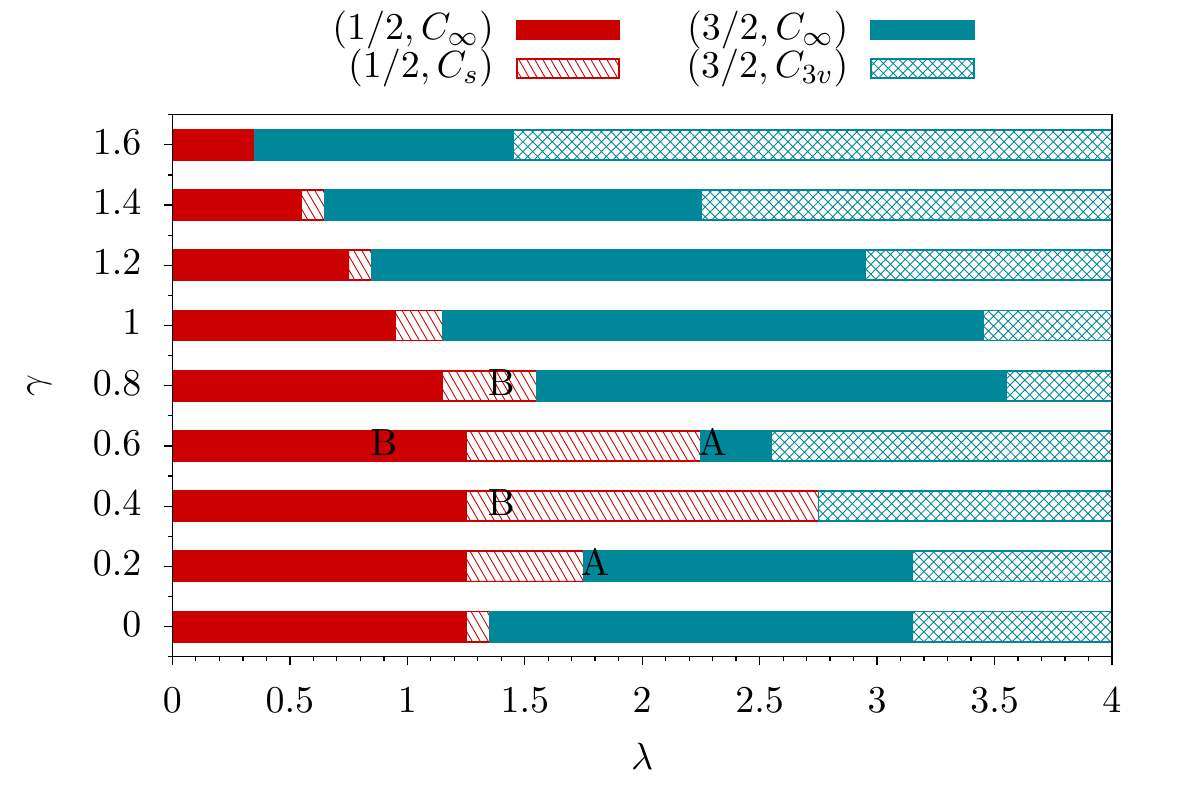}
\end{center}
\caption{\label{fig:lithiumphase}
  Sections of the UHF phase diagram of quantum dot lithium ($N=3$) as a function of the magnetic field parameter $\gamma$
  and the coupling strength $\lambda$.
  Labels A and B indicate PIMC parameters in Figs.~\ref{fig:mfp3}~and~\ref{fig:mpp}, respectively.
}
\end{figure}

For $B>0$, all four structures are relevant in the UHF approximation.
Figure \ref{fig:lithiumphase} gives the UHF phase diagram.
At fixed $\gamma$, the four states typically occur in the sequence $(\frac{1}{2},C_\infty)$,
$(\frac{1}{2},C_{s})$, $(\frac{3}{2},C_\infty)$, $(\frac{1}{2},C_{s})$ in increasing order of $\lambda$,
with two exceptions.
(i) For $0.4\lesssim\gamma\lesssim0.5$,
the $S_z=\frac{3}{2}$ system is already symmetry-broken when the ground state becomes spin-polarized,
thus the spin-flip at $\lambda=2.8$ connects two symmetry-broken states.
(ii) For $\gamma\ge1.5$ the range of $(\frac{1}{2},C_s)$ disappears, i.e.,
the symmetry-breaking occurs only within the fully spin-polarized range;
this behavior is typical for large $\gamma$ for all system sizes.

\subsubsection{The $S_z=\frac{3}{2}$ subspace of $N=3$}

The reentrance of rotational symmetry with the spin-flip transition deserves further attention.
Hence we performed systematic PIMC simulations for all $\gamma>0$ where this occurred.

\begin{figure*}[htbp]
\begin{center}
  \includegraphics[width=.64\textwidth, keepaspectratio]{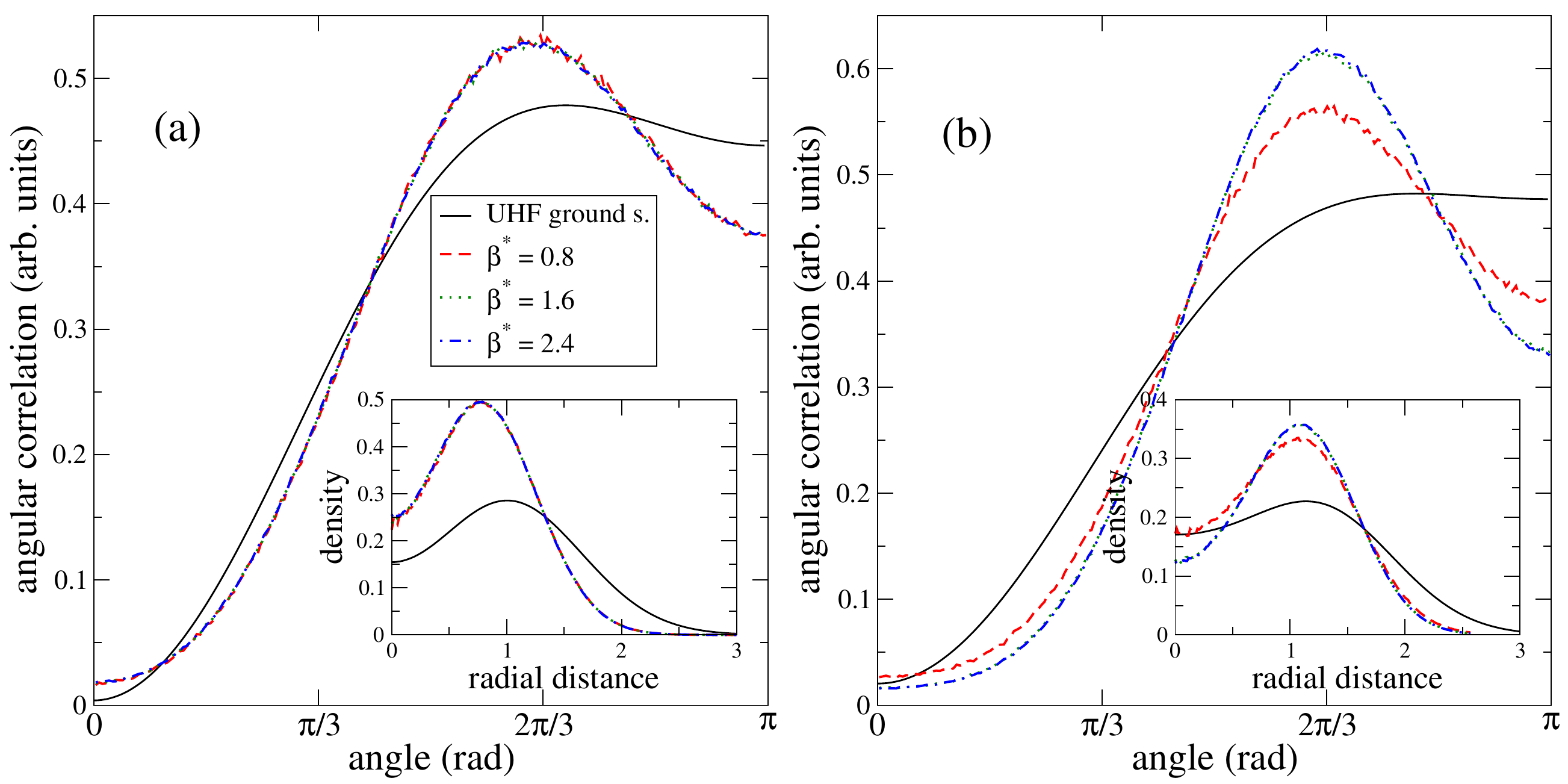}
\end{center}
\caption{\label{fig:mfp3}
  (a) The angular correlation function $g(\theta)$ for quantum dot lithium ($N=3$) in the fully
  spin-polarized sector at $\gamma=0.2$ and $\lambda=1.8$.
  (b) The same at $\gamma=0.6$ and $\lambda=2.3$.
  The radial density is shown in the insets. Label A in Fig.~\ref{fig:lithiumphase}.
}
\end{figure*}

Figure \ref{fig:mfp3} shows that the angular correlation is strongly peaked near $\theta=2\pi/3$ already at the
smallest coupling $\lambda$ where the spins are fully polarized,
i.e., in the vicinity of the $(\frac{1}{2},C_s)$-$(\frac{3}{2},C_\infty)$ phase boundary.
This correlation peak grows further with increasing $\lambda$, and it is accompanied by the relative depletion of the
density near the origin, in comparison with the UHF ground state.
The correlated motion of electrons also allows for a smaller diameter of the cloud.
The angular correlation evolves continuously across the $\lambda$ value where the $C_{3v}$ structure
becomes the UHF ground state: no sudden change is observed, even though for larger $\lambda$'s we take the guiding phase from
the symmetry-breaking UHF ground state.
We conclude that the threefold ordering that corresponds to a triangular Wigner molecule
is present throughout the spin-polarized regime; the restored rotational symmetry is a misleading result of the UHF approximation.
Note that $g(\theta)$ is independent of the inverse temperature $\beta^\ast$ for $\gamma=0.2$,
but for $\gamma=0.6$ it saturates for $\beta^\ast\ge1.6$ only.
Hence the energy scale associated with threefold ordering in the later case can be estimated.
Using $\hbar\omega_0=3.32$ meV, commonly assumed when analyzing Ref.~\onlinecite{Ashoori92}, this corresponds to
$T=\frac{2\hbar\omega_0}{\beta^\ast k_\text{B}\sqrt{4-\gamma^2}}\lesssim 25$ K,
much larger than the experimental temperature.

As increasing temperature washes out correlations, and mean-field theory neglects correlations, the high temperature PIMC
results become more similar to the UHF results, as see in Fig.~\ref{fig:mfp3}(b).
This observation holds in all cases we have checked.


\subsubsection{The $S_z=\frac{1}{2}$ subspace of $N=3$}

\begin{figure*}[htbp]
\begin{center}
\includegraphics[width=0.36\textwidth, keepaspectratio]{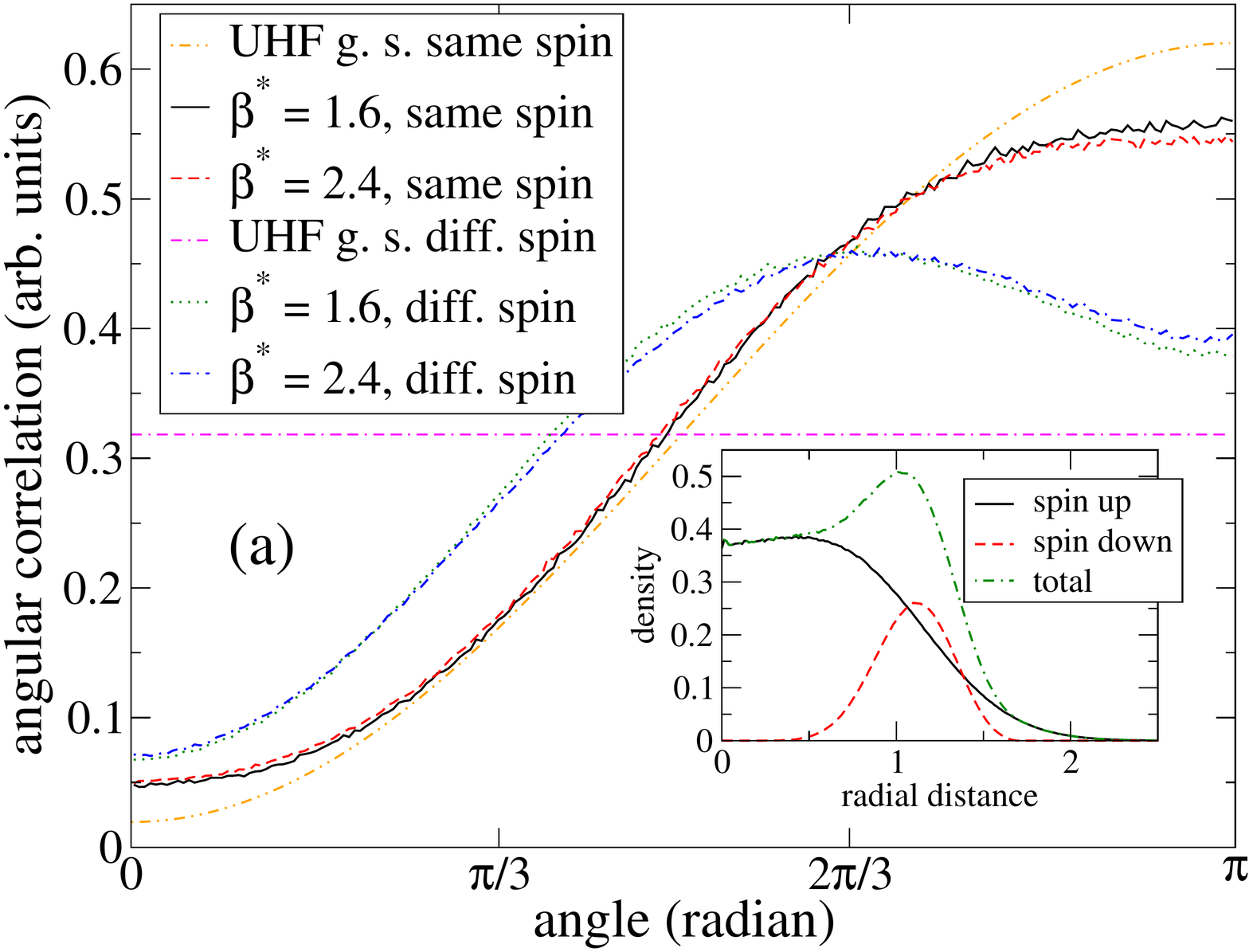}
\includegraphics[width=0.36\textwidth, keepaspectratio]{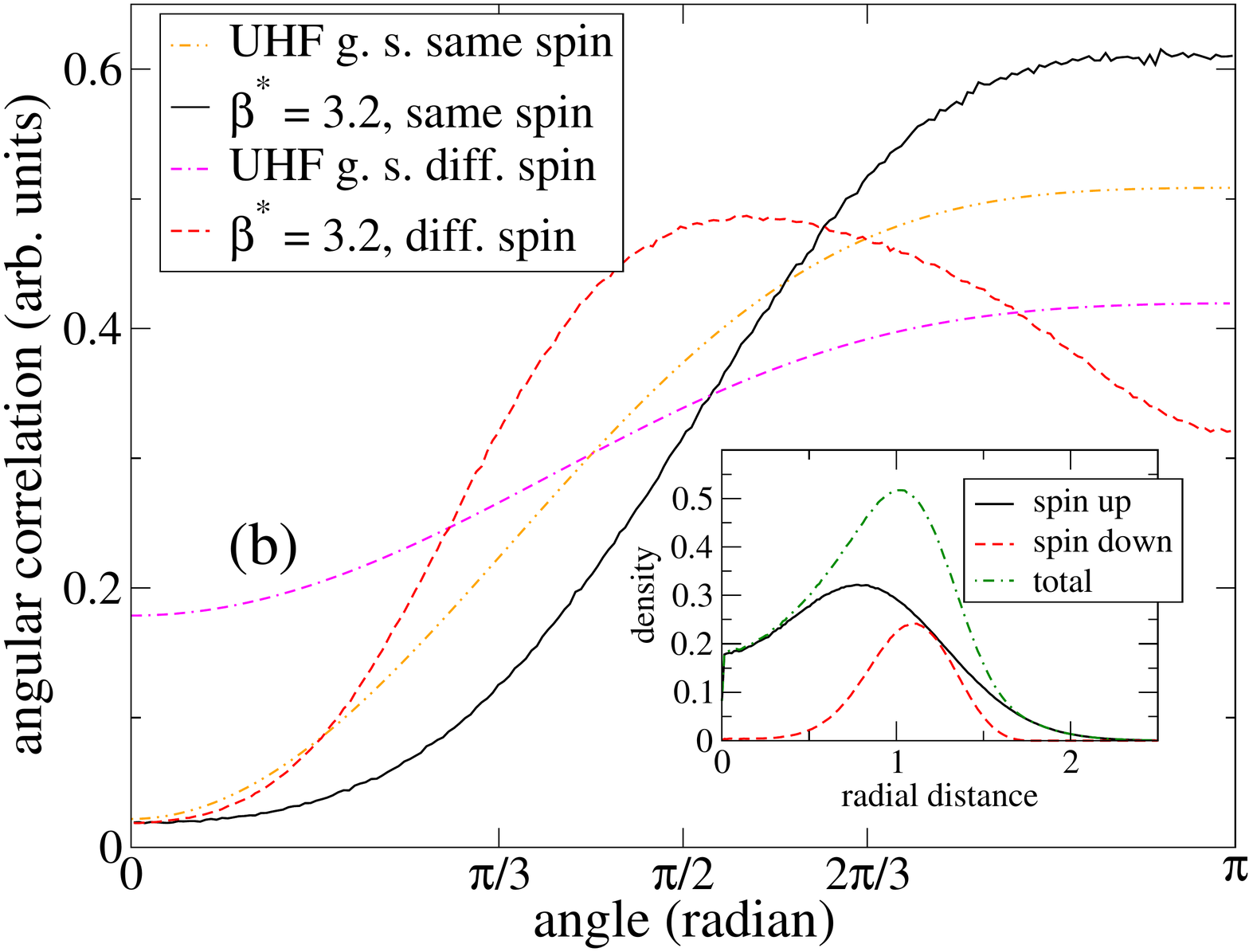}
\includegraphics[width=0.36\textwidth, keepaspectratio]{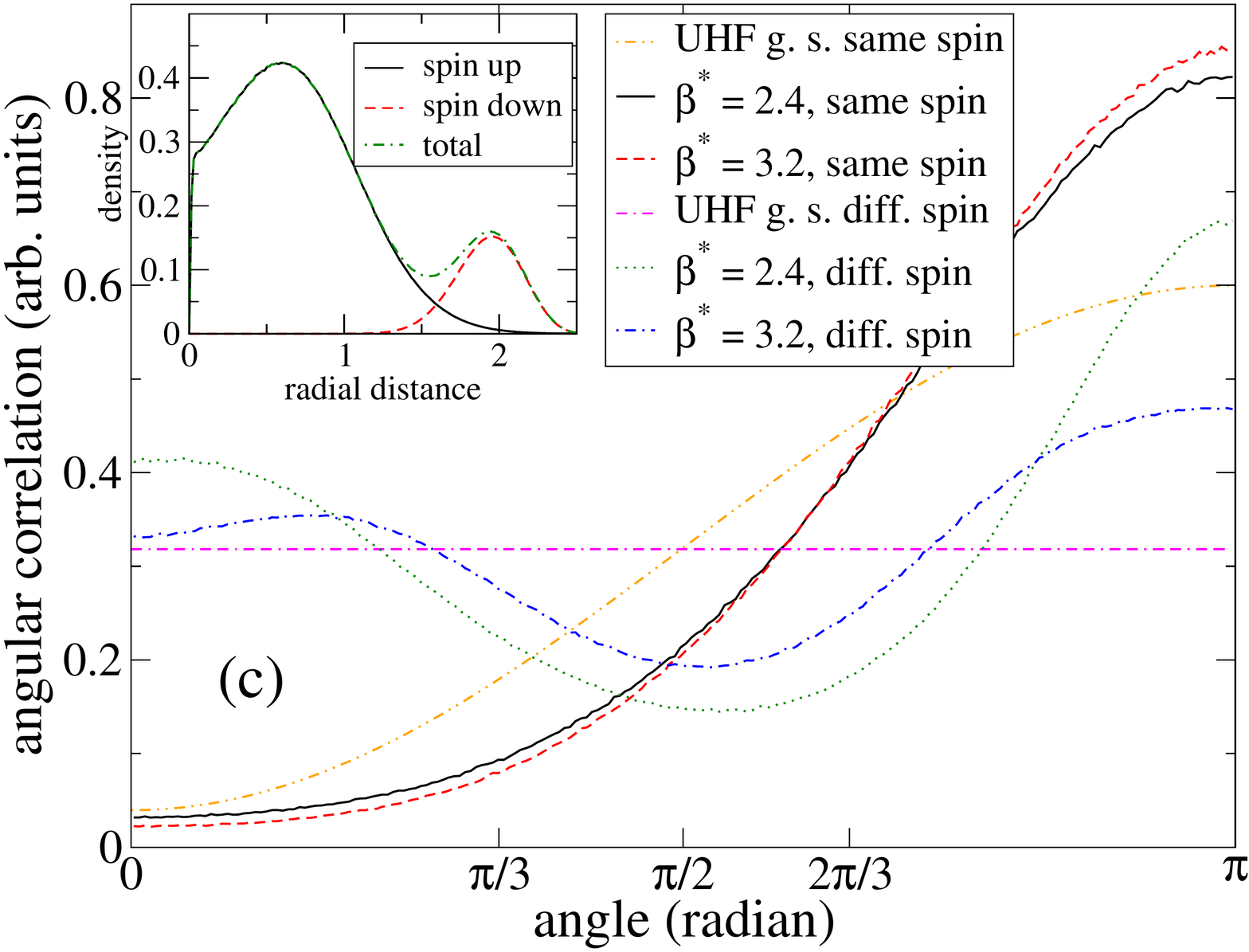}
\includegraphics[width=0.3\textwidth, keepaspectratio]{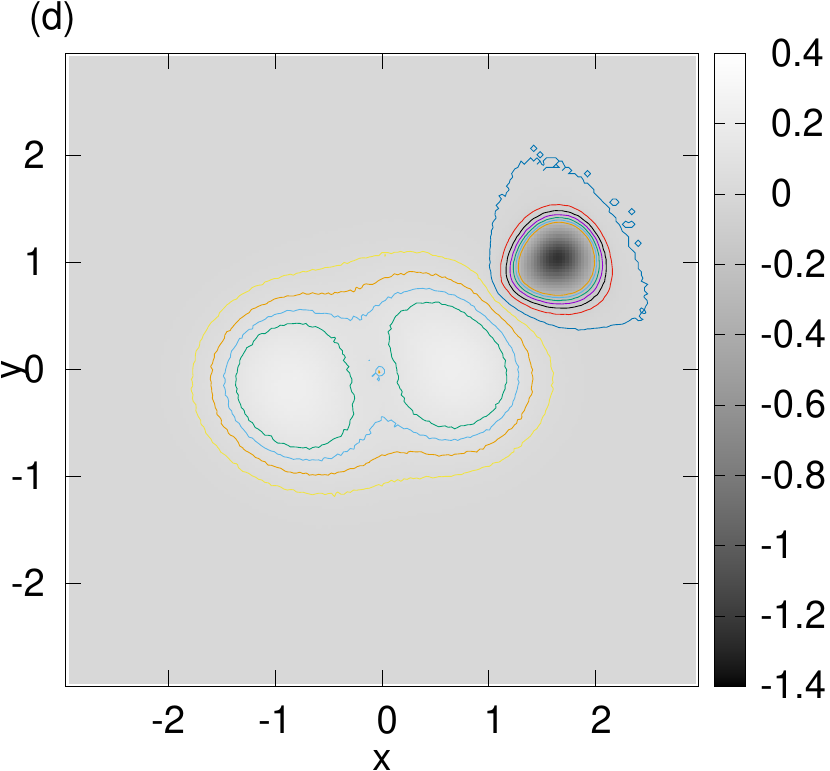}
\end{center}
\caption{\label{fig:mpp}
  [(a)-(c)] The angular correlation function $g(\theta)$ for partially spin-polarized ($S_z=\frac{1}{2}$) quantum dot lithium ($N=3$)
  for (a) $\gamma=0.6$ and $\lambda=0.9$,
  (b) $\gamma=0.8$ and $\lambda=1.4$,
  (c) $\gamma=1.4$ and $\lambda=0.4$.
  (d) Real-space spin density histogram for the parameters of (c), $\beta^\ast=3.2$.  Label B in Fig.~\ref{fig:lithiumphase}.
}
\end{figure*}

We performed PIMC on both sides of the $(\frac{1}{2},C_\infty)\to(\frac{1}{2},C_s)$ transition.
For moderate magnetic fields the correlations characteristic for the $C_s$ state --- identical spin
at large angle about the origin, opposite spins at moderate angles, testifying an isosceles triangle configuration --- already
appear at small coupling, in the $C_\infty$ range; see Fig.~\ref{fig:mpp}(a) for an example.
These correlations intensify without any sudden transformation when the UHF ground state changes to $(\frac{1}{2},C_s)$;
cf.\ Fig.~\ref{fig:mpp}(b).
On the other hand, at larger fields $\gamma=1.4$ and 1.6 we find that the minority spin electron is localized off
the more tightly bound majority spin molecule,
and tends to be on the same line as the majority spin electrons, see the $\theta=0$ and $\theta=\pi$ peaks of the different-spin ACF
in Fig.~\ref{fig:mpp}(c).
In these cases, we have not been able to reach the asymptotic behavior at the highest inverse temperature $\beta^\ast$.

\subsection{Quantum dot beryllium, $N=4$}

\begin{figure}[htbp]
\begin{center}
\includegraphics[width=\columnwidth, keepaspectratio]{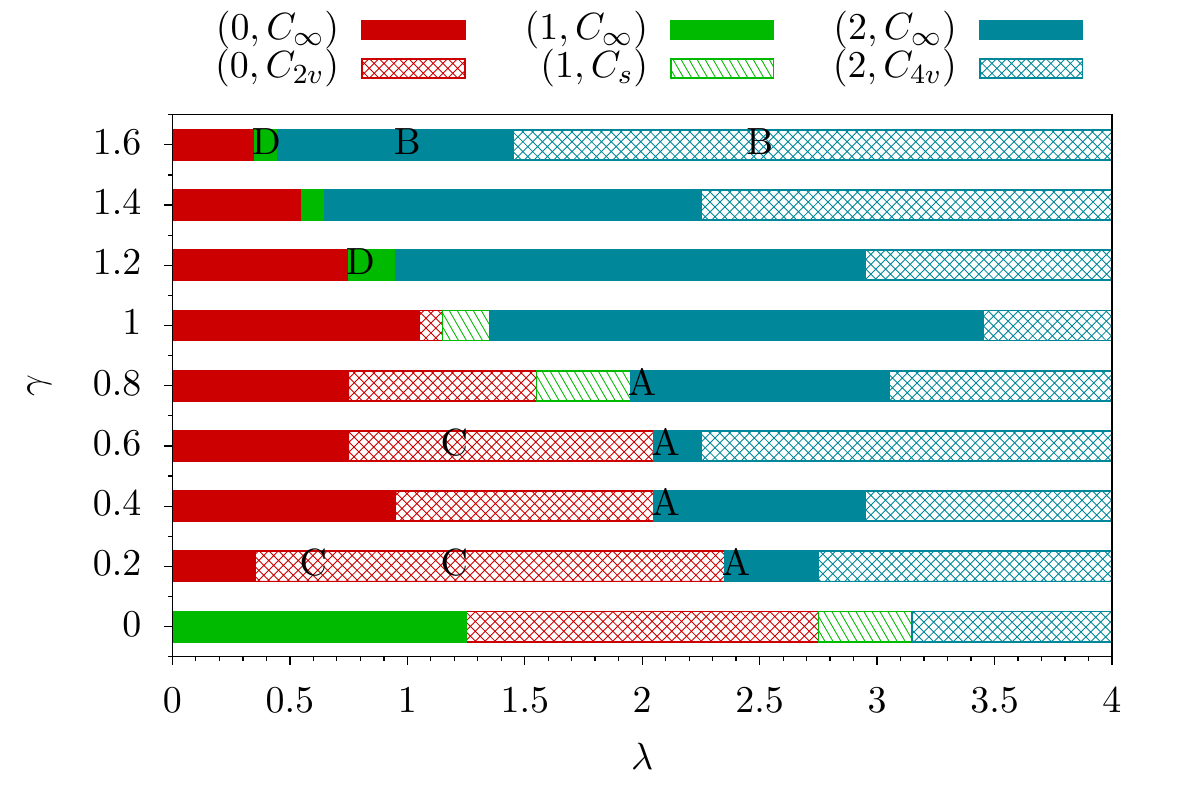}
 \def\svgwidth{.8\columnwidth}
 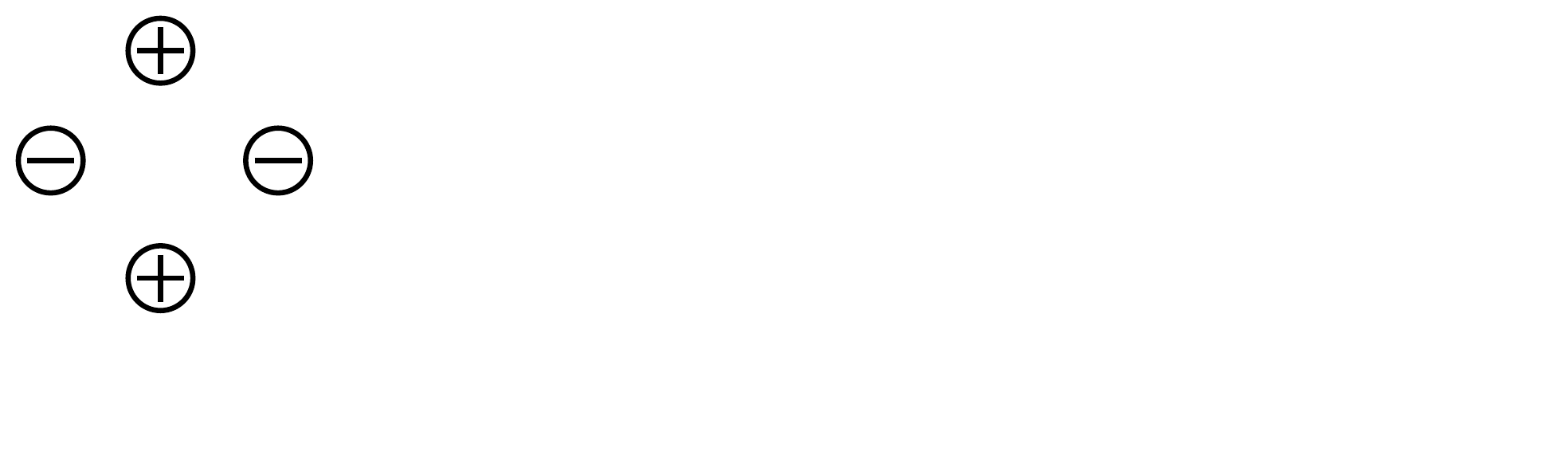
\end{center}
\caption{\label{fig:berylliumphase}
  (Top)
  Sections of the UHF phase diagram of quantum dot beryllium ($N=4$) as a function of the magnetic field parameter $\gamma$
  and the coupling strength $\lambda$.
  The $\gamma=0$ line is exceptional; here, the sequence of transitions is
  $(1,C_\infty)\stackrel{\lambda=1.3}{\to}(0,C_{2v})\stackrel{\lambda=2.8}{\to}(1,C_{s})\stackrel{\lambda=3.2}{\to}(2,C_{4v})$.
  Labels A, B, C, and D indicate PIMC parameters in Figs.~\ref{fig:mfp4}, \ref{fig:mfp4b}, \ref{fig:msz0_4}~and~\ref{fig:msz1_4},
  respectively.
  (Bottom)
  Sketch of the particle and spin density peak configurations in the relevant symmetry-breaking states of quantum dot beryllium.
  Note that $(0,C_s)$ is relevant in PIMC only.
}
\end{figure}

The relevant UHF ground-state structures are the rotationally invariant $(0,C_\infty)$, $(1,C_\infty)$, $(2,C_\infty)$ states,
and the symmetry-broken states $(0,C_{2v})$, $(1,C_s)$, $(2,C_{4v})$,
whose particle and spin density configurations are sketched in Fig.~\ref{fig:berylliumphase}.
A fourth symmetry-breaking structure $(0,C_s)$ appears in PIMC only.

For $B>0$, the weak-coupling limit has $S_z=0$, as the degenerate shell of Fock-Darwin states is destroyed by any
magnetic field, and Hund's rule no longer selects the ground state.
For relatively weak fields, $\gamma\le0.6$, the subsequent transitions follow the path
$(0,C_\infty)\to(0,C_{2v})\to(2,C_\infty)\to(2,C_{4v})$.
The $S_z=1$ states play no role here.
For $\gamma\ge0.8$, there is an intervening $S_z=1$ state between the unpolarized and the fully polarized states.
For $\gamma=0.8$ and 1, this is symmetry-broken, while for $\gamma\ge1.2$, it is rotationally invariant.
The interval where $(0,C_{2v})$ is the ground state shrinks with increasing $\gamma$ and completely disappears for $\gamma\ge1.2$.
See Fig.~\ref{fig:berylliumphase}.

\begin{figure}[htbp]
\begin{center}
\includegraphics[width=\columnwidth, keepaspectratio]{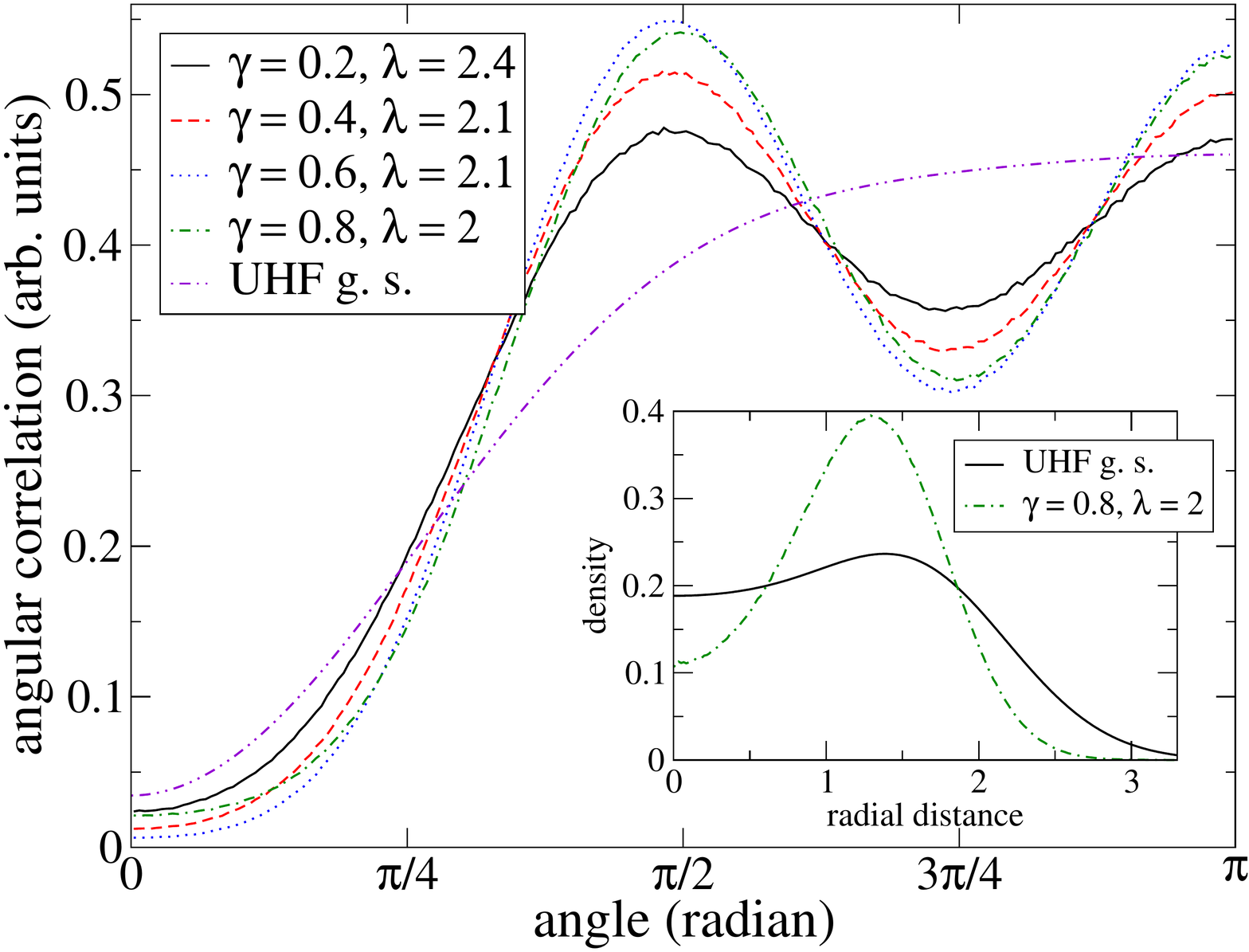}
\end{center}
\caption{\label{fig:mfp4}
  The angular correlation function $g(\theta)$ for fully spin-polarized quantum dot beryllium ($N=4$) for $\gamma=0.2$ to 0.8,
  at the coupling strength values just following the value where UHF predicts full spin-polarization.
  The radial density is shown for comparison in the inset for one case.
  $\beta^\ast=2.4$ for all PIMC results.  Label A in Fig.~\ref{fig:berylliumphase}.
}
\end{figure}
 
 
\subsubsection{The $S_z=2$ subspace of $N=4$}

\begin{figure*}[htbp]
\begin{center}
\includegraphics[width=0.36\textwidth, keepaspectratio]{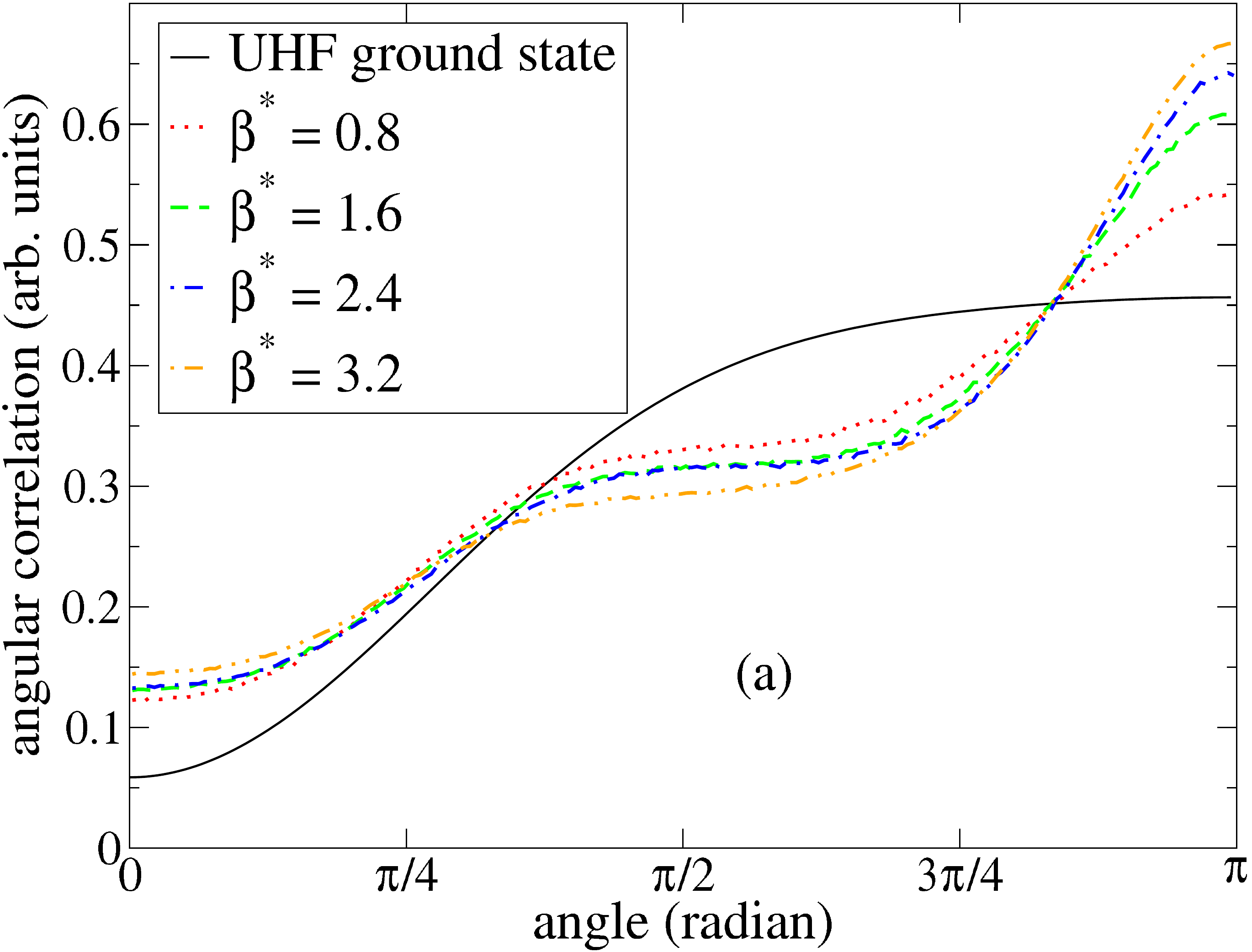}
\includegraphics[width=0.3\textwidth, keepaspectratio]{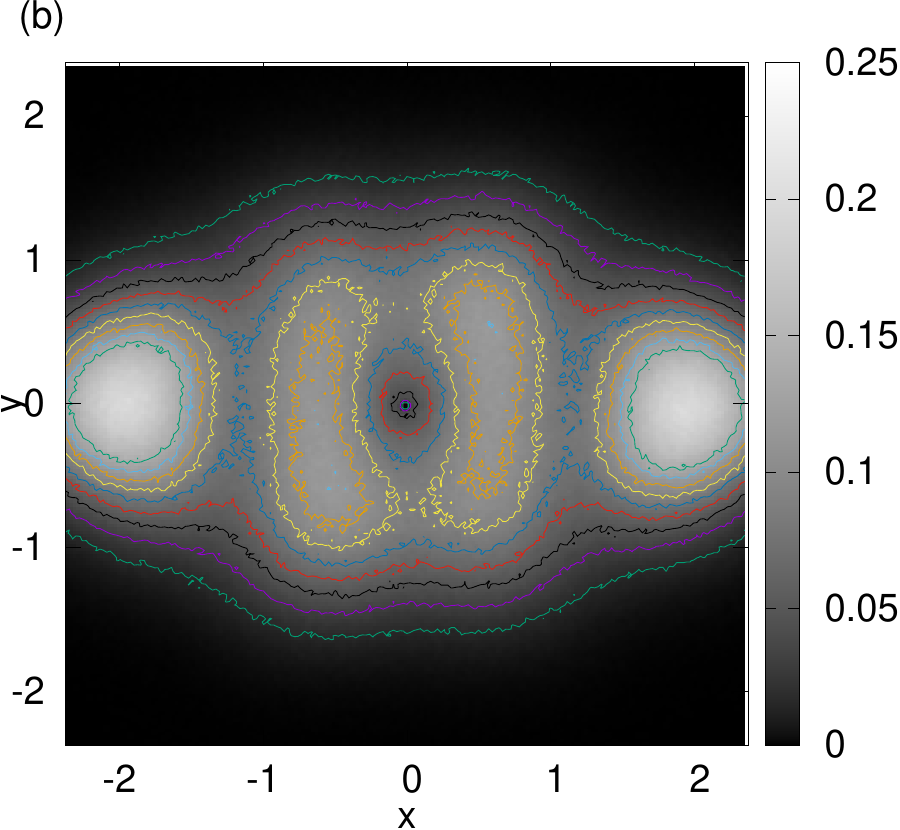}
\includegraphics[width=0.36\textwidth, keepaspectratio]{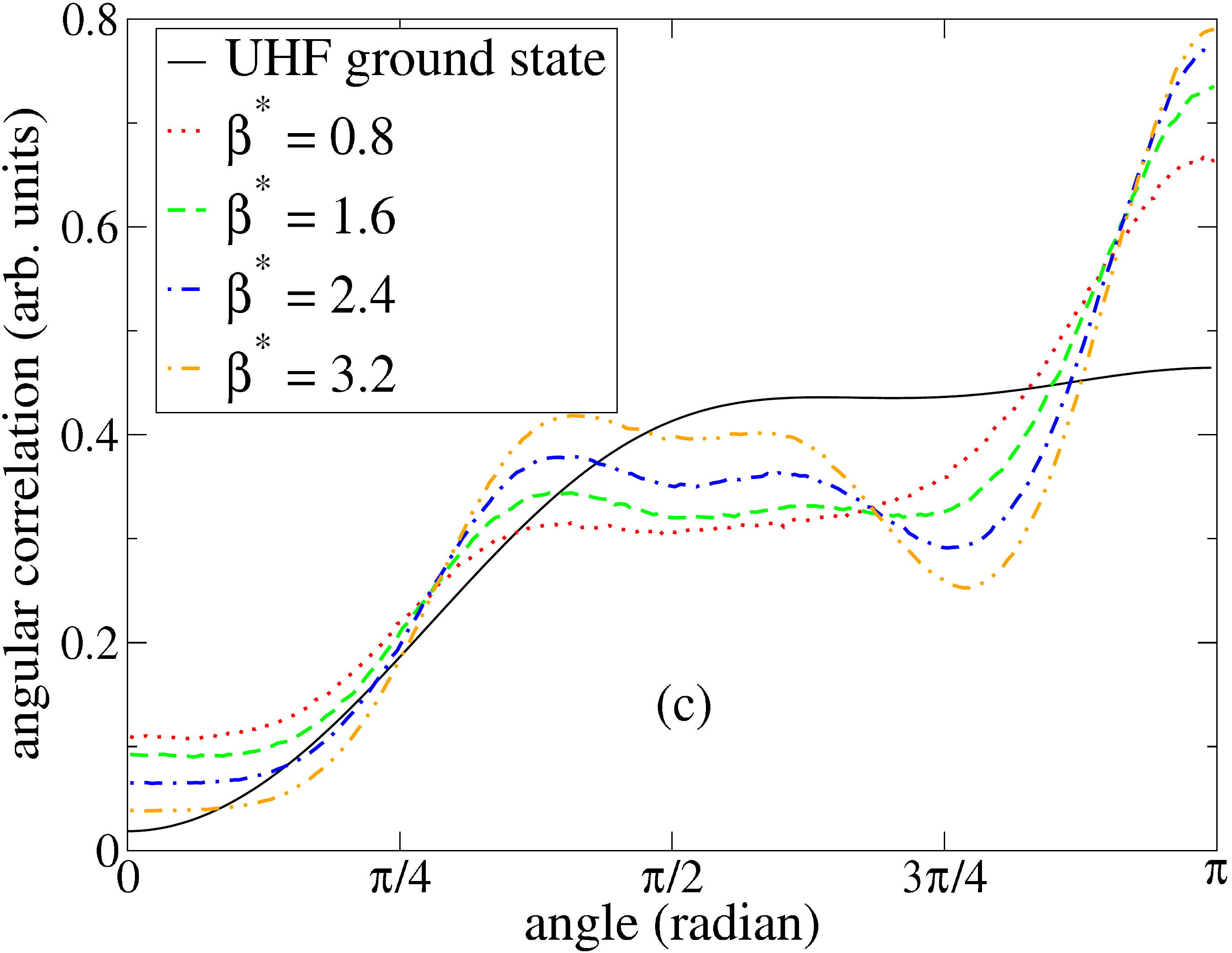}
\includegraphics[width=0.3\textwidth, keepaspectratio]{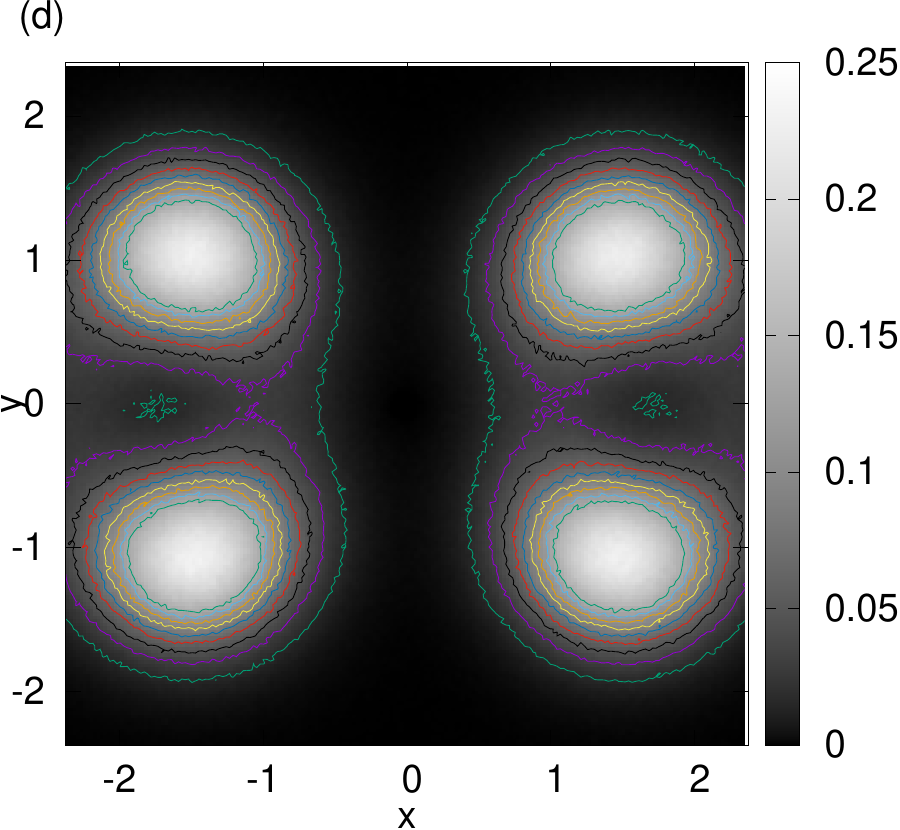}
\end{center}
\caption{\label{fig:mfp4b}
  [(a) and (b)] The angular correlation function $g(\theta)$ and the real-space particle density, respectively,
  in a PIMC simulation of fully polarized quantum dot beryllium ($N=4$) at $\gamma=1.6$ and $\lambda=1$.
  [(c) and (d)] Corresponding results at $\lambda=2.5$.  Label B in Fig.~\ref{fig:berylliumphase}.
}
\end{figure*}

We performed PIMC simulations near the coupling strength where UHF predicts the last spin flip,
in order to check the UHF prediction that the electron system recovers rotational invariance in this range.
As Fig.~\ref{fig:mfp4} demonstrates, the angular correlations characteristic of a $C_{4v}$ type are clearly
present at the lowest coupling strength where UHF predicts full spin-polarization.
At higher $\lambda$'s these correlations increase slightly (not shown).
For $\gamma=0.2$ and $0.4$, the temperature does not affect the result, indicating that it is below
the energy scale of ordering, while for $\gamma\ge0.6$ this is not the case.
Again, no sudden change is seen in the correlation when the UHF ground state changes to $(2,C_{4v})$.
Similar correlations are present for somewhat higher magnetic fields.
For $\gamma=1.6$, however, the picture changes.
At small couplings we see peaks in the angular correlation near $\pi/2$ and $\pi$, but the latter
peak is definitely sharper; cf. Fig.~\ref{fig:mfp4b}(a).
The particle density in real space [Fig.~\ref{fig:mfp4b}(b)] shows a unidirectional modulation, but we note that such a picture
is unavoidably washed out in longer simulations because of the orientational random walk of the system.
At larger couplings, Figs.~\ref{fig:mfp4b}(c) and \ref{fig:mfp4b}(d), we observe an elongated system with two clearly separated
localized peaks.
This behavior is also present in the angular correlation, where the peak $\theta=\pi/2$ splits into two
nearby peaks.
This gradual reordering does not coincide with the $(2,C_\infty)\to(2,C_{4v})$ transition of the UHF ground state;
it occurs at higher $\lambda$.

\subsubsection{The $S_z=0$ subspace of $N=4$}

\begin{figure*}[htbp]
\begin{center}
\includegraphics[width=0.32\textwidth, keepaspectratio]{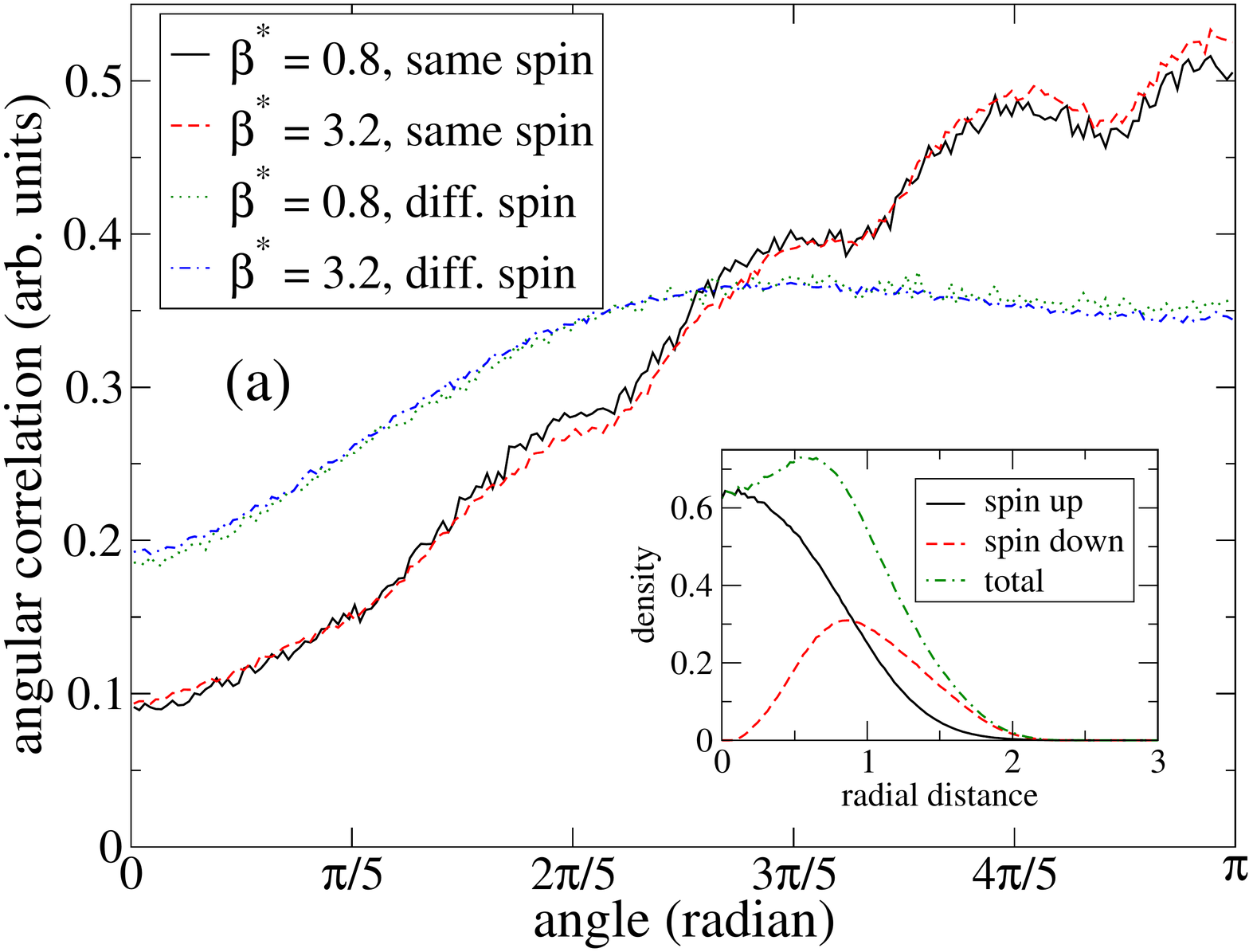}
\includegraphics[width=0.32\textwidth, keepaspectratio]{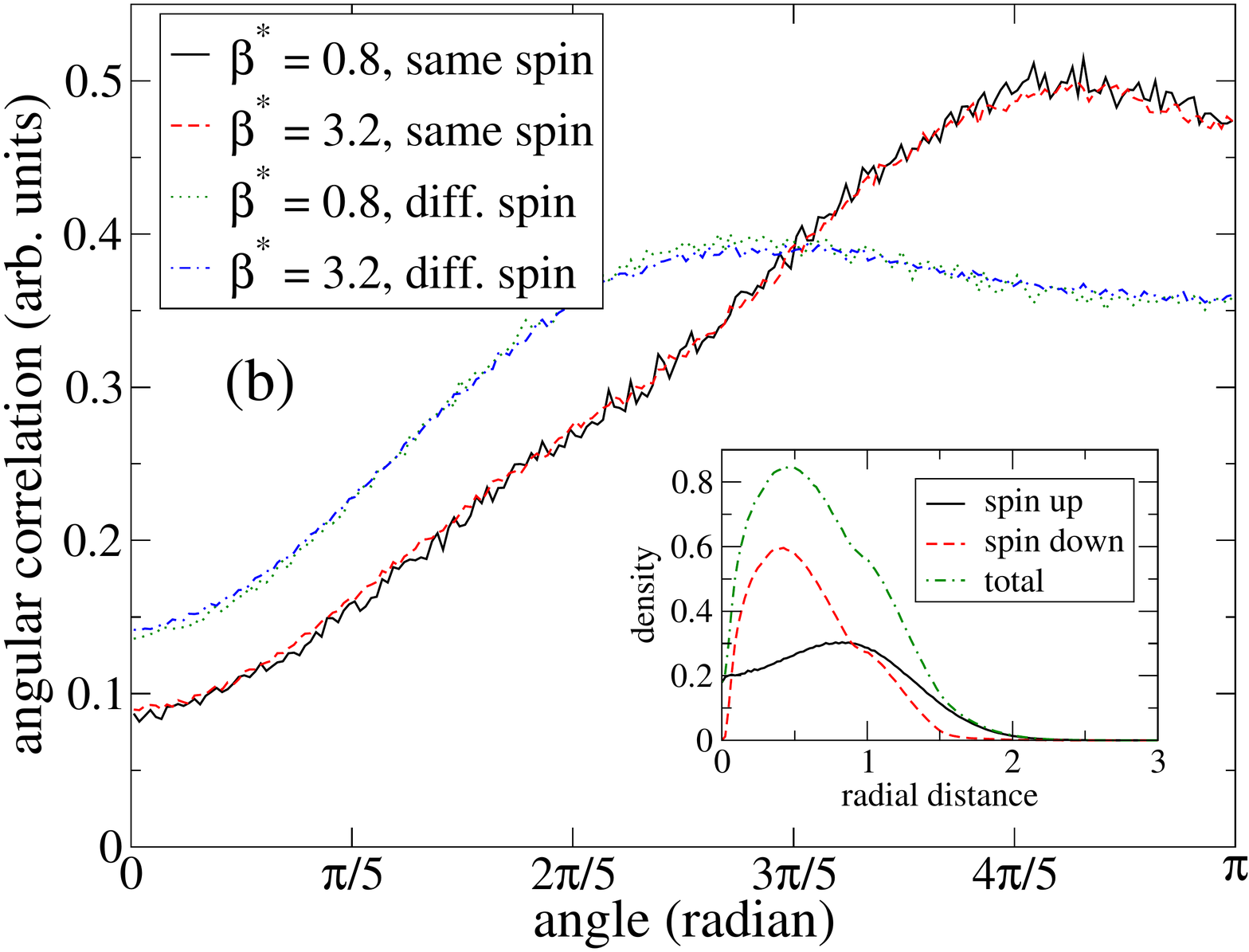}
\includegraphics[width=0.32\textwidth, keepaspectratio]{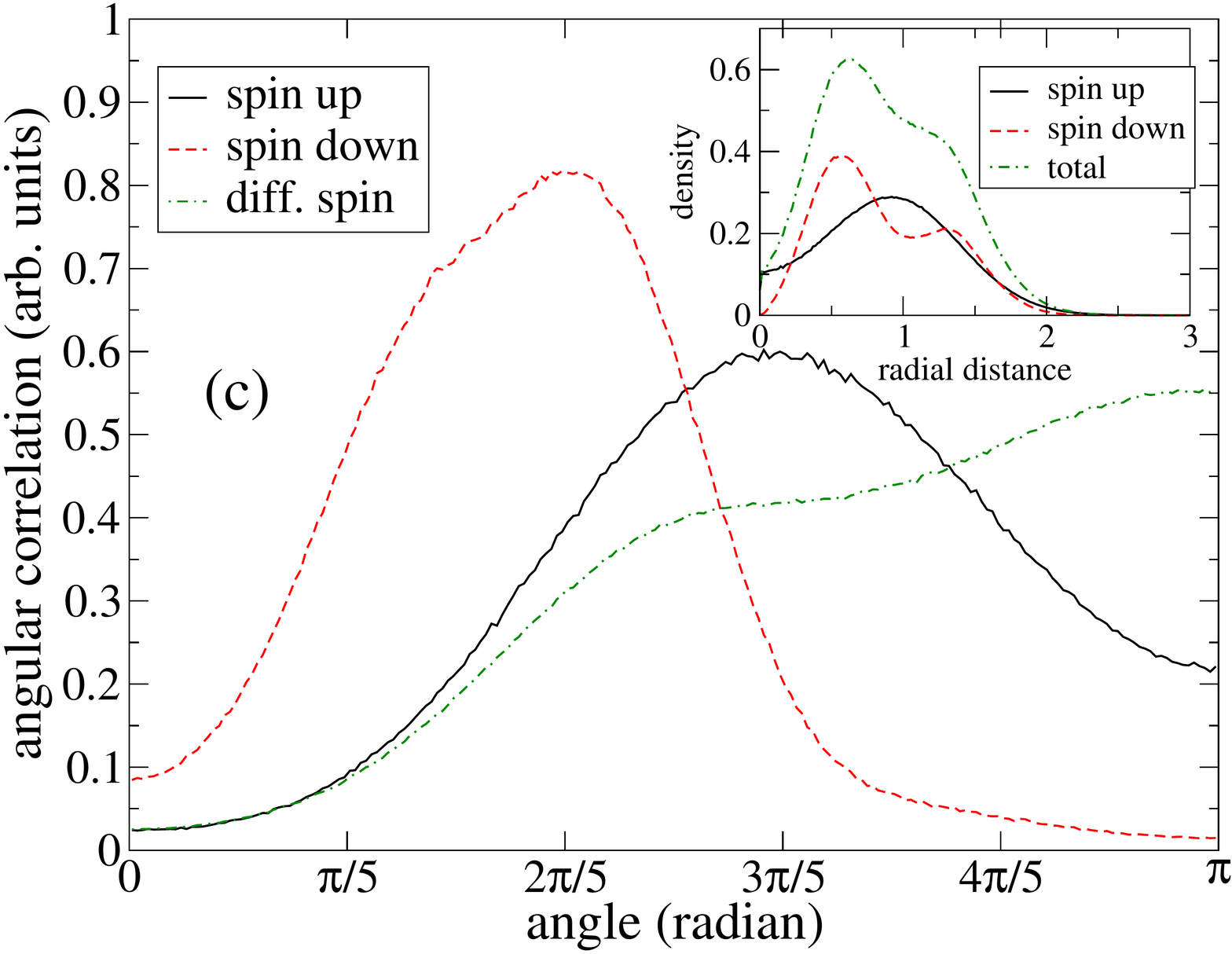}
\end{center}
\caption{\label{fig:msz0_4}
  The angular correlation functions $g_\text{same}(\theta)$ and $g_{\uparrow\downarrow}(\theta)$
  (a) at $\gamma=0.2$, $\lambda=0.6$,
  (b) at $\gamma=0.2$, $\lambda=1.2$,
  (c) at $\gamma=0.6$, $\lambda=1.2$ (here, $\beta^\ast=3.2$), respectively,
  in the $S_z=0$ sector of quantum dot beryllium ($N=4$).
  The radial densities are shown in the insets.  Label C in Fig.~\ref{fig:berylliumphase}.
}
\end{figure*}

Here we focus on the small-$\lambda$ region, where UHF predicts a transition from $(0,C_\infty)$ to $(0,C_{2v})$.
In most cases we observe a gradual build-up of different-spin correlations near $\theta=\pi/2$,
a feature of the $C_{2v}$-type spin modulation.
No sudden change occurs at the $\lambda$-values where UHF predicts the $C_\infty$ to $C_{2v}$ transition,
c.f.\ the first two panels in Fig.~\ref{fig:msz0_4}.
Notice that for $\lambda=0.6$, the same-spin correlation has a step-like fine structure.
We cannot offer an interpretation of these steps, but note that they are very well reproducible and
temperature-independent as seen in Fig.~\ref{fig:msz0_4}(a).

A very interesting situation arises at $\gamma=0.6$.
For small $\lambda$, we see ring-like structures with slightly increased correlations with respect to the UHF state.
For $\lambda\ge1.2$, within the $(0,C_{2v})$ region of UHF, we see a correlation of like spins at $\theta\approx\pi/2$,
which corresponds to a dipolar spin modulation, not a quadrupolar one of $C_{2v}$ symmetry.
See Fig.~\ref{fig:msz0_4}(c).
Schematically, this state would correspond to the $(0,C_s)$ configuration in Fig.~\ref{fig:berylliumphase}, slightly distorted,
a molecule that never occurs as a UHF ground state.


\subsubsection{The $S_z=1$ subspace of $N=4$}

\begin{figure*}[htbp]
\begin{center}
\includegraphics[width=0.36\textwidth, keepaspectratio]{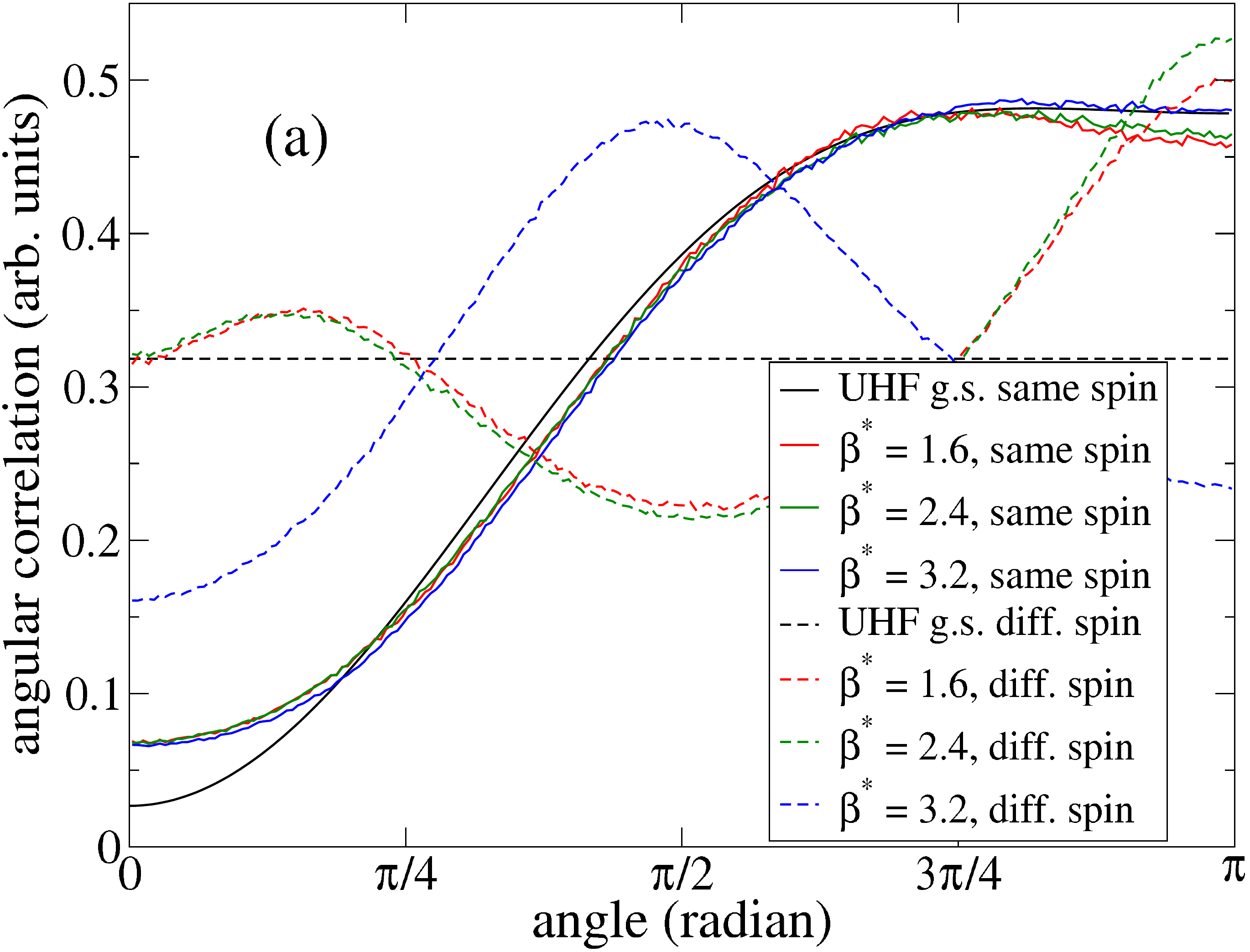}
\includegraphics[width=0.36\textwidth, keepaspectratio]{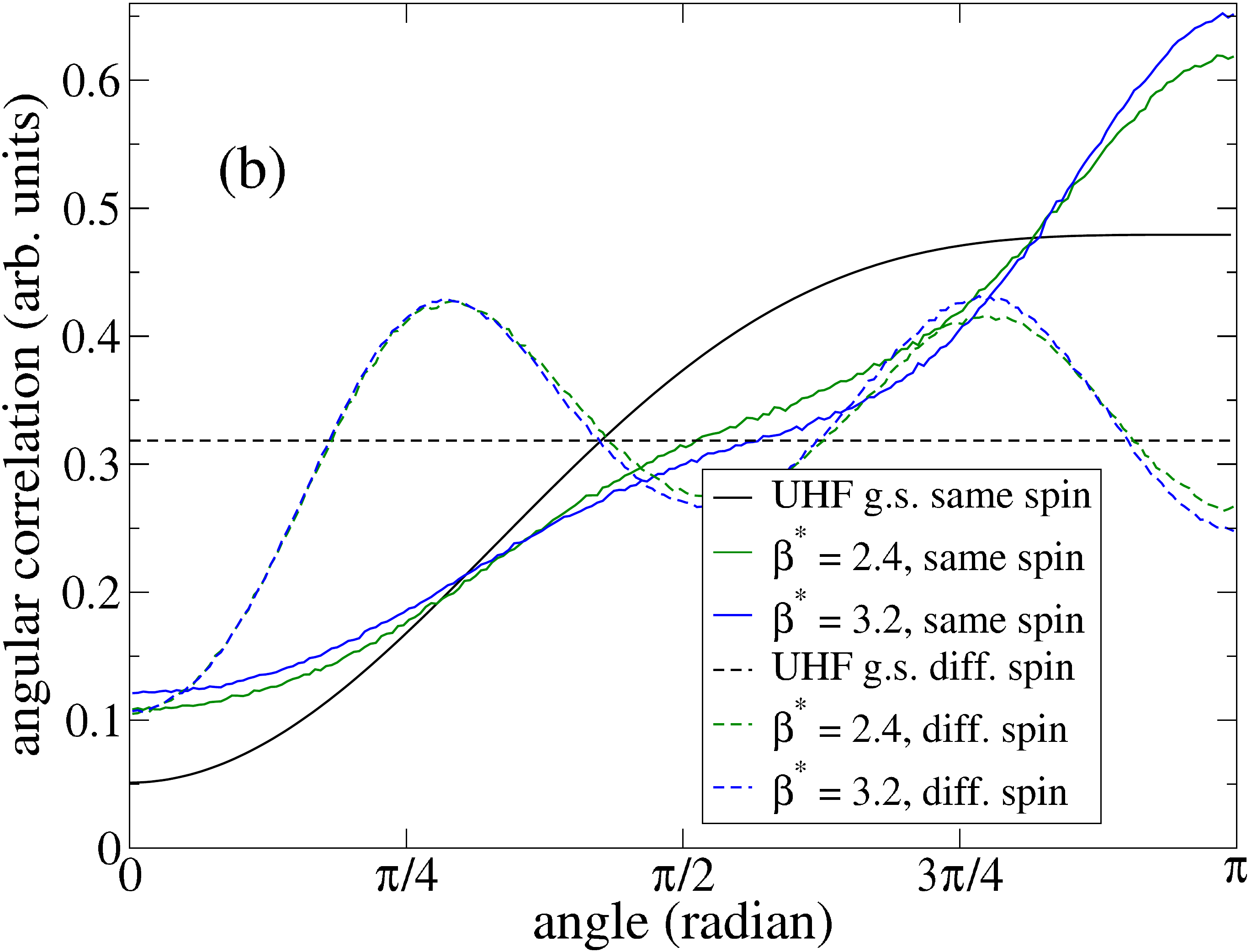}
\includegraphics[width=0.3\textwidth, keepaspectratio]{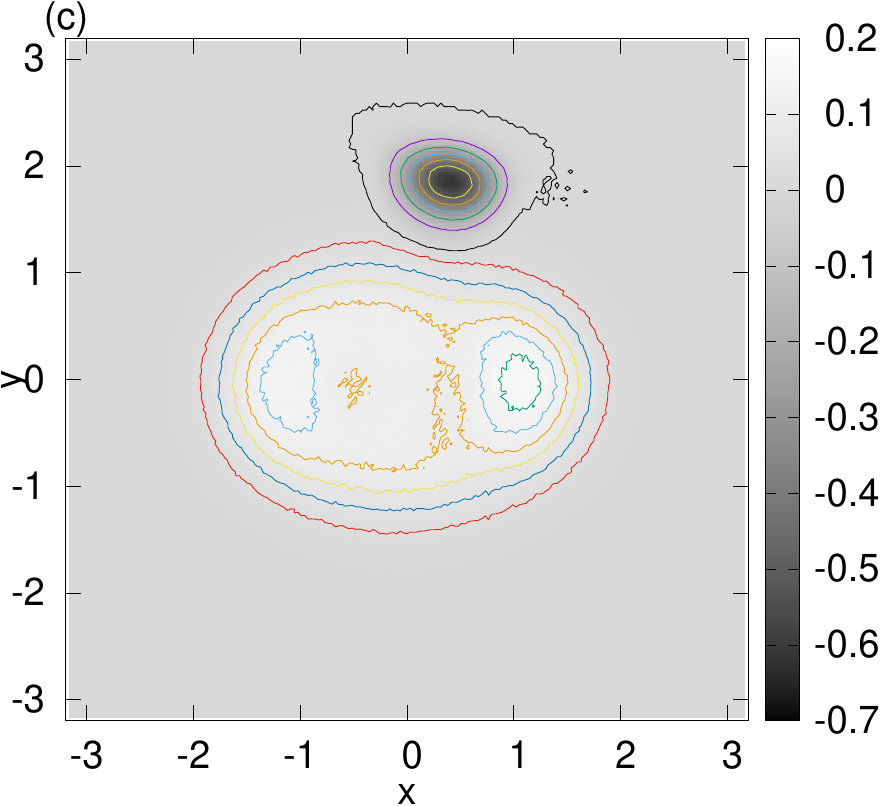}
\includegraphics[width=0.3\textwidth, keepaspectratio]{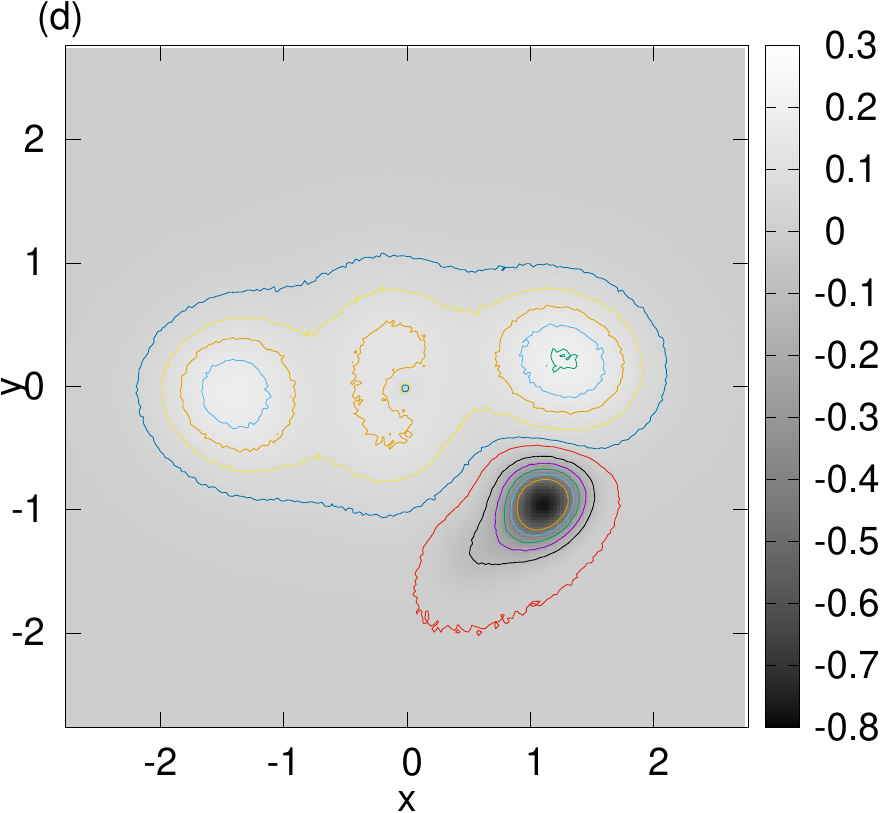}
\end{center}
\caption{\label{fig:msz1_4}
  PIMC results in the $S_z=1$ sector of quantum dot beryllium ($N=4$).
  [(a) and (c)] The angular correlation function $g(\theta)$ and the real-space spin density histogram using $\beta^\ast=3.2$
  at $\gamma=1.2$ and $\lambda=0.8$.
  [(b) and (d)] The same at $\gamma=1.6$ and $\lambda=0.4$.  Label D in Fig.~\ref{fig:berylliumphase}.
}
\end{figure*}

This subspace is relevant in UHF in small intervals for $\gamma\ge0.8$.
No symmetry-breaking is predicted by UHF for $\gamma\ge1.2$.
PIMC is performed at the smallest $\lambda$ where the UHF ground state has $S_z=1$.
In all cases we find that at sufficiently low temperature the minority spin electron increasingly avoids the cloud of
majority spin electrons, cf.\ Fig.~\ref{fig:msz1_4}.
At $\gamma=1.2$, $\lambda=0.8$ and $\gamma=1.6$ and $\lambda=0.4$, the majority spins prefer opposite angles.
In real space histograms this appears as a striped pattern in the majority spin cloud, cf.\ Figs.~\ref{fig:msz1_4}(c) and \ref{fig:msz1_4}(d).
The strongly localized minority spin electron is located in an interstitial.
At $\gamma=1.2$, $\lambda=0.8$, there are two such locations,
as the peak near $\pi/2$ in the different-spin angular correlation demonstrates.
At $\gamma=1.6$, $\lambda=0.4$ there are four such locations,
the different-spin angular correlation having peaks near $\theta=\pi/4$ and $3\pi/4$.

\subsection{Quantum dot boron, $N=5$}

\begin{figure}
\begin{center}
  \includegraphics[width=\columnwidth, keepaspectratio]{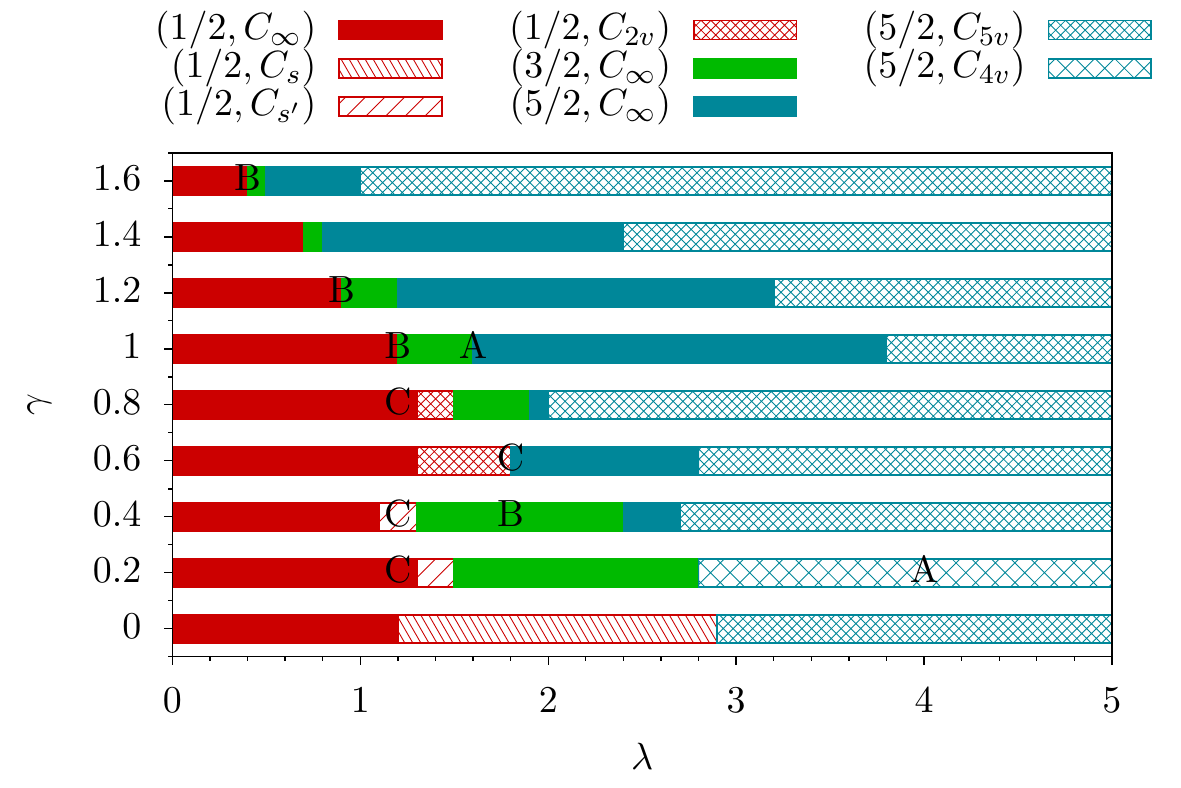}
   \def\svgwidth{\columnwidth}
 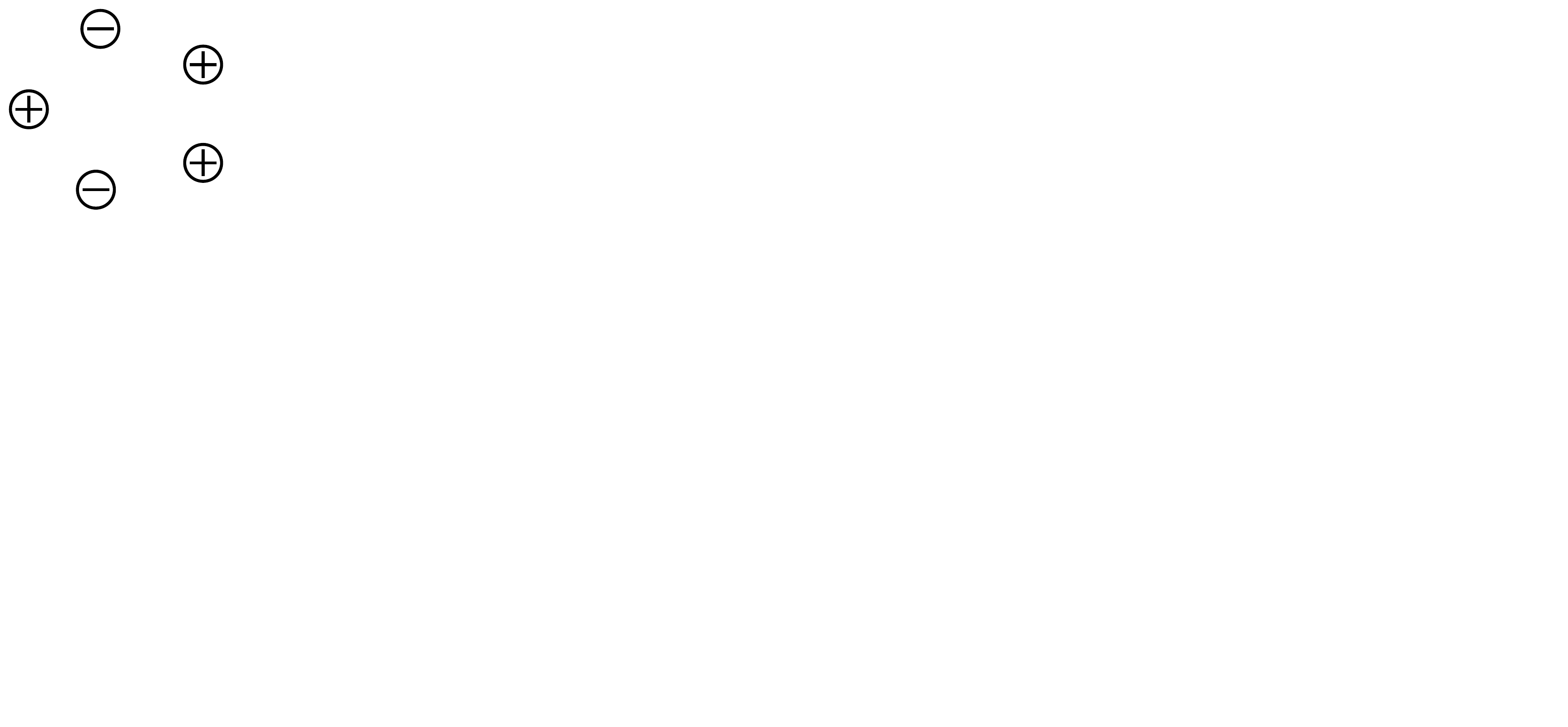
\end{center}
\caption{\label{fig:boron}
  (Top)
  Sections of the UHF phase diagram of quantum dot boron ($N=5$) at the magnetic field parameter $\gamma=\frac{\omega_c}{\omega}$
  fixed at multiples of 0.2.
  We do not attempt to draw phase boundaries by connecting the transition points because of the limited
  resolution in the $\gamma$ direction.
  (Bottom)
  Sketch of the spin density peaks in the relevant symmetry-breaking states of quantum dot boron.
  Note that $(\frac{3}{2},C_s)$, $(\frac{3}{2},C_{4v})$, and $(\frac{1}{2},C_s''')$ are relevant in PIMC only.
  Labels A, B, and C indicate PIMC parameters in Figs.~\ref{fig:mfp5}, \ref{fig:msz1.5_5}, and \ref{fig:msz0.5_5}, respectively.
}
\end{figure}

Apart from the three rotationally invariant states $(S_z,C_\infty)$ (where $S_z=\frac{1}{2}$, $\frac{3}{2}$, or $\frac{5}{2}$),
the symmetry-breaking states that occur as a mean-field ground state or a correlated state
are sketched in Fig.~\ref{fig:boron}.

As seen in Fig.~\ref{fig:boron},
UHF predicts different behavior for small magnetic fields $0<\gamma\le0.8$ and larger ones $\gamma\ge1$.
In the former case the spatial symmetry breaks early, then full polarization is reached through a sequence of spin flips,
which restore rotational symmetry before eventually giving way to a spin-polarized symmetry-broken state.
In the latter case, spin is polarized first, and the only symmetry-broken state is fully polarized.
Note the exceptional nature of the strong coupling limit at $\gamma=0.2$: here the symmetry of the particle density
becomes $C_{4v}$ instead of $C_{5v}$.


\subsubsection{The $S_z=\frac{5}{2}$ subspace of $N=5$}

\begin{figure*}[htbp]
\begin{center}
  \includegraphics[width=.64\textwidth, keepaspectratio]{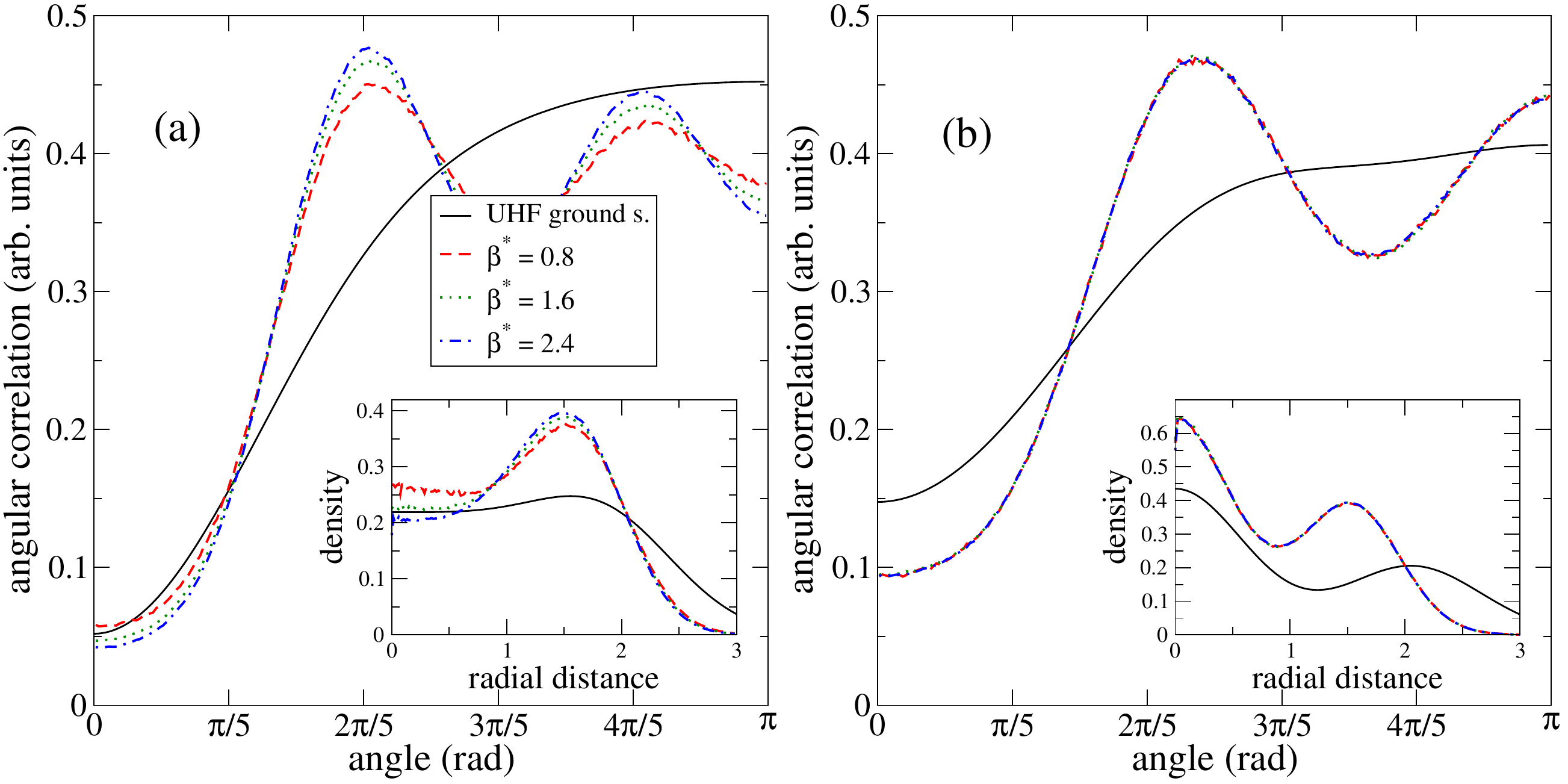}
\end{center}
\caption{\label{fig:mfp5}
  The angular correlation function in the fully spin-polarized sector of quantum dot boron ($N=5$).
  The insets show the radial density.
  (a) Typical pentagonal structure at $\gamma=1$, $\lambda=1.6$, as found everywhere $\gamma\neq0.2$
  (b) Square-coordinated structure at $\gamma=0.2$, $\lambda=4$; for $\gamma=0.2$ such correlations
  are present for all $\lambda$. Label A in Fig.~\ref{fig:boron}.
}
\end{figure*}

Figure \ref{fig:mfp5} shows PIMC simulations in the fully polarized sector on both
sides of the symmetry-breaking transition as predicted by UHF.
Again, we find that the angular correlation evolves continuously on both sides; the angular correlation is
already strong at the $\lambda$ where the last spin flips.
Most notably, PIMC confirms the UHF prediction that the electrons are correlated in a centered square structure at $\gamma=0.2$.

\subsubsection{The $S_z=\frac{3}{2}$ subspace of $N=5$}

\begin{figure*}[htbp]
  \begin{center}
    \includegraphics[width=0.36\textwidth, keepaspectratio]{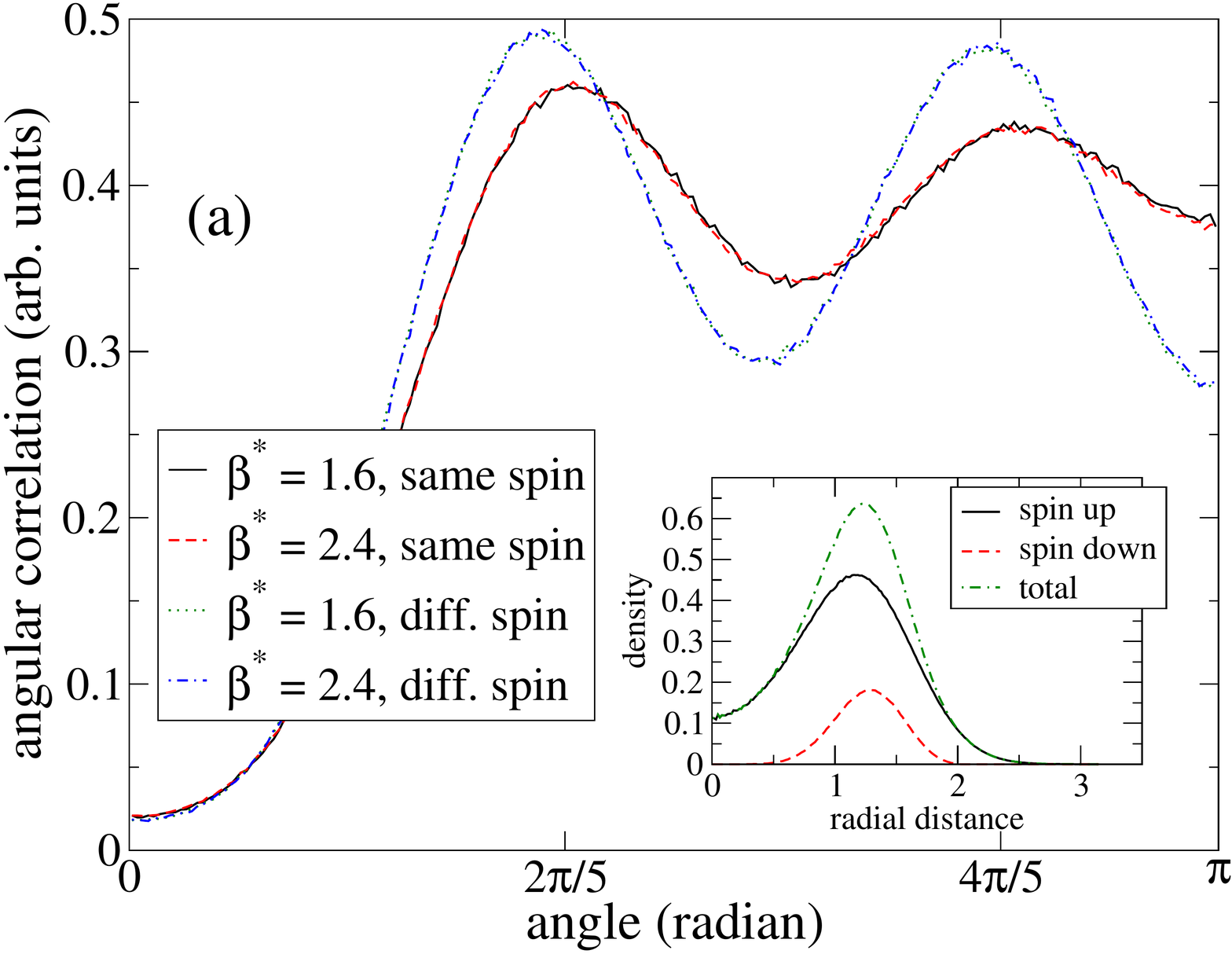}
    \includegraphics[width=0.36\textwidth, keepaspectratio]{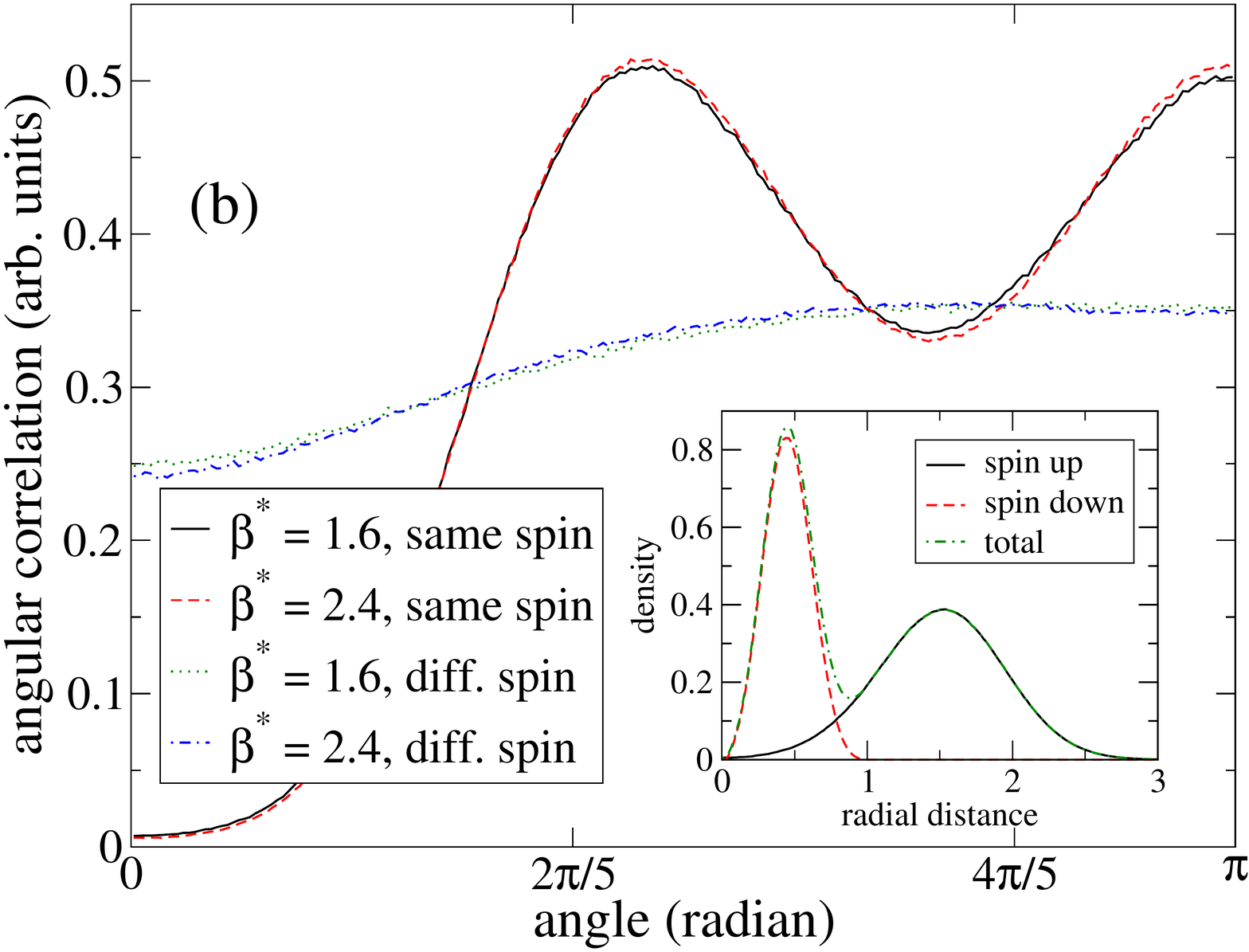}

    \includegraphics[width=0.36\textwidth, keepaspectratio]{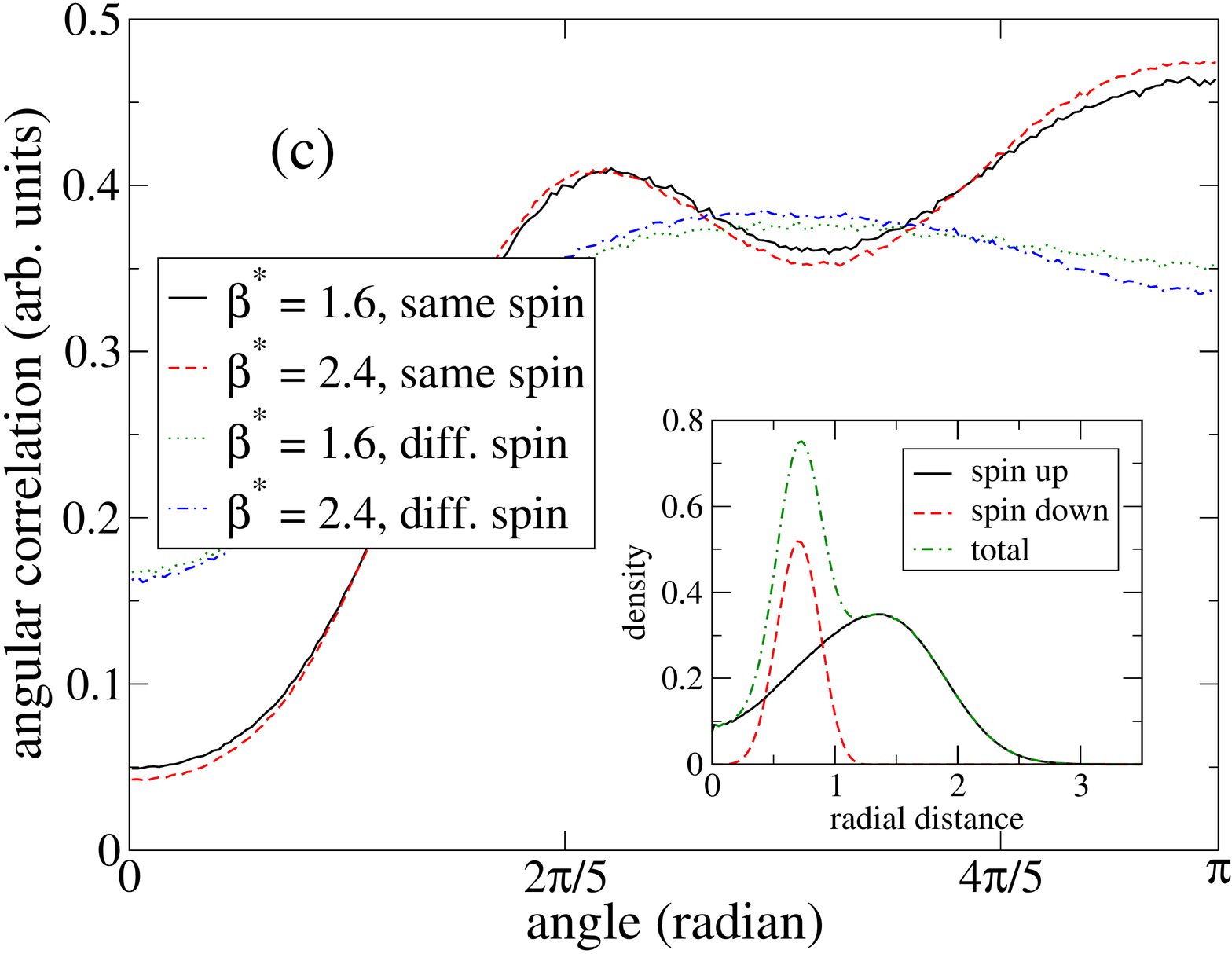}
    \includegraphics[width=0.36\textwidth, keepaspectratio]{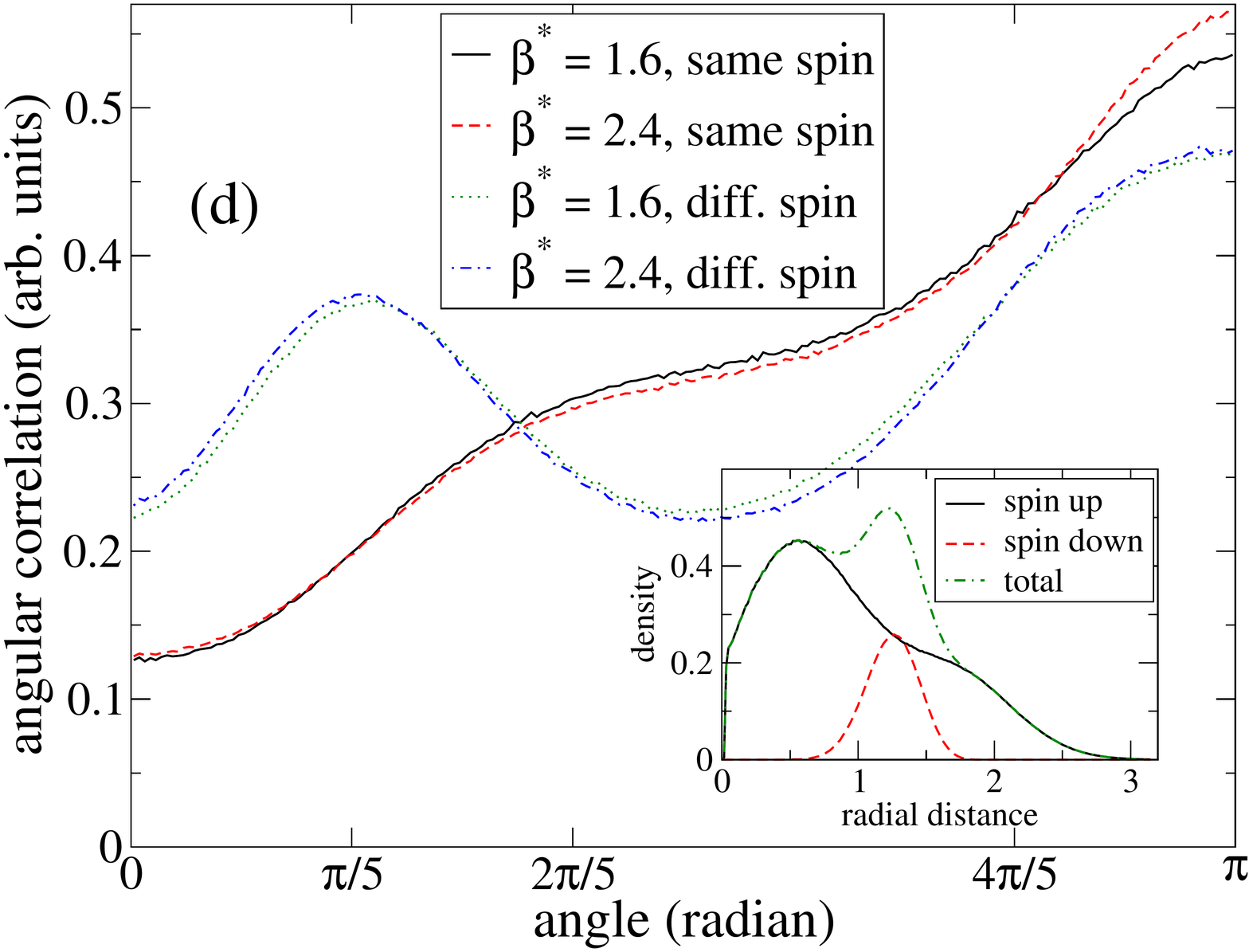}
\end{center}
\caption{\label{fig:msz1.5_5}
  The angular correlation function $g(\theta)$ in the $S_z=\frac{3}{2}$ sector of quantum dot boron ($N=5$)
  (a) at $\gamma=0.4$ and $\lambda=1.8$,
  (b) at $\gamma=1$ and $\lambda=1.2$,
  (c) at $\gamma=1.2$ and $\lambda=0.9$, and
  (d) at $\gamma=1.6$ and $\lambda=0.4$.
  The inset shows the radial particle densities, distinct for each spin and in total, at $\beta^\ast=2.4$.  Label B in Fig.~\ref{fig:boron}.
}
\end{figure*}

We focus on the $(\frac{3}{2},C_\infty)$ UHF region.
As Fig.~\ref{fig:msz1.5_5}(a) demonstrates, ordering into pentagonal structure is clear at $\lambda=1.8$ for a
weak magnetic field $\gamma=0.4$: the center of the dot is depleted, while there are correlation peaks
near $\theta=2\pi/5$ and  $\theta=4\pi/5$ both for identical and different spins.
This state, which we call $(\frac{3}{2},C_s)$, never occurs as an UHF ground state.
The same structure emerges at higher couplings (tested at $\lambda=2.3$).
At a lower coupling, $\lambda=1.3$, the spin-down electron is localized at a somewhat greater radius than
the ring of majority spin electrons; the same happens at $\gamma=0.8$ and $\lambda=1.5$, 1.8.
On the other hand, at $\gamma=1$, the spin-down electron is near the origin,
surrounded by a square of spin-up electrons; cf.\ Fig.~\ref{fig:msz1.5_5}(b).
We will call this state $(\frac{3}{2},C_{4v})$.
Note that the down-spin electron actually occupies a small ring around the center of the trap.
The spin-up square and the down-spin electron find it advantageous to revolve around the center of mass in a correlated
manner; a situation we see quite often.
Increasing the magnetic field further, the spin-down ring widens and approaches the ring of spin-up electrons,
as seen in Fig.~\ref{fig:msz1.5_5}(c).
At the same time, the angular correlation peaks of spin-up electrons move to slightly smaller angles.
We interpret this scenario as a gradual distortion of $(\frac{3}{2},C_{4v})$ into $(\frac{3}{2},C_s)$.
Finally, at $\gamma=1.6$, we see in Fig.~\ref{fig:msz1.5_5}(d) that the radius of the spin-down ring becomes
larger than that of the up-spins; the latter no longer shows strong angular ordering.

\begin{figure*}[htbp]
  \begin{center}
    \includegraphics[width=0.36\textwidth, keepaspectratio]{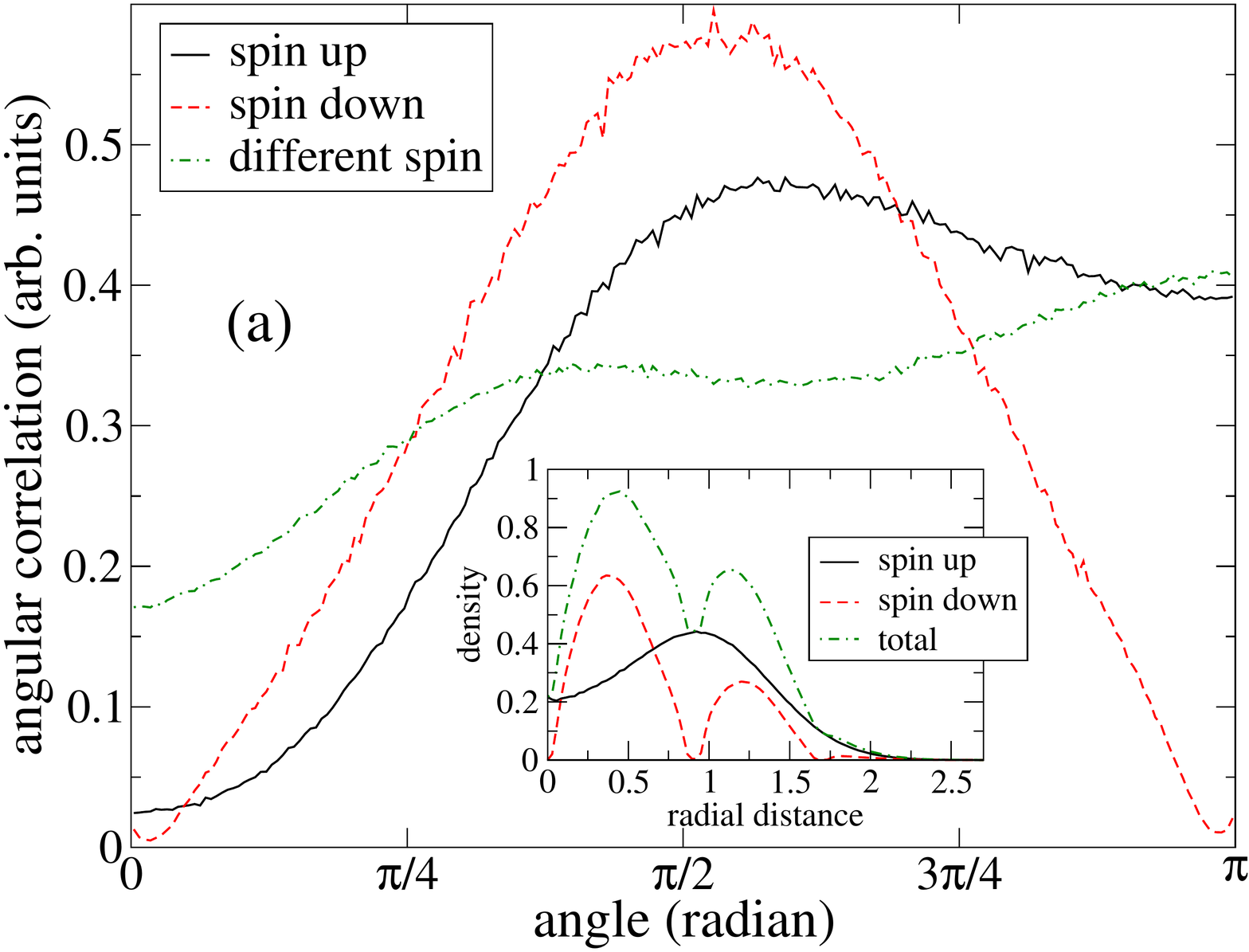}
    \includegraphics[width=0.36\textwidth, keepaspectratio]{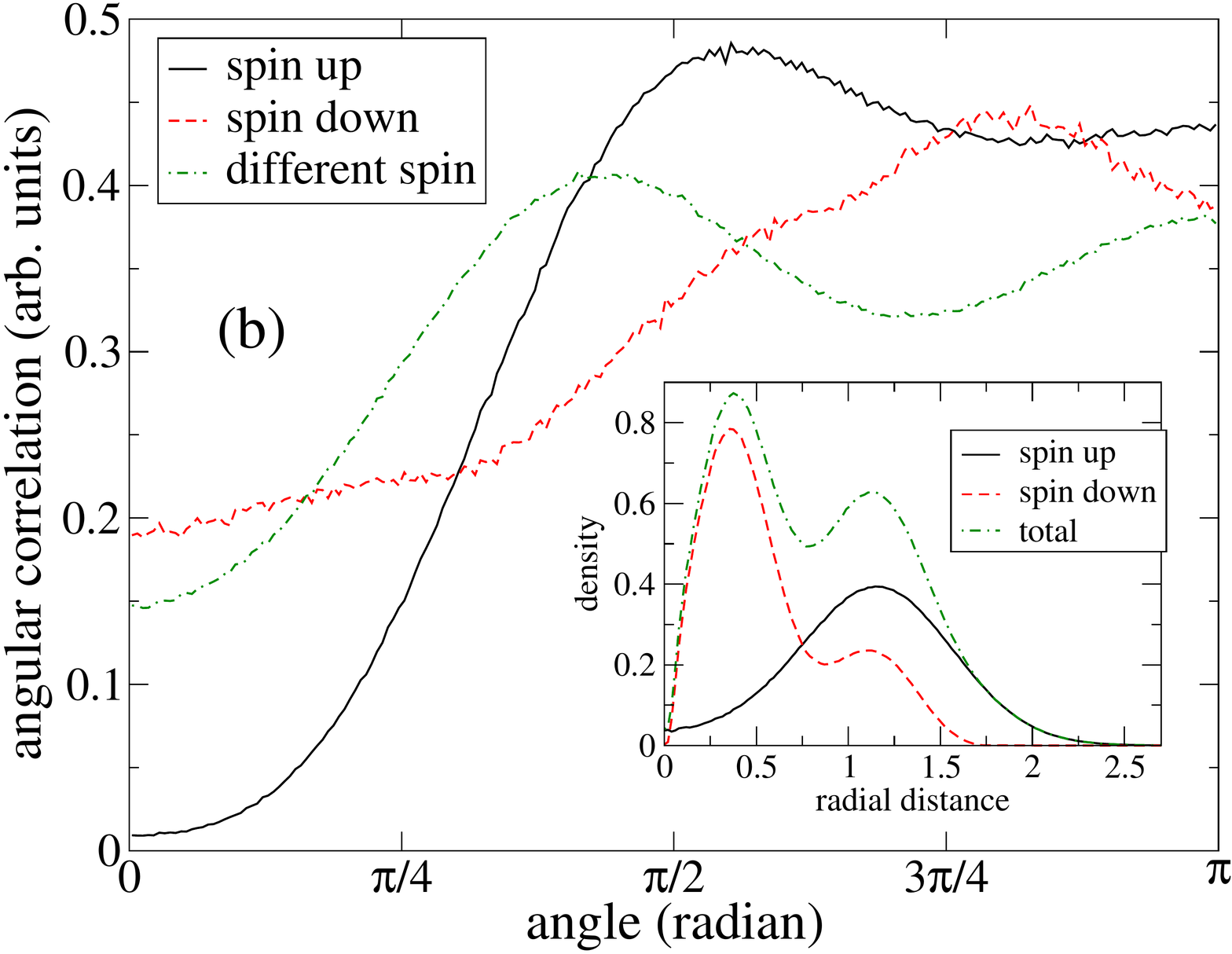}

    \includegraphics[width=0.36\textwidth, keepaspectratio]{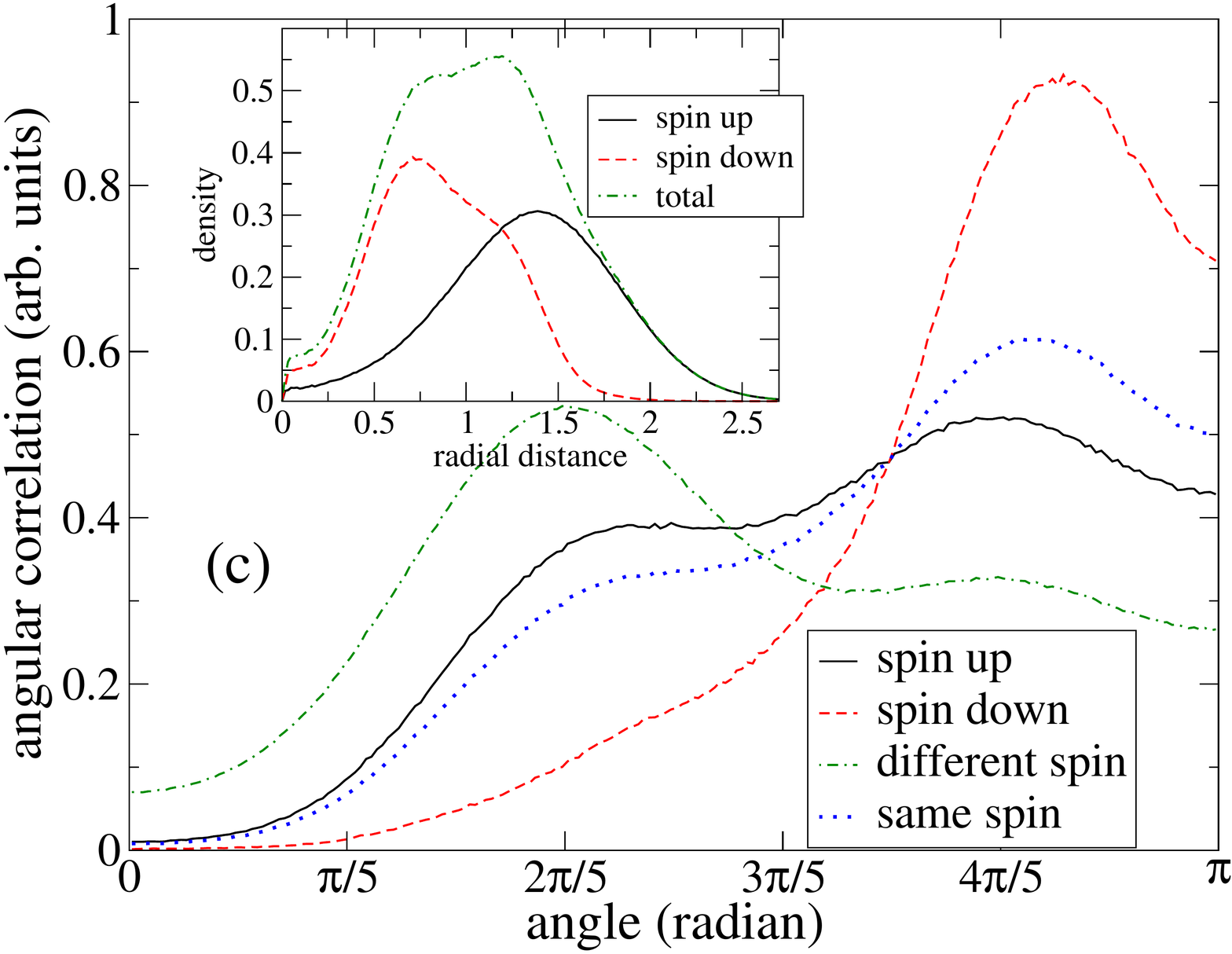}
    \includegraphics[width=0.36\textwidth, keepaspectratio]{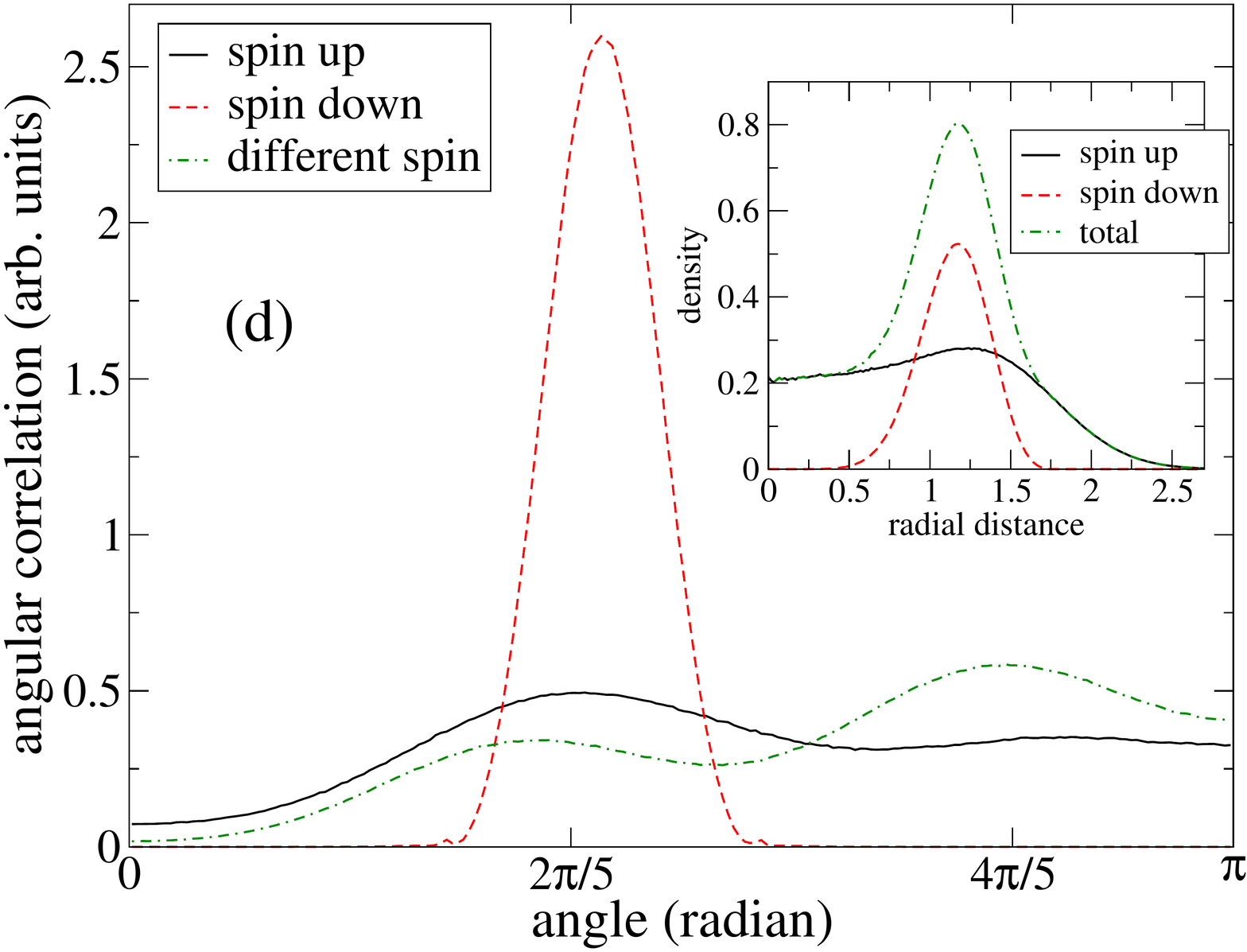}
\end{center}
\caption{\label{fig:msz0.5_5}
  Angular correlation functions and radial densities (insets) in the $S_z=\frac{1}{2}$ sector of quantum dot boron ($N=5$) at
  (a) for $\gamma=0.2$ and $\lambda=1.2$,
  (b) for $\gamma=0.4$ and $\lambda=1.2$,
  (c) for $\gamma=0.6$ and $\lambda=1.8$,
  and (d) $\gamma=0.8$ and $\lambda=1.2$.
  $\beta^\ast=2.4$ in all panels.  Label C in Fig.~\ref{fig:boron}.
}
\end{figure*}


\subsubsection{The $S_z=\frac{1}{2}$ subspace of $N=5$}

Here PIMC demonstrates the emergence of correlated structures that have no precursor in the UHF calculation.
For a small magnetic field $\gamma=0.2$, we find that a double ring structure emerges,
see the inset of Fig.~\ref{fig:msz0.5_5}(a).
For majority spins, we observe a weak avoidance, only slightly stronger than by UHF,
but the angular correlation peak is near $\theta=\pi/2$.
The minority spin angular correlation is also peaked at $\theta=\pi/2$.
We identify this as the new structure $(\frac{1}{2},C_s''')$ in the lower row of Fig.~\ref{fig:boron}.
A similar effect is observed at $\gamma=0.4$, c.f.\ Fig.~\ref{fig:msz0.5_5}(b) for $\lambda=1.2$.
Here, the minority spins are correlated at a greater angle, suggesting a distorted structure.
The central spin-up electron is slightly off the origin.

Quite different correlations arise at a greater field, $\gamma=0.6$, as shown in Fig.~\ref{fig:msz0.5_5}(c).
The center of the trap is depleted, and only a single ring is manifest.
The different-spin angular correlation is peaked near $\theta=2\pi/5$,
and a minor peak is discernible near $\theta=4\pi/5$.
The same-spin angular correlation also shows a double-peak structure,
but now the dominant peak is at the greater, the minor peak at the smaller angle.
This is consistent with the structure of the $(\frac{1}{2},C_s)$ state, now encoded in the correlations.
Note that the pentagon of localized electrons does not need to be regular,
as the presence of two spins unavoidably reduces the symmetry.
These findings at $\gamma=0.6$ are at odds with the UHF prediction of a $(\frac{1}{2},C_{2v})$ state.

At a still greater field, $\gamma=0.8$, we find yet another behavior.
As Fig.~\ref{fig:msz0.5_5}(d) demonstrates, the total charge density has a dip at the center,
the minority spins are strongly correlated near $\theta=2\pi/5$
and the majority ones near multiples of $\theta=2\pi/5$.
We consider this as a modification of the $(\frac{1}{2},C_s'')$ structure, now encoded in correlations.
The different sharpness of the correlation peaks for the two spins is a surprise to us.


\subsection{Quantum dot carbon, $N=6$}

\begin{figure*}
\begin{center}
  \includegraphics[width=\textwidth, keepaspectratio]{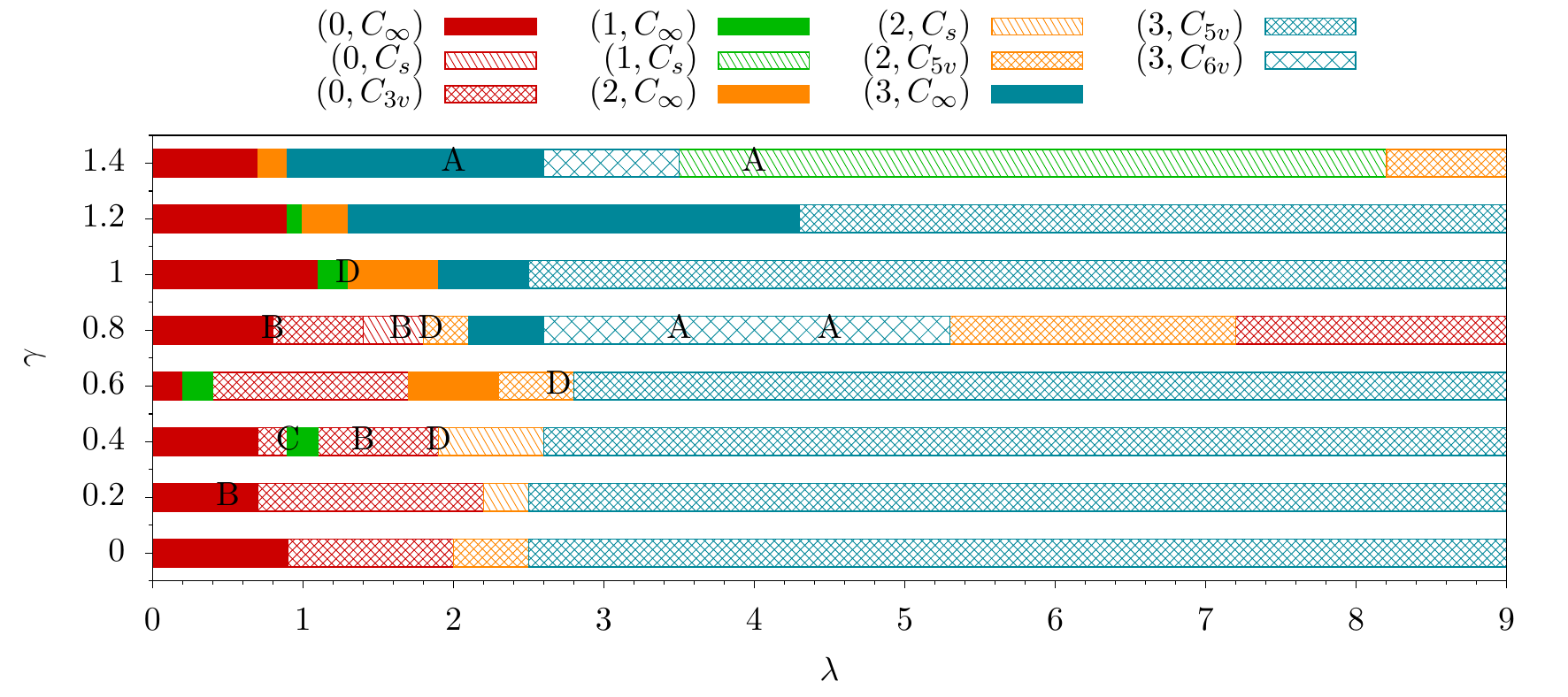}
 \def\svgwidth{\textwidth}
 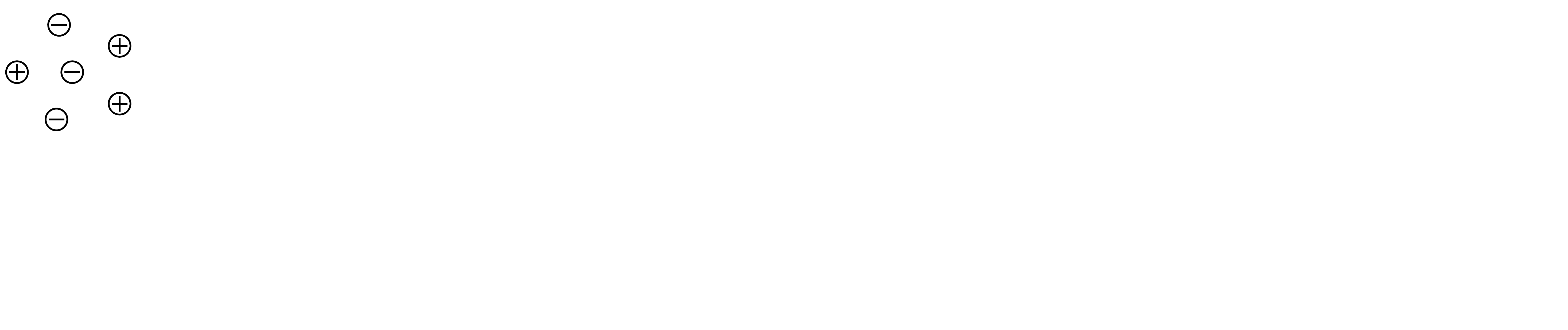
\end{center}
\caption{\label{fig:carbon}
  (Top)
  Sections of the UHF phase diagram of quantum dot carbon ($N=6$) at the magnetic field parameter $\gamma=\frac{\omega_c}{\omega}$
  fixed at multiples of 0.2.
  We do not attempt to draw phase boundaries by connecting the transition points because of the limited
  resolution in the $\gamma$ direction.
  Labels A, B, C, and D indicate PIMC parameters in Figs.~\ref{fig:mfp6}, \ref{fig:msz0_6}, \ref{fig:msz1_6}~and~\ref{fig:msz2_6},
  respectively.
  (Bottom)
  Sketch of the spin density peaks in the relevant symmetry-breaking states of quantum dot carbon.
  Note that $(2,C_s')$ is never an UHF ground state, but a state with such correlations occurs in PIMC.
  For $S_z=0$, equivalent structures are obtained by interchanging spin-up and spin-down.
}
\end{figure*}

Apart from the four rotationally invariant states $(S_z,C_\infty)$ (where $S_z=0,1,2,3$),
the relevant symmetry-breaking states are sketched in Fig.~\ref{fig:carbon}.

Sections of the UHF phase diagram at fixed magnetic field $\gamma$ are also shown in Fig.~\ref{fig:carbon}.
We find several remarkable features which should be critically assessed.
At $\gamma=0.6$, rotational symmetry is recovered after the spin changes from 0 to 2 at $\lambda=1.7$;
the same happens at $\gamma=1$ and $\lambda\ge1.3$ and  at $\gamma=1.2$ and $\lambda\ge1$.
At $\gamma=0.4$, spatial ordering is predicted with a subsequent jump in the total spin; but this evolution
is interrupted by a rotationally symmetric $S_z=1$ phase at $\lambda=0.9$ and 1.
Most remarkably, the strong coupling limit is not fully polarized in the UHF approximation at $\gamma=0.8$ and 1.4.
At these magnetic field strengths, the fully polarized state has sixfold symmetry with no density peak at the center of the trap.
For $\gamma\ge1$, gradual spin polarization precedes spatial ordering.

\subsubsection{The $S_z=3$ subspace of $N=6$}

\begin{figure*}[htbp]
\begin{center}
\includegraphics[width=0.64\textwidth, keepaspectratio]{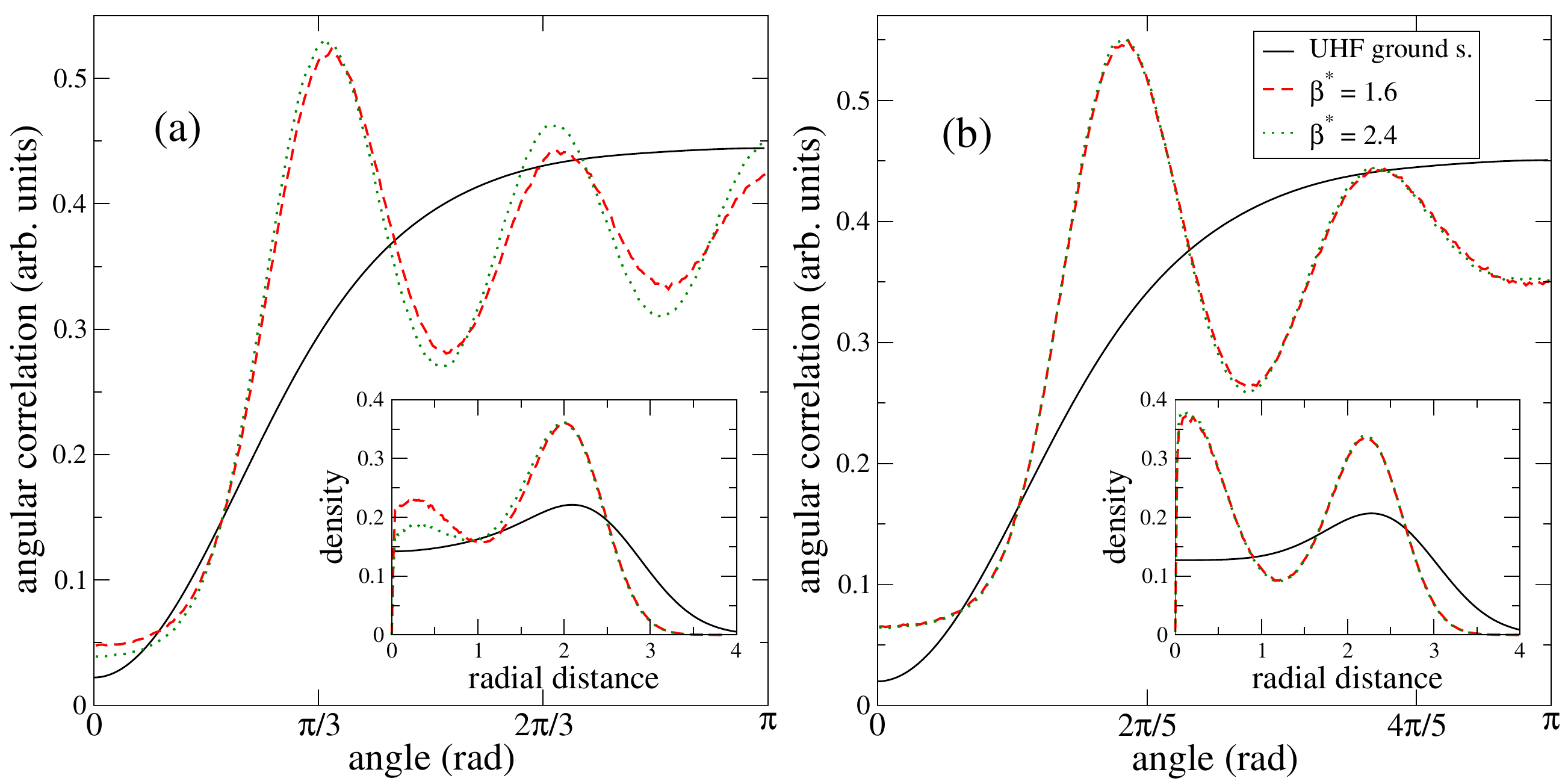}
\includegraphics[width=0.64\textwidth, keepaspectratio]{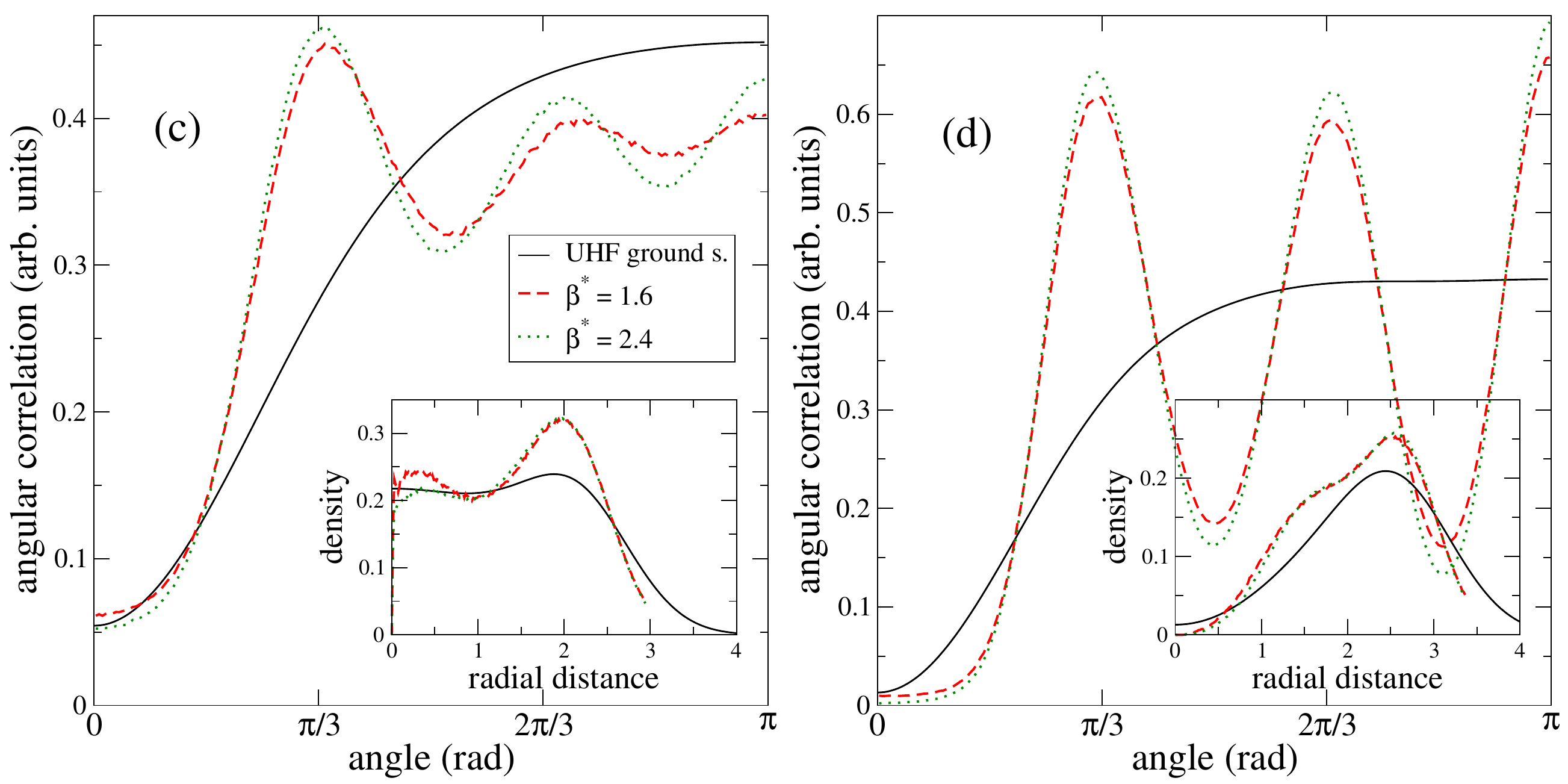}
\end{center}
\caption{\label{fig:mfp6}
  (a) The angular correlation function $g(\theta)$ for fully spin-polarized quantum dot carbon ($N=6$) at $\gamma=0.8$ and $\lambda=3.5$.
  PIMC confirms the sixfold configuration in a ring.
  (b) As the coupling increases to $\lambda=4.5$, the angular correlation shows fivefold symmetry.
  At the same time, a particle density peak forms in the center.
  (c) The same at $\gamma=1.4$ and $\lambda=2$.
  (d) The same at $\gamma=1.4$ and $\lambda=4$, which is slightly beyond the upper end of the
  $S_z=3$ interval in UHF. Label A in Fig.~\ref{fig:carbon}.
}
\end{figure*}

To determine the spatial structure in the large $\lambda$ limit, we performed PIMC simulations for $\gamma=0.8$ and 1.4.
The results are shown in Fig.~\ref{fig:mfp6}.
At $\gamma=0.8$, PIMC demonstrates that the hexagonal ring configuration gives way to a pentagonal structure $(3,C_{5v})$,
although the peaks in the ACF are not exactly at multiples of $2\pi/5$.
By contrast, at $\gamma=1.4$ no such transition is observed, and the symmetry remains sixfold up to the highest couplings studied.
The radial density shows two nearby rings, cf.\ the inset of Fig.~\ref{fig:mfp6}(c,d);
the two radii approach each other with increasing $\lambda$.
We interpret this as a distortion of the hexagon in the last item of Fig.~\ref{fig:carbon};
the symmetry is $C_{3v}$, but the strong coupling limit approaches $C_{6v}$.
It would be interesting to know if these optimized structures make the large-$\lambda$ limit fully polarized;
unfortunately, our current method is unable to decide this issue.

\subsubsection{The $S_z=0$ subspace of $N=6$}

\begin{figure*}[htbp]
\begin{center}
\includegraphics[width=0.36\textwidth, keepaspectratio]{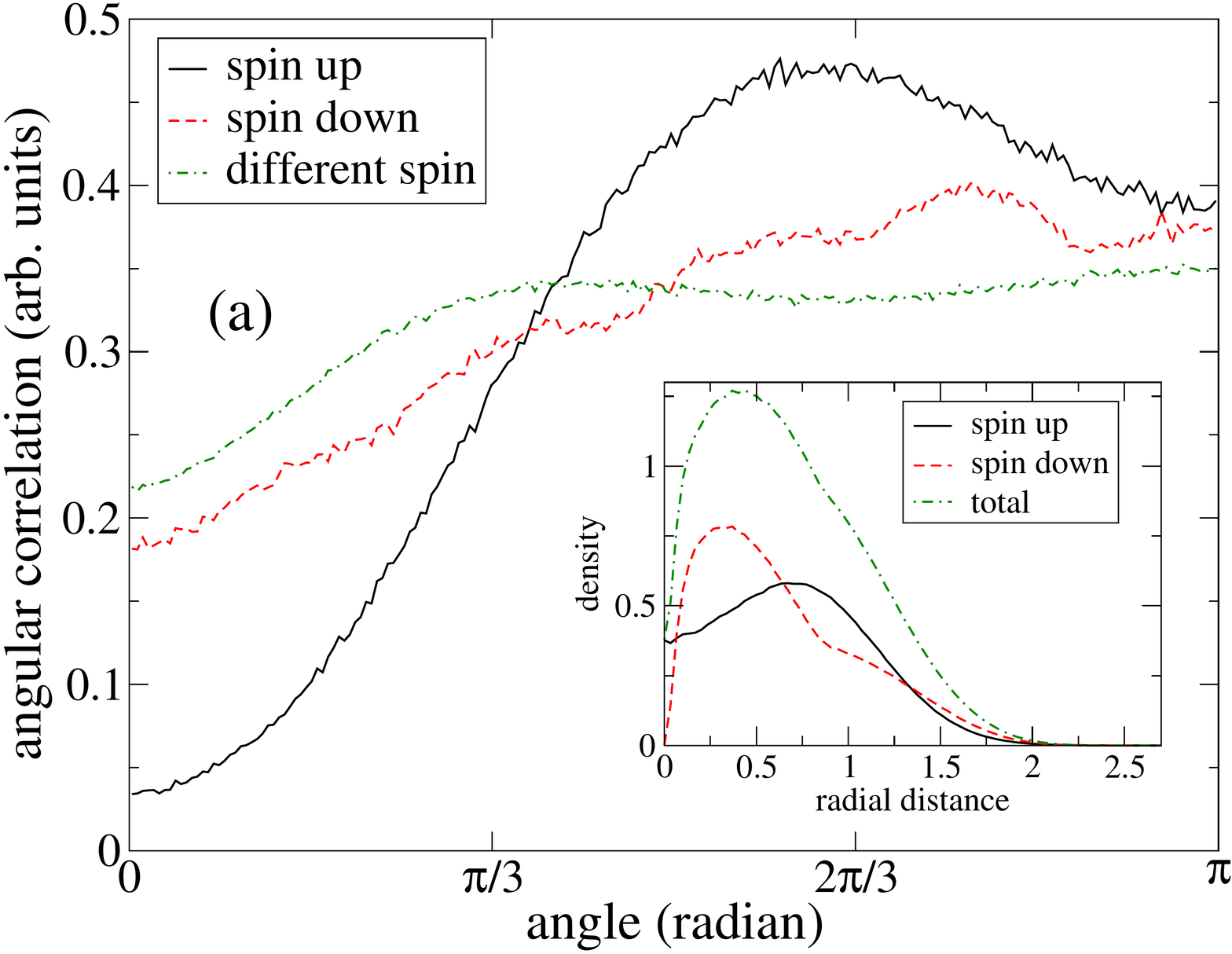}
\includegraphics[width=0.36\textwidth, keepaspectratio]{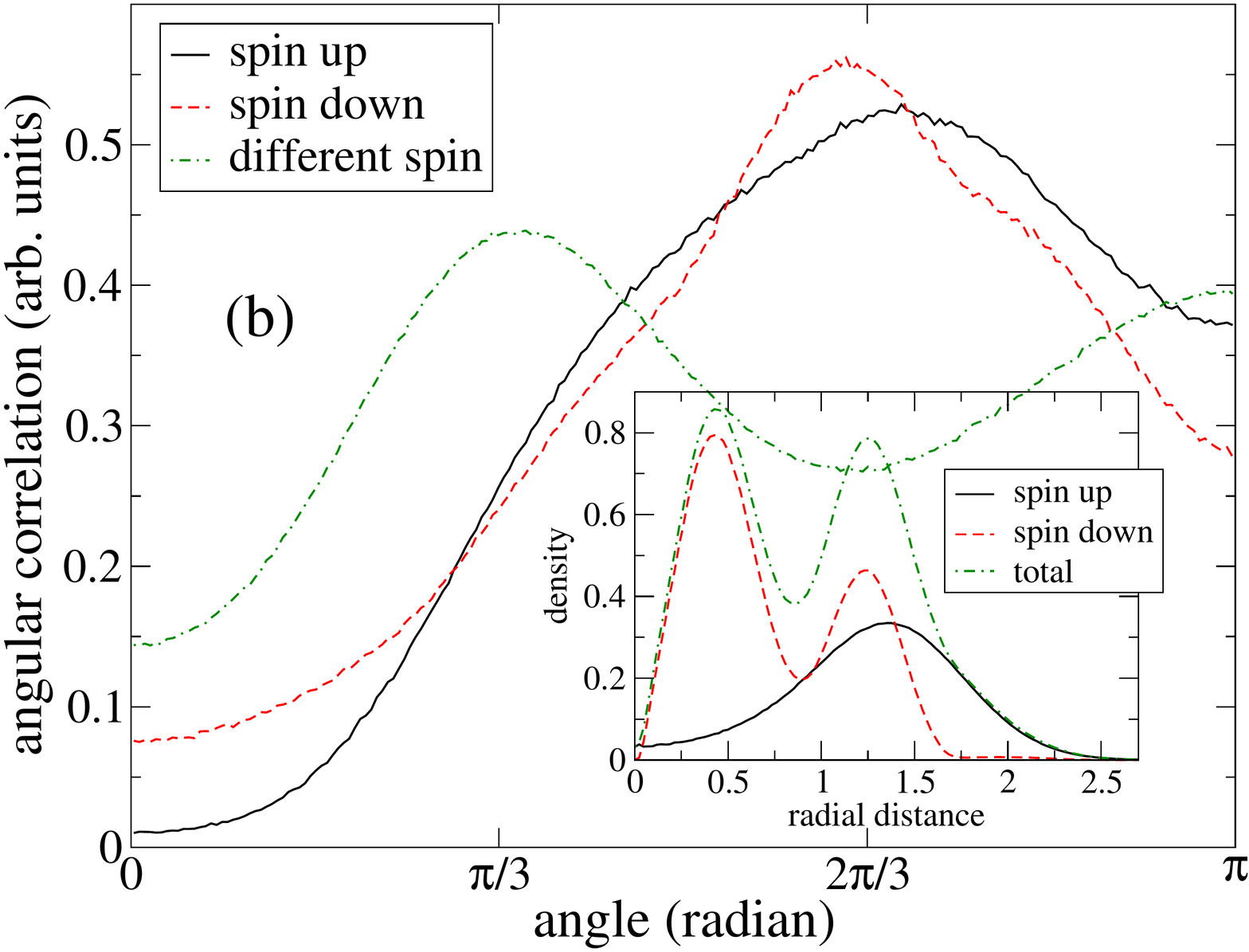}
\includegraphics[width=0.36\textwidth, keepaspectratio]{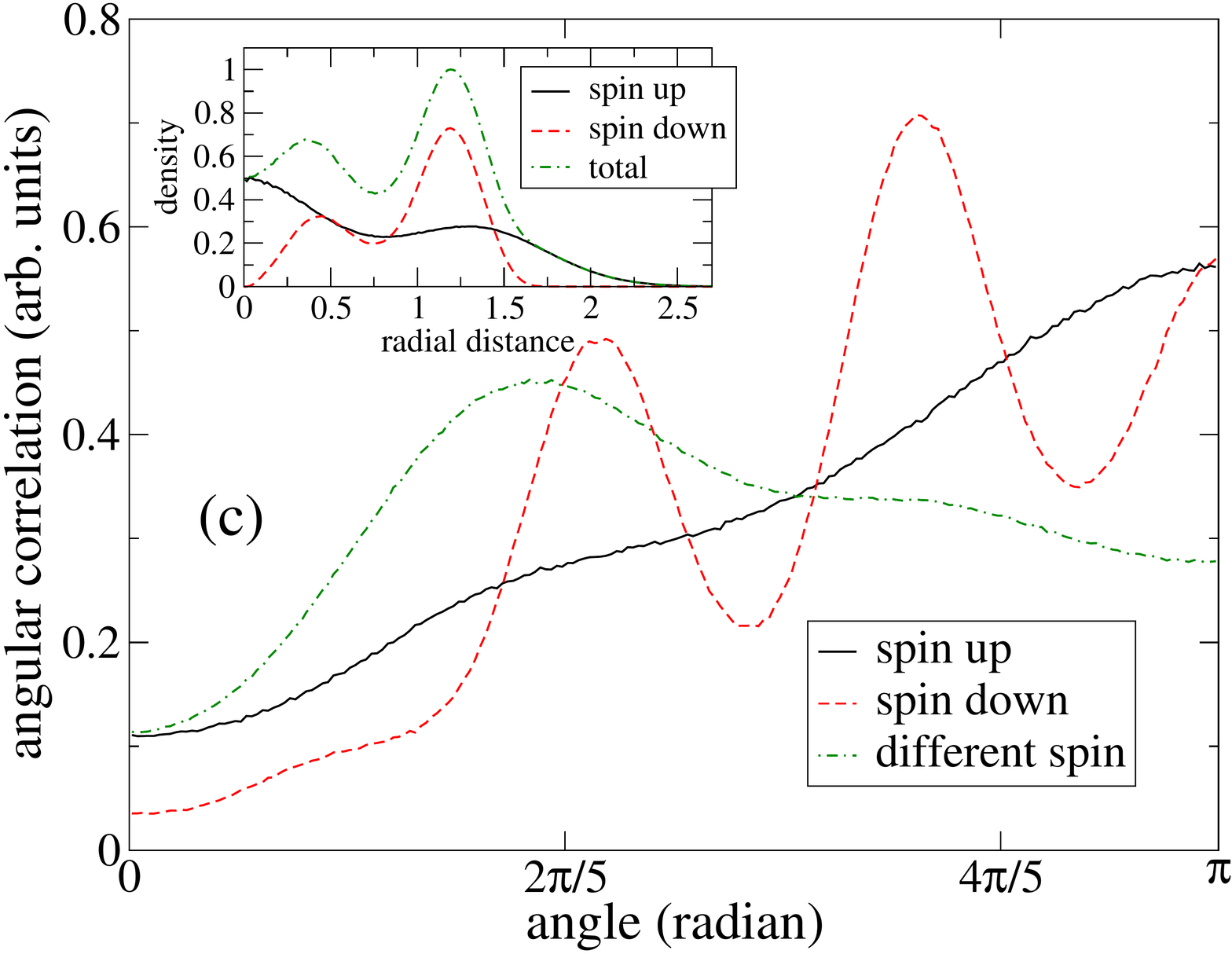}
\includegraphics[width=0.36\textwidth, keepaspectratio]{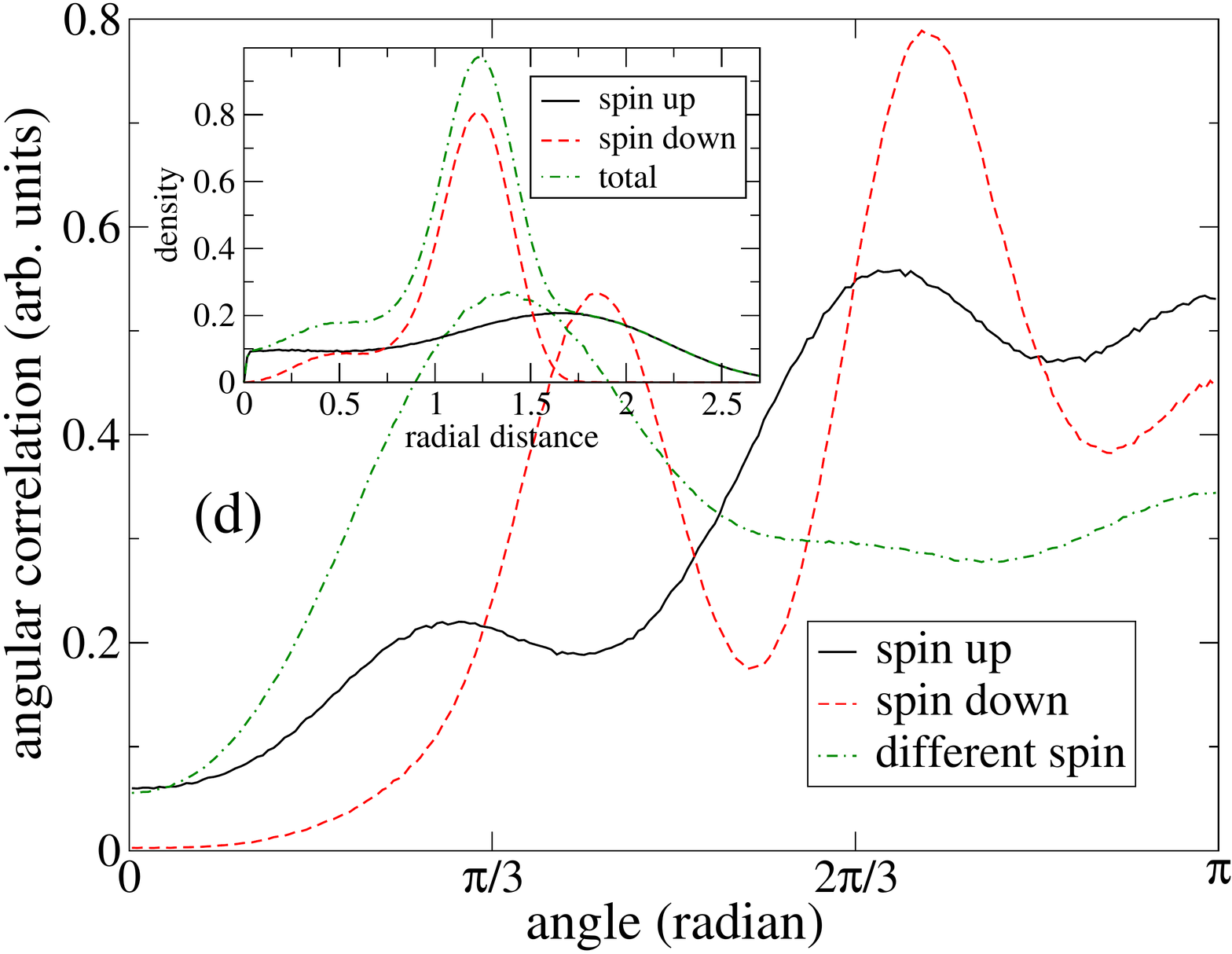}
\end{center}
\caption{\label{fig:msz0_6}
  The angular correlation functions and the radial density (insets) for quantum dot carbon ($N=6$) of spin $S_z=0$.
  (a) At $\gamma=0.2$ and $\lambda=0.5$, PIMC shows $(0,C_{3v})$ correlations while the UHF ground state is $(0,C_\infty)$.
  (b) At $\gamma=0.4$ and $\lambda=1.4$, UHF and PIMC agree on the $(0,C_{3v})$ state.
  (c) At $\gamma=0.8$ and $\lambda=0.8$, $(0,C_s)$ type correlations emerge on the UHF $(0,C_\infty)$-$(0,C_{3v})$ phase
  boundary.
  (d) At $\gamma=0.8$ and $\lambda=1.7$, UHF predicts $(0,C_s)$ but PIMC indicates $(0,C_{3v})$ type correlations.
  $\beta^\ast=2.4$ in all panels. Label B in Fig.~\ref{fig:carbon}.
}
\end{figure*}

In this sector PIMC predicts remarkable deviations from the UHF structure.
For small fields, $\gamma=0.2$ and 0.4 shown in Figs.~\ref{fig:msz0_6}(a) and \ref{fig:msz0_6}(b),
a multiple ring structure gradually emerges in the density with increasing $\lambda$.
The radius of the greater spin-down ring is approximately the same as the radius of the single spin-up ring,
but there is a smaller spin-down ring as well.
The same-spin angular correlation suggests triangular ordering, and
the different-spin correlations are peaked near $\theta=\pi/3$.
Apparently, this is a symmetry-reduced version of the $(0,C_{3v})$ state, with the spin-down triangle being isosceles,
and possibly reduced in size.
As these may involve some loss of exchange energy, UHF never favors such a low-symmetry structure,
but PIMC appreciates the energy gain from better correlations.
At a higher field $\gamma=0.8$, we find cases where the correlations are characteristic of a $(0,C_s)$
state, as seen from the density peak at the center, the almost opposite angle of spin-up electrons,
and the correlation of spin-down electrons at two angles defining an isosceles triangle;
different spins are correlated near $\theta\approx2\pi/5$, see Fig.~\ref{fig:msz0_6}(c).
On the other hand, still at $\gamma=0.8$ but at higher $\lambda$, PIMC detects a $C_{3v}$ structure
where the UHF ground state was $(0,C_s)$, see Fig.~\ref{fig:msz0_6}(d).
At intermediate coupling values, the behavior is quite mixed, precluding a transparent interpretation.

\subsubsection{The $S_z=1$ subspace of $N=6$}

\begin{figure}[htbp]
\begin{center}
\includegraphics[width=\columnwidth, keepaspectratio]{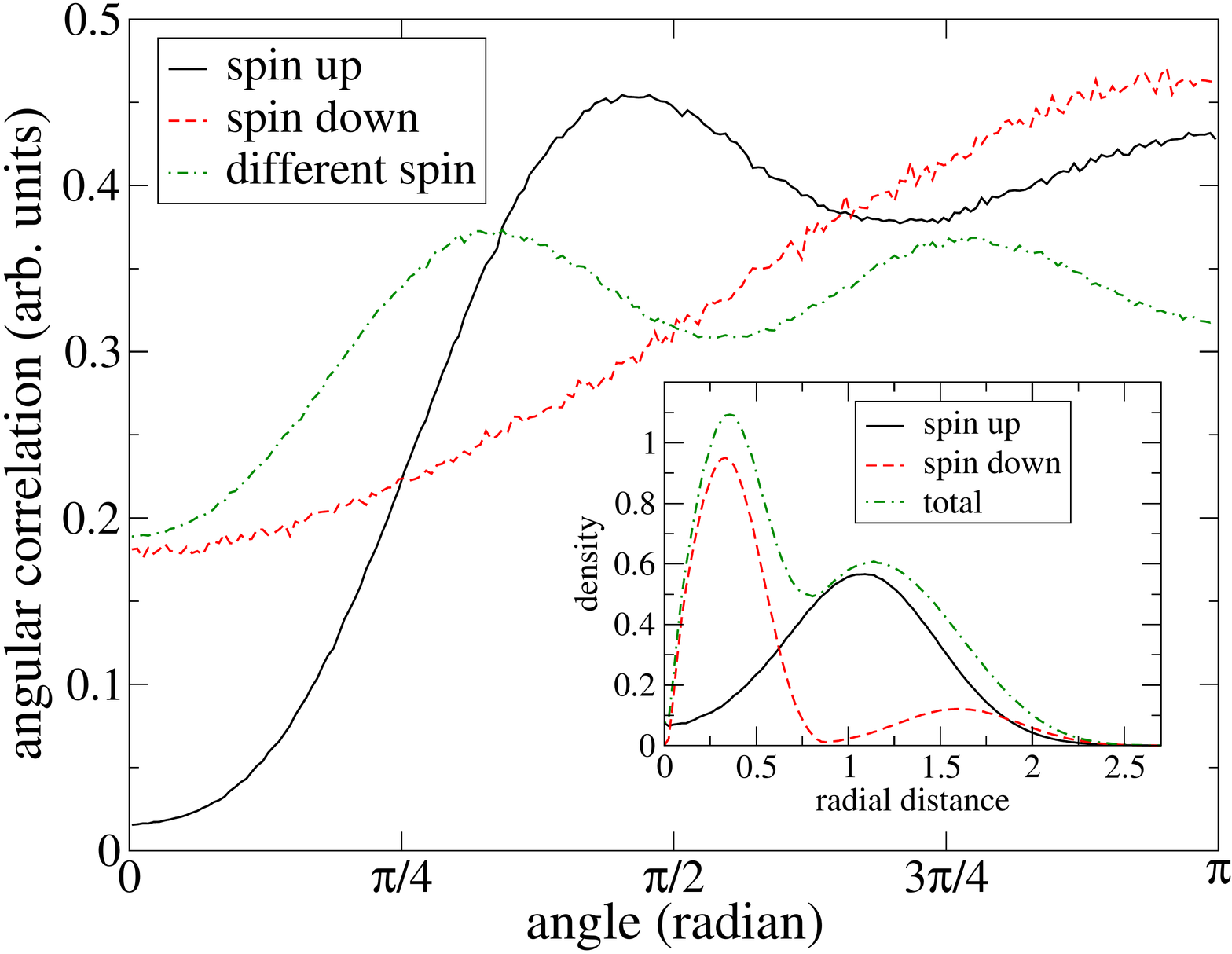}
\end{center}
\caption{\label{fig:msz1_6}
  The angular correlation functions $g_{\uparrow\uparrow}(\theta)$, $g_{\downarrow\downarrow}(\theta)$ and $g_{\uparrow\downarrow}(\theta)$
  in the $S_z=1$ sector of quantum dot carbon ($N=6$) at $\gamma=0.4$ and $\lambda=0.9$.
  (Inset) The total and the spin-resolved radial density at $\beta^\ast=1.6$; similar results are obtained for $\beta^\ast=0.8$ and 2.4. Label C in Fig.~\ref{fig:carbon}.
}
\end{figure}

Here PIMC finds correlated structures where UHF predicts $(1,C_\infty)$ in the $0.4\le\gamma\le 1.2$ regime.
The four up-spin electrons are located in a square configuration, as seen in the angular correlation function and the
spin-resolved radial density in Fig.~\ref{fig:msz1_6}.
One of the two down-spin electrons is located at the center of the trap, the other is off-center, avoiding the
vertices of the up-spin square structure,
as seen in the peaks at $\theta=\pi/4$ and $3\pi/4$ in the different spin angular correlation function.
At higher $\gamma$, there is an additional feature: the down-spin electrons tend to occur at opposite sides of the trap,
and, at the same time, the density has a dip at the center.

\subsubsection{The $S_z=2$ subspace of $N=6$}

\begin{figure*}[htbp]
\begin{center}
\includegraphics[width=0.32\textwidth, keepaspectratio]{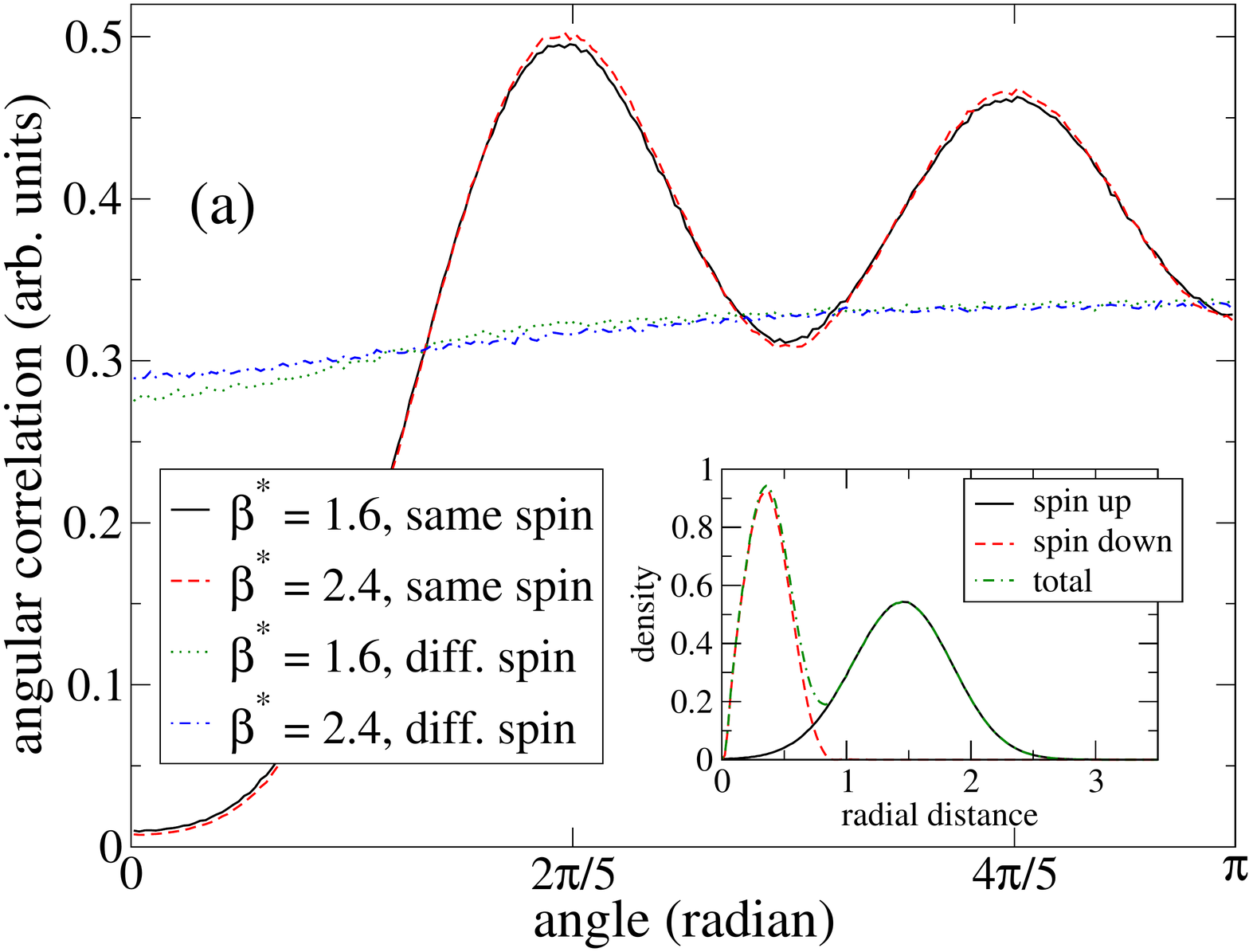}
\includegraphics[width=0.32\textwidth, keepaspectratio]{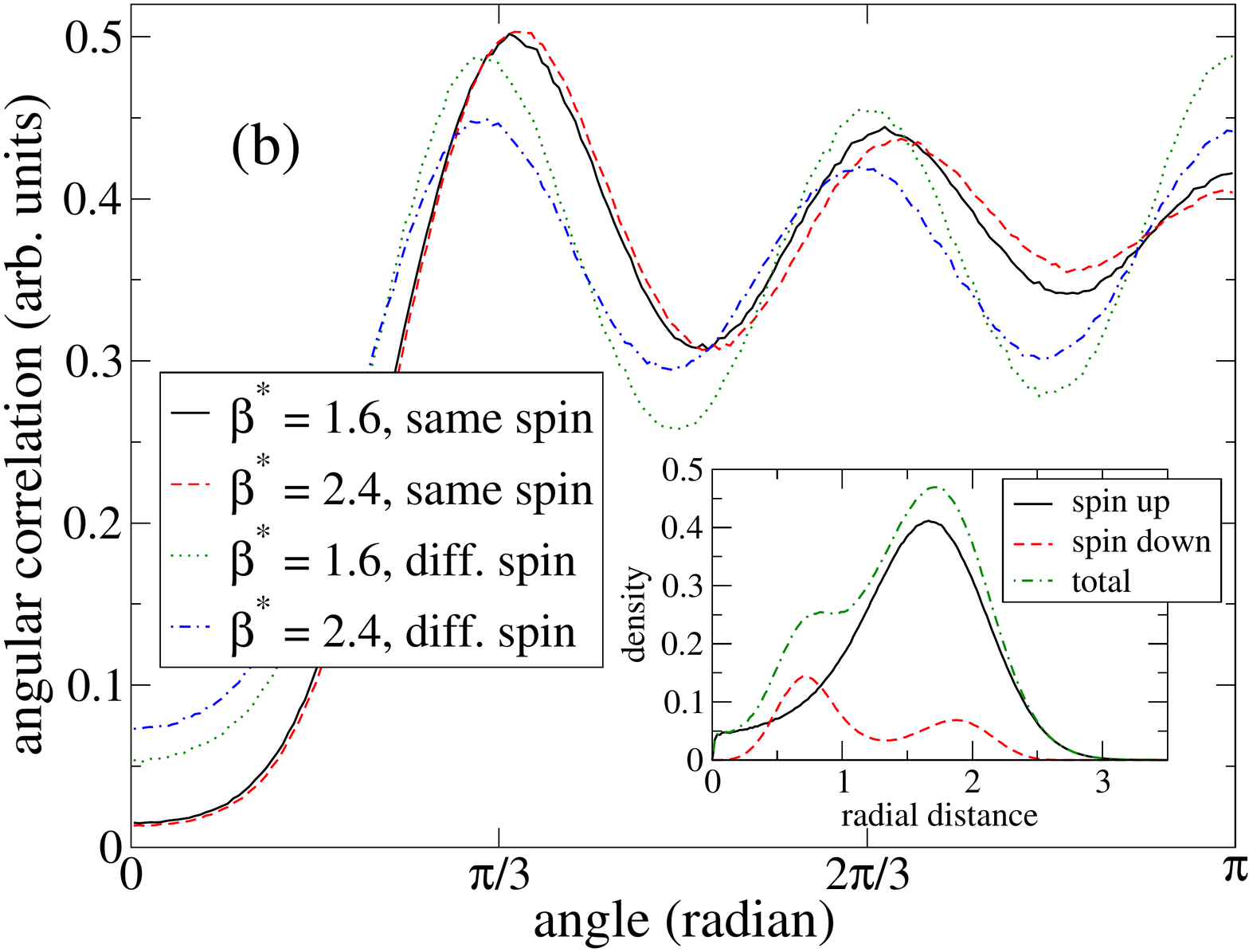}
\includegraphics[width=0.32\textwidth, keepaspectratio]{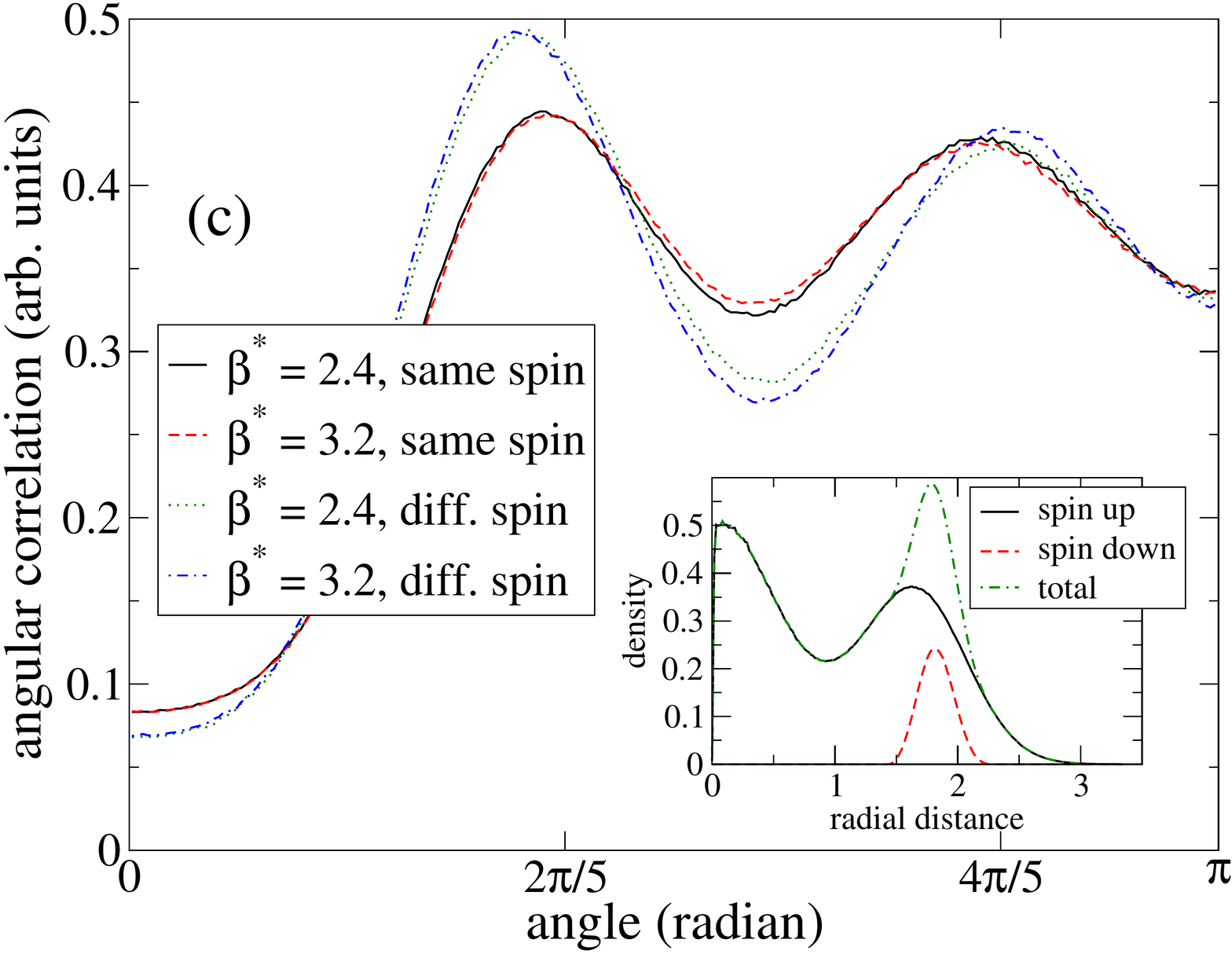}

\includegraphics[width=0.32\textwidth, keepaspectratio]{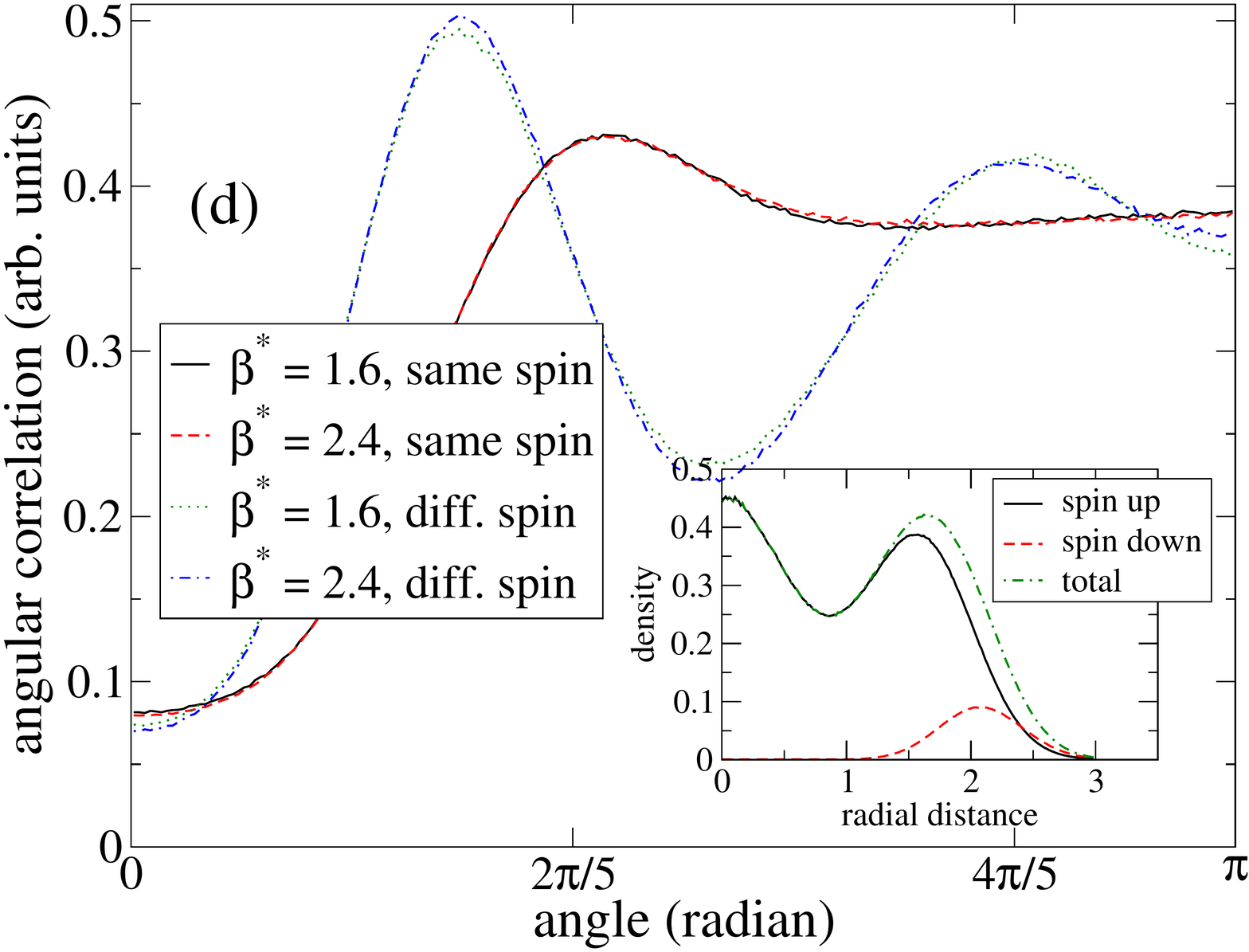}
\includegraphics[width=0.27\textwidth, keepaspectratio]{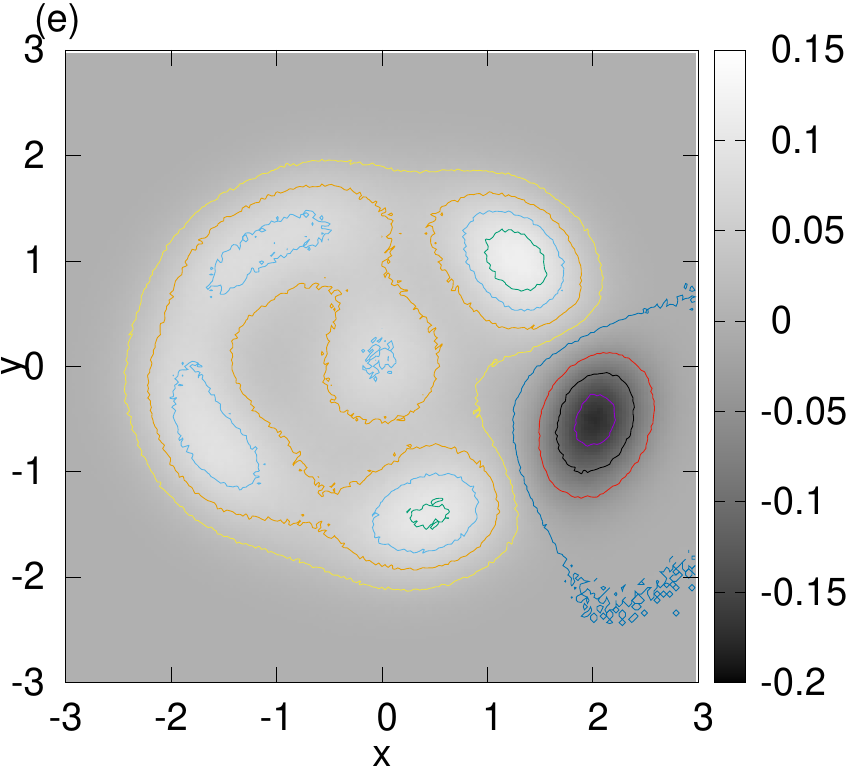}
\end{center}
\caption{\label{fig:msz2_6}
  (a) The angular correlation function $g(\theta)$ for quantum dot carbon ($N=6$) of spin $S_z=2$ at $\gamma=0.4$ and $\lambda=1.9$.
  (b) The radial density and $g(\theta)$ at $\gamma=0.6$ and $\lambda=2.7$.
  (c) The angular correlation function at $\gamma=0.8$ and $\lambda=1.8$.
  [(d) and (e)] The same at $\gamma=1$ and $\lambda=1.3$, and the spin density histogram for $\beta^\ast=2.4$. Label D in Fig.~\ref{fig:carbon}.
}
\end{figure*}

In this sector, whose $C_\infty$, $C_s$, and $C_{5v}$ states are relevant in UHF in intervals for the magnetic fields we studied,
PIMC often and unequivocally overrides the UHF structures.
At $\gamma=0.2$ the correlations of the $C_s$ state are present as a weak modulation of the angular correlation.
At $\gamma=0.4$, however, PIMC shows a transition from the $C_s$ to the $C_{5v}$ state:
as Fig.~\ref{fig:msz2_6}(a) demonstrates for $\lambda=1.9$, the same-spin angular correlation shows a pentagonal structure but the
single down-spin electron is not correlated with the majority spins, and it is at the center of the dot in the spin density (inset).
This applies in the whole interval of $S_z=2$ at $\gamma=0.4$, although for higher $\lambda$'s only at the lowest temperature we studied.
At $\gamma=0.6$, on the other hand, a ring structure emerges in the density, and both the same-spin and the different-spin
angular correlations show peaks at multiples of $\theta=\pi/3$ [see Fig.~\ref{fig:msz2_6}(b)];
this overrides both the $C_\infty$ and $C_{5v}$ predictions of UHF.
We identify this state as $(2,C_s')$, electrons localized at the vertices of a hexagon, one of them with a down spin.
On the other hand, PIMC confirms the $(3,C_{5v})$ structure in the high-$\lambda$ limit.
At $\gamma=0.8$, PIMC predicts $C_s$ symmetry instead of $C_{5v}$, as seen from the similarity of the same-spin and the different-spin
angular correlations, c.f.\ Fig.~\ref{fig:msz2_6}(c).
In the intervals of the $(2,C_\infty)$ state at $\gamma\ge1$ PIMC clearly shows the emergence of the correlations of
the $(2,C_s)$ structure as seen in Figs.~\ref{fig:msz2_6}(d) and \ref{fig:msz2_6}(e).

\subsection{Quantum dot nitrogen, $N=7$}

\begin{figure*}
\begin{center}
  \includegraphics[width=\textwidth, keepaspectratio]{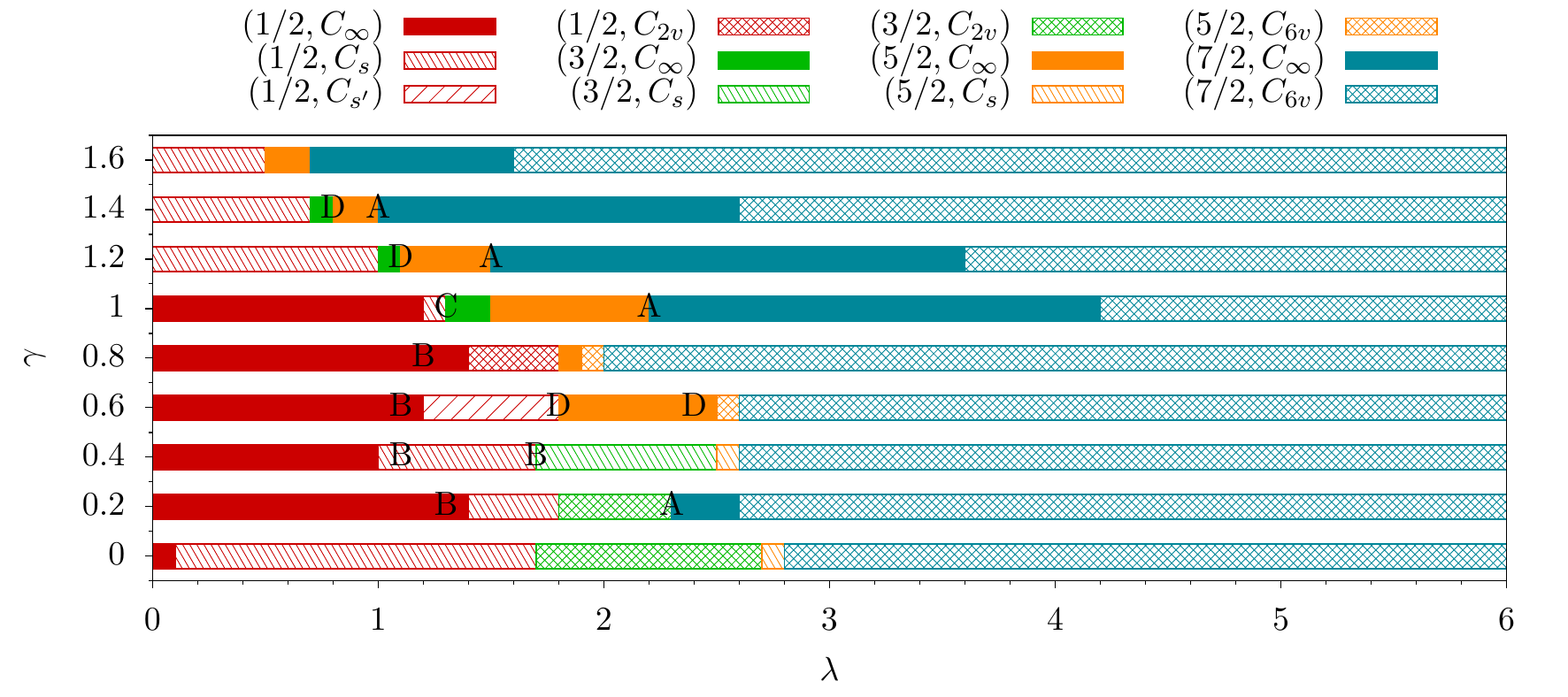}
  \def\svgwidth{\textwidth}
 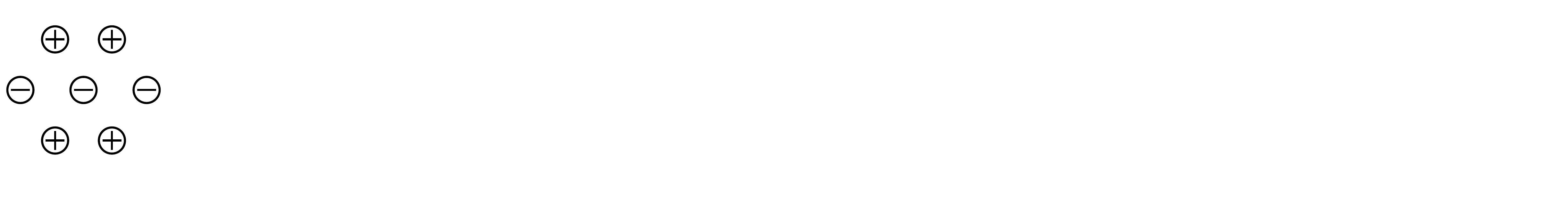
\end{center}
\caption{\label{fig:nitrogen}
  (Top)
  Sections of the UHF phase diagram of quantum dot nitrogen ($N=7$) at the magnetic field parameter $\gamma=\frac{\omega_c}{\omega}$
  fixed at multiples of 0.2.
  Labels A, B, C, and D indicate PIMC parameters in Figs.~\ref{fig:mfp7}, \ref{fig:msz0.5_7}, \ref{fig:msz1.5_7}, and~\ref{fig:msz2.5_7},
  respectively.
  (Bottom)
  Sketch of the spin density peaks in the relevant symmetry-breaking states of quantum dot nitrogen.
}
\end{figure*}

Apart from the four rotationally invariant states $(S_z,C_\infty)$ (where $S_z=\frac{1}{2}$, $\frac{3}{2}$,
$\frac{5}{2}$, or $\frac{7}{2}$),
the relevant symmetry-breaking states are sketched in Fig.~\ref{fig:nitrogen}.

A hexagon with an electron at the center is just a finite piece of a Wigner crystal, and as Fig.~\ref{fig:nitrogen}
shows, almost all symmetry-breaking UHF ground states are of this type, with different spin structures.
The only exception is the $(\frac{1}{2},C_s)$ state, which has a small interval of relevance for $\gamma=0.2$ and 0.4.
For strong coupling, UHF always predicts full spin polarization, a $(\frac{7}{2},C_{6v})$ state.

\subsubsection{The $S_z=\frac{7}{2}$ subspace of $N=7$}

\begin{figure}[htbp]
\begin{center}
\includegraphics[width=\columnwidth, keepaspectratio]{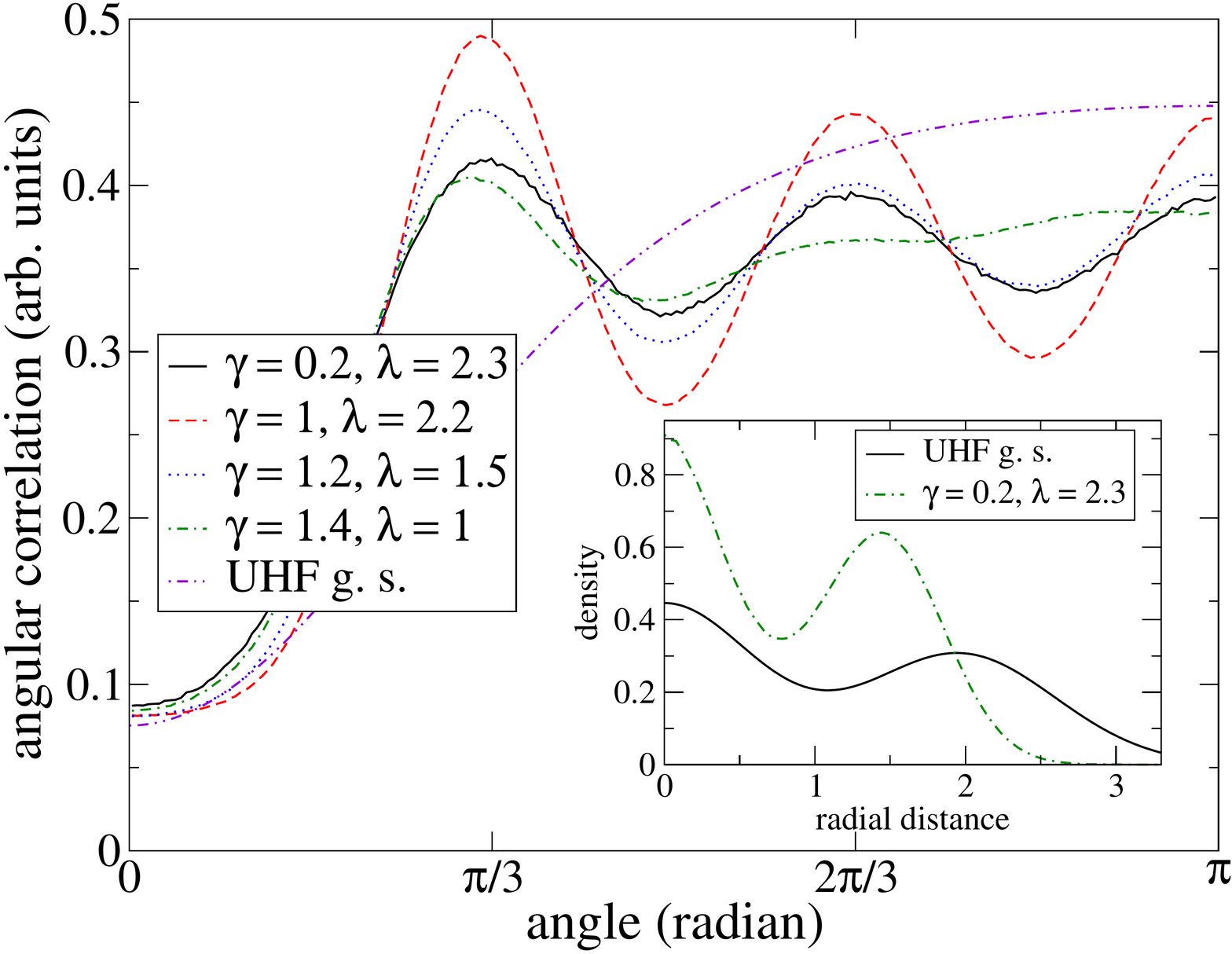}
\end{center}
\caption{\label{fig:mfp7}
  The angular correlation function $g(\theta)$ for fully spin-polarized quantum dot nitrogen ($N=7$) for $\gamma=0.2$, 1, 1.2, and 1.4,
  at the coupling strength values just following the value where UHF predicts full spin-polarization.
  The inverse temperature is  $\beta^\ast=2.4$.
  The inset compares the radial density for $\gamma=0.2$ and $\lambda=2.3$ from PIMC and UHF. Label A in Fig.~\ref{fig:nitrogen}.
  }
\end{figure}

Again, we can prove that the recurrence of rotational
symmetry upon flipping the last spin is a mistake of UHF.
As seen in Fig.~\ref{fig:mfp7}, for $\gamma=0.2$, 1, 1.2, and 1.4 the angular correlation function indicates a hexagonal
arrangement, even though for $\gamma=1.4$ this correlation is rather weak at first, but gets stronger with increasing $\lambda$.

\subsubsection{The $S_z=\frac{1}{2}$ subspace of $N=7$}

\begin{figure*}[htbp]
\begin{center}
\includegraphics[width=0.32\textwidth, keepaspectratio]{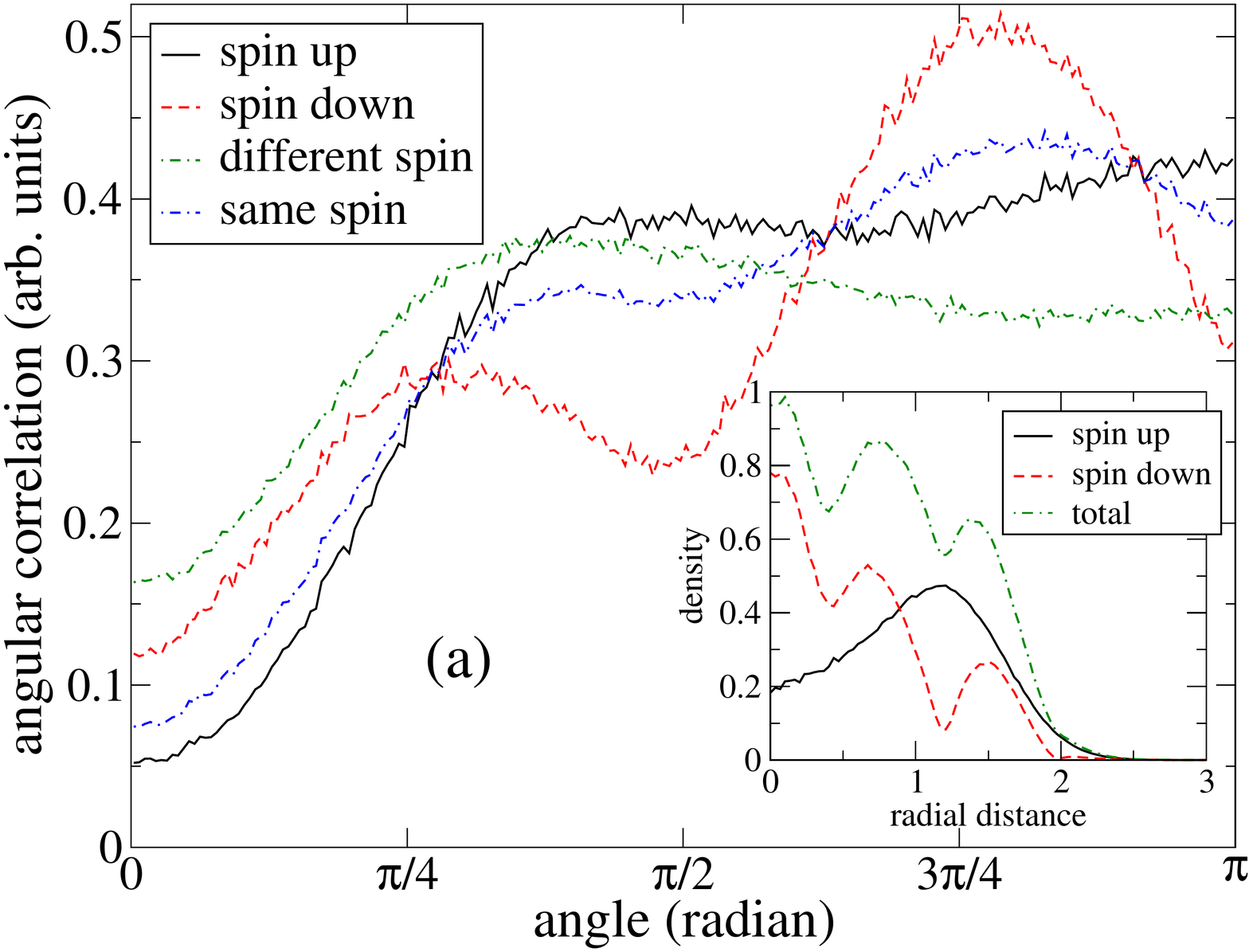}
\includegraphics[width=0.32\textwidth, keepaspectratio]{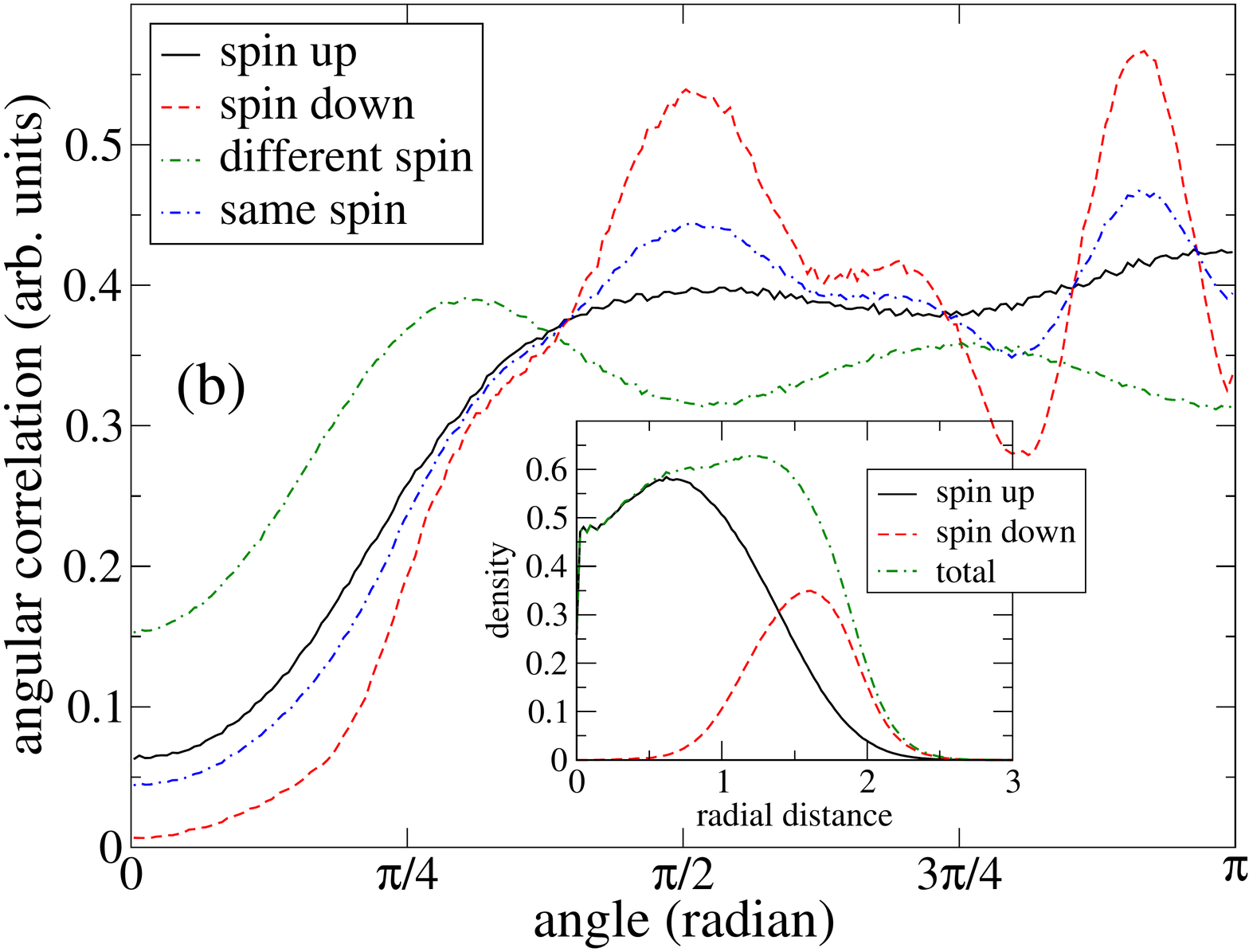}
\includegraphics[width=0.32\textwidth, keepaspectratio]{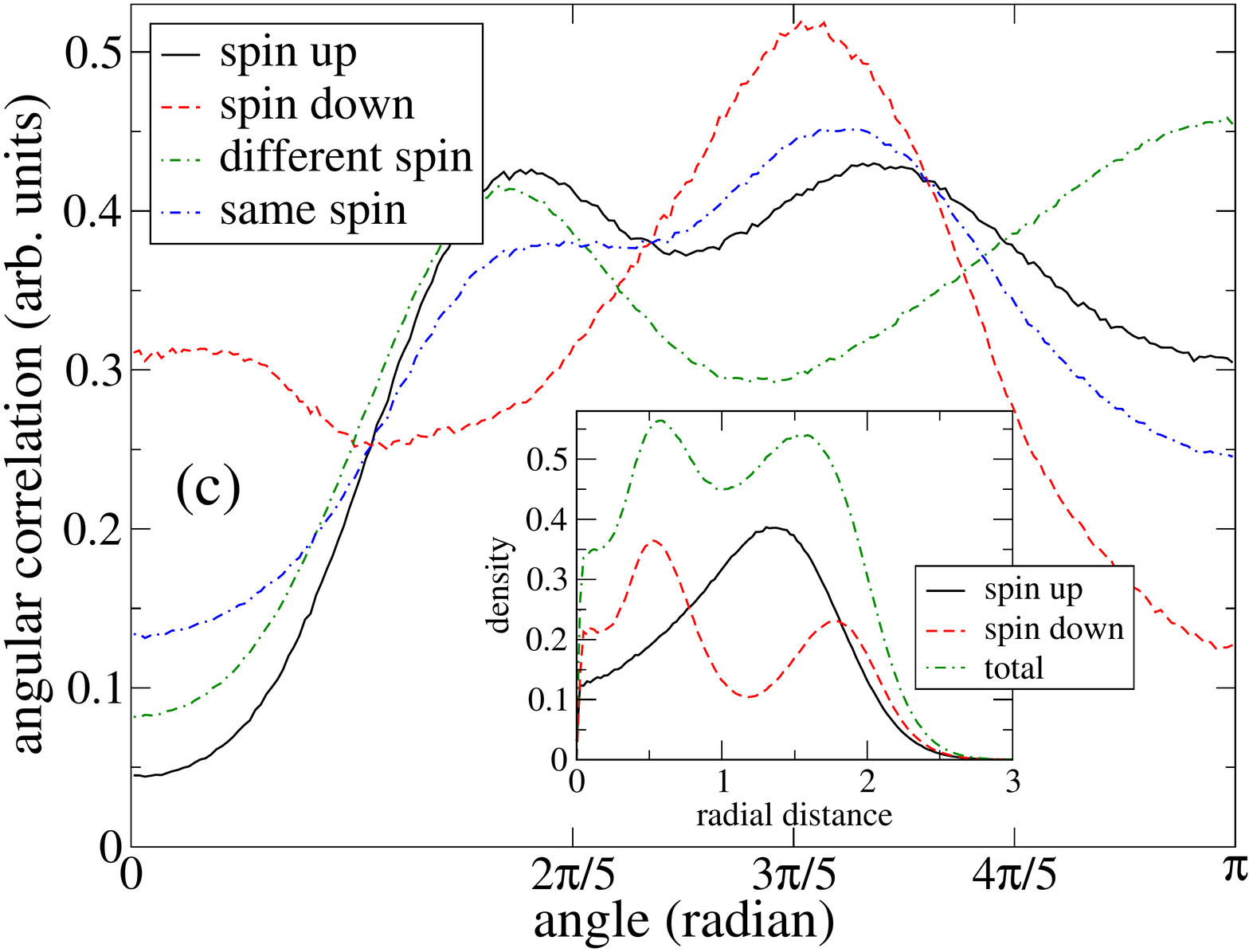}

\includegraphics[width=0.32\textwidth, keepaspectratio]{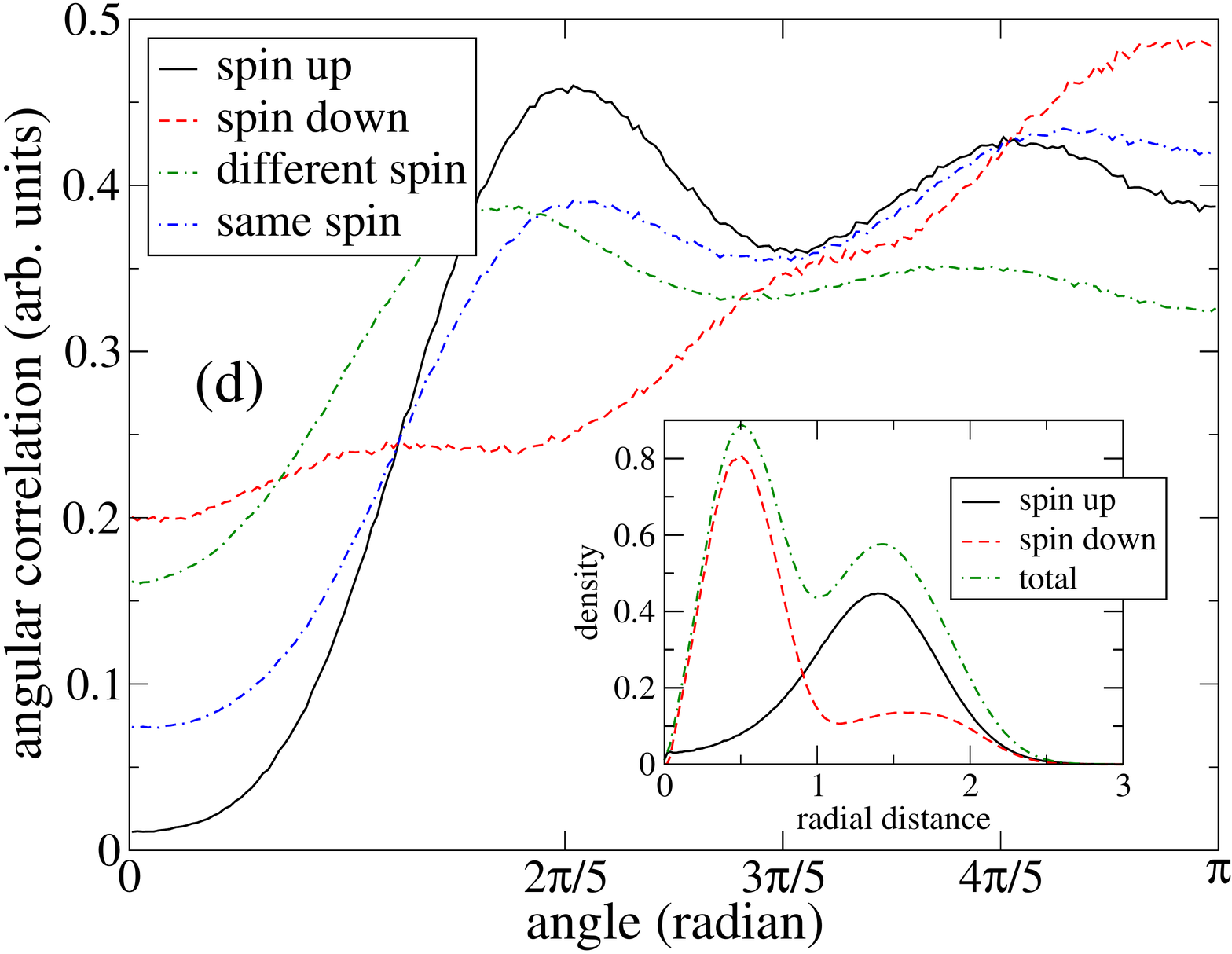}
\includegraphics[width=0.32\textwidth, keepaspectratio]{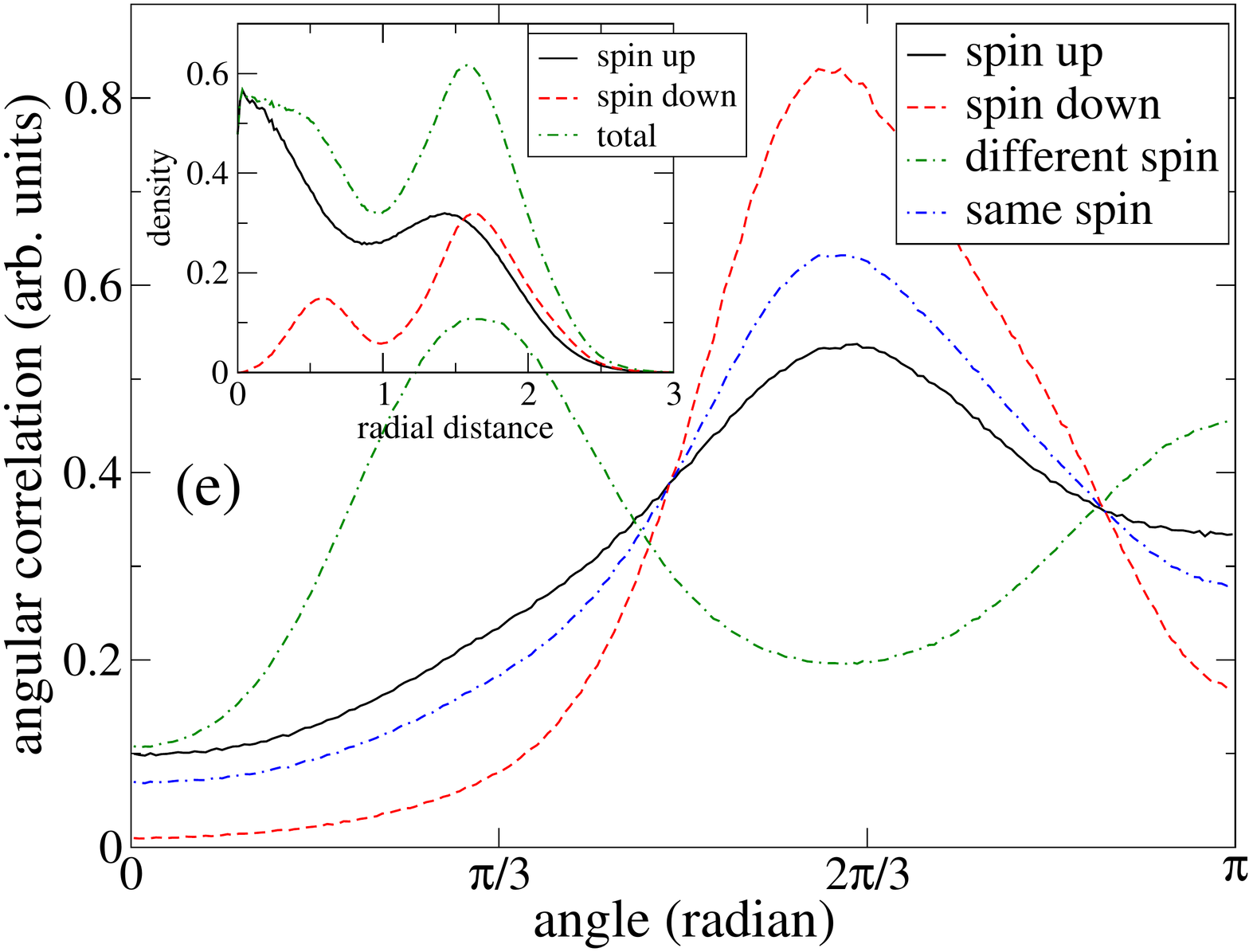}
\end{center}
\caption{\label{fig:msz0.5_7}
  Angular correlation and radial density (insets) by PIMC in the $S_z=\frac{1}{2}$ sector of quantum dot nitrogen ($N=7$).
  The parameters are
  (a) $\gamma=0.2$ and $\lambda=1.3$,
  (b) $\gamma=0.4$ and $\lambda=1.1$,
  (c) $\gamma=0.4$ and $\lambda=1.7$,
  (d) $\gamma=0.6$ and $\lambda=1.1$, and
  (e) $\gamma=0.8$ and $\lambda=1.2$.
  Label B in Fig.~\ref{fig:nitrogen}.
}
\end{figure*}

In the least polarized sector, which is relevant for the UHF ground state at small coupling,
UHF predicts a transition from the rotationally symmetric state to the $C_s$, $C_s'$, or $C_{2v}$ molecules.
PIMC, by contrast, discovers a rich variety of correlated structures.
For $\gamma=0.2$ and $\lambda=1.3$, still in the $(\frac{1}{2},C_\infty)$ range by UHF,
the spin-resolved radial density finds a spin-down at the center, a ring of four spin-up electrons,
and two spin-down ones at slightly greater and at slightly smaller radii than the radius of the ring; cf.\ Fig.~\ref{fig:msz0.5_7}(a).
The spin-up ACF shows a very weak angular correlation, consistent with unlocalized, liquid-like
behavior of the spin-up electrons in the ring.
The down-spin ACF has a strong peak at $\theta=3\pi/4$ and a weaker one near $\theta=\pi/4$;
this fairly rigid structure of presumably two spin-down electrons (the central electron is typically
uncorrelated with the rest) in the background of unlocalized
spin-ups in the ring is difficult to identify with a localized structure.
Moreover, this behavior is independent of temperature in the $0.8\le\beta^\ast\le2.4$ range, and
it is quite robust: we find it with only quantitative differences for other $\lambda$'s at $\gamma=0.2$.
The change of the UHF ground state from $C_\infty$ to $C_s$ at $\lambda=1.4$ has no effect.
At $\gamma=0.4$ we find something else: for intermediate couplings $\lambda=0.9$ and 1.1,
a square arrangement of the spin-up electrons is manifest, but the spin-down electrons form a ring
at a slightly greater radius; cf.\ Fig.~\ref{fig:msz0.5_7}(b).
The spin-down electrons are close to three side centers of the square, as seen form the $g_{\downarrow\downarrow}(\theta)$ and
$g_{\uparrow\downarrow}(\theta)$ correlation functions.
As seen in Fig.~\ref{fig:msz0.5_7}(c), at the upper end of the $S_z=\frac{1}{2}$ interval, $\lambda=1.7$, a different
order emerges: a spin-down electron near the center is surrounded by a slightly distorted pentagon that contains
one spin-down electron; the other spin-down electron is outside; we find similar order for $\gamma=0.8$ and $\lambda=1.7$.
On the other hand, the $(\frac{1}{2},C_s)$ structure is clearly discernible in PIMC at $\gamma=0.6$,
as seen in Fig.~\ref{fig:msz0.5_7}(d): there is a central ring of two spin-down electrons,
and the remaining spin-down electron and the four spin-up ones are arranged in a pentagon.
Such correlations continue in the small-$\lambda$ region where UHF finds a rotationally symmetric ground state.
At $\gamma=0.8$ and intermediate couplings $\lambda=1.2$ and 1.4 (on two sides of the $C_\infty\to C_{2v}$ UHF
transition), there is a central spin-up electron, and the angular correlation functions
show a hexagonal arrangement of electrons of alternating spin around it,
as the spin-resolved ACF's are peaked near $\theta=2\pi/3$ but the different-spin ACF is peaked at $\theta=\pi/3$.
However, the radial density in Fig.~\ref{fig:msz0.5_7}(e) demonstrates that one of the spin-down electrons is at a slightly
smaller radius than the rest.

\subsubsection{The $S_z=\frac{3}{2}$ subspace of $N=7$}

\begin{figure}[htbp]
\begin{center}
\includegraphics[width=\columnwidth, keepaspectratio]{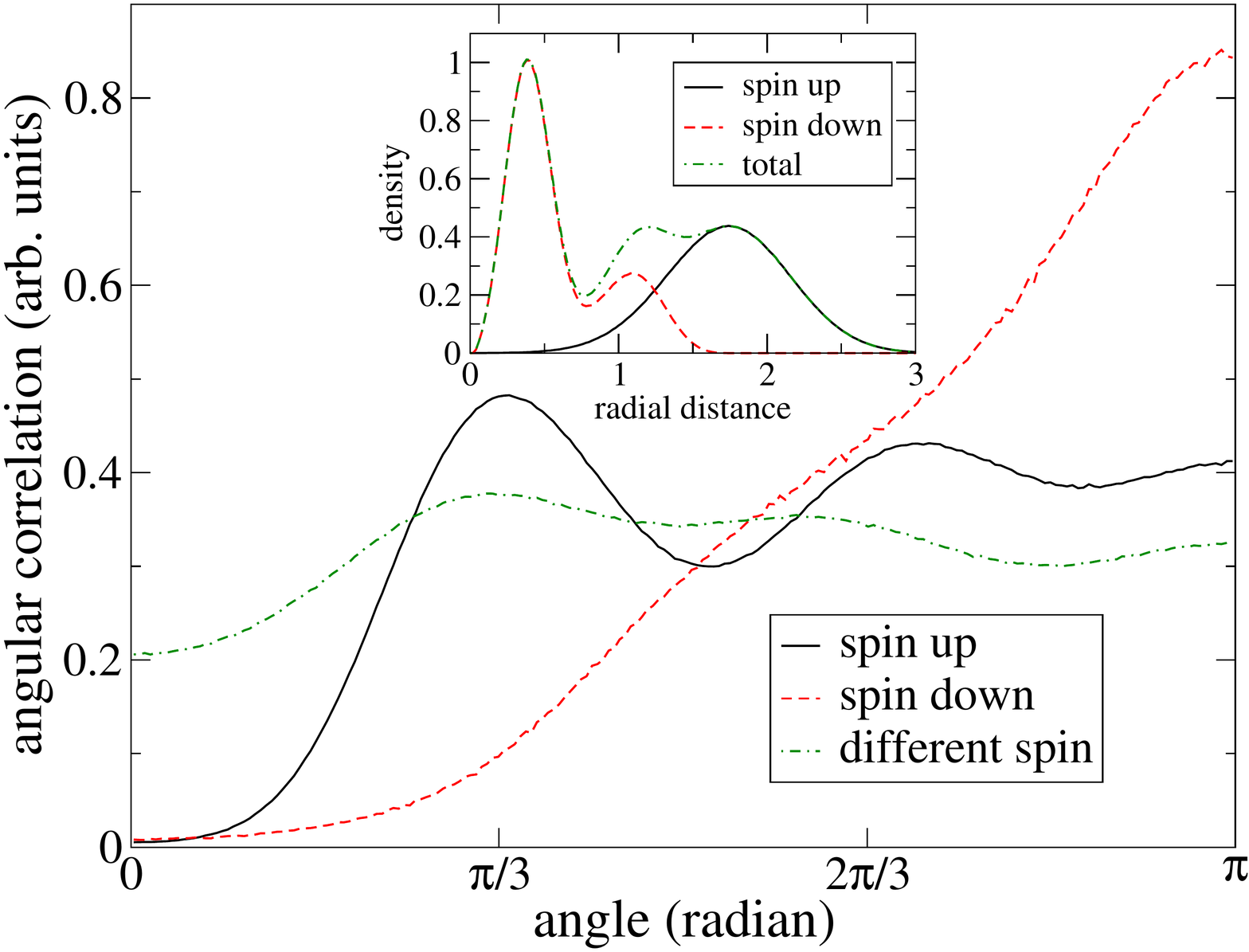}
\end{center}
\caption{\label{fig:msz1.5_7}
  The angular correlation functions and the radial particle density (inset)
  in the $S_z=\frac{3}{2}$ sector of quantum dot nitrogen ($N=7$) for $\gamma=1$, $\lambda=1.3$, and $\beta^\ast=3.2$.
  Label C in Fig.~\ref{fig:nitrogen}.
}
\end{figure}

We find remarkable correlated states where the $(\frac{3}{2},C_\infty)$ UHF ground state is relevant.
As Fig.~\ref{fig:msz1.5_7} shows for $\gamma=1$ and $\lambda=1.3$,
the density assumes a three-ring structure.
Each of the two inner rings contains a spin-down electron.
The outer ring has five spin-up electrons in a pentagonal arrangement at high temperature (not shown).
As the temperature is decreased, the angle between spin-up neighbors gradually changes
from about $2\pi/5$ to $\pi/3$, while the spin-down electron approaches the empty slot in the outer ring;
the figure shows this latter situation.
The spin-down electrons are most of the time on opposite sides of the origin.
The different-spin angular correlation function shows
that the spin-down electrons try to avoid the spin-up pentagon, as far as their correlated motion permits.
We identify this state as a distorted variant of the $(\frac{3}{2},C_s)$ structure.
At higher magnetic fields ($\gamma=1.2$, $\lambda=1$ and $\gamma=1.4$, $\lambda=0.7$),
the two down-spin rings gradually merge, and the features in the same-spin angular correlation
functions weaken, suggesting a liquid-like character.

\subsubsection{The $S_z=\frac{5}{2}$ subspace of $N=7$}

\begin{figure*}[htbp]
\begin{center}
\includegraphics[width=0.36\textwidth, keepaspectratio]{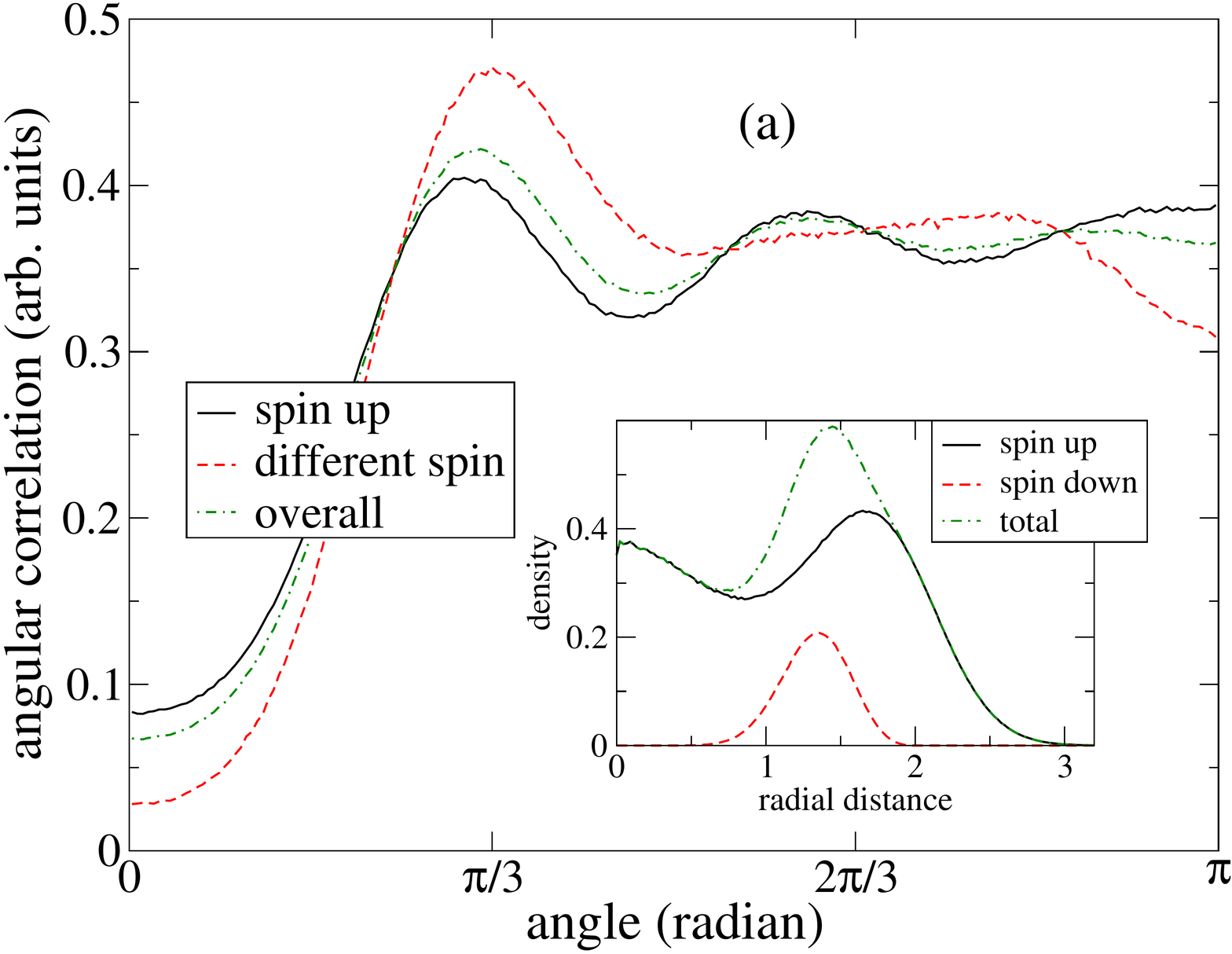}
\includegraphics[width=0.36\textwidth, keepaspectratio]{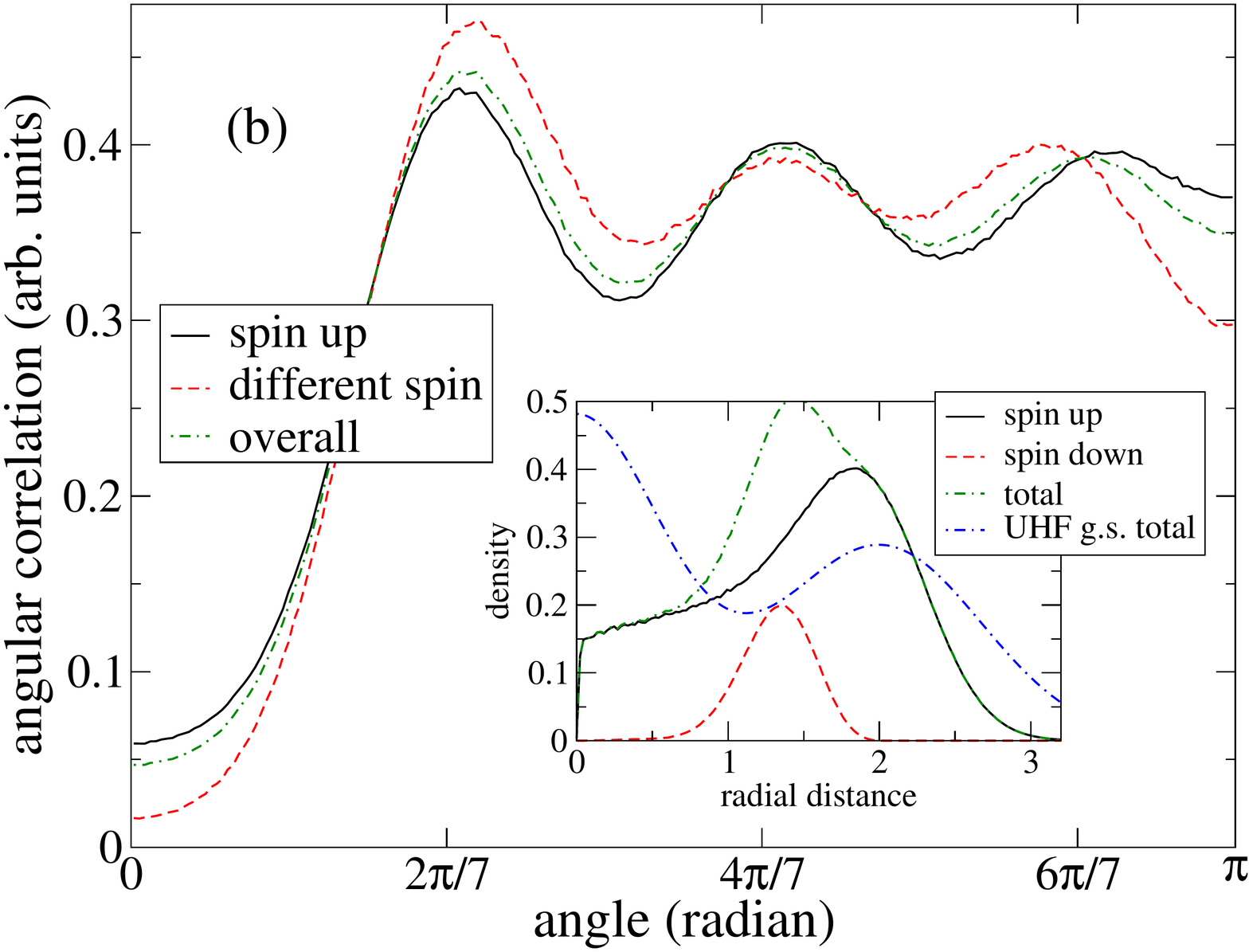}

\includegraphics[width=0.36\textwidth, keepaspectratio]{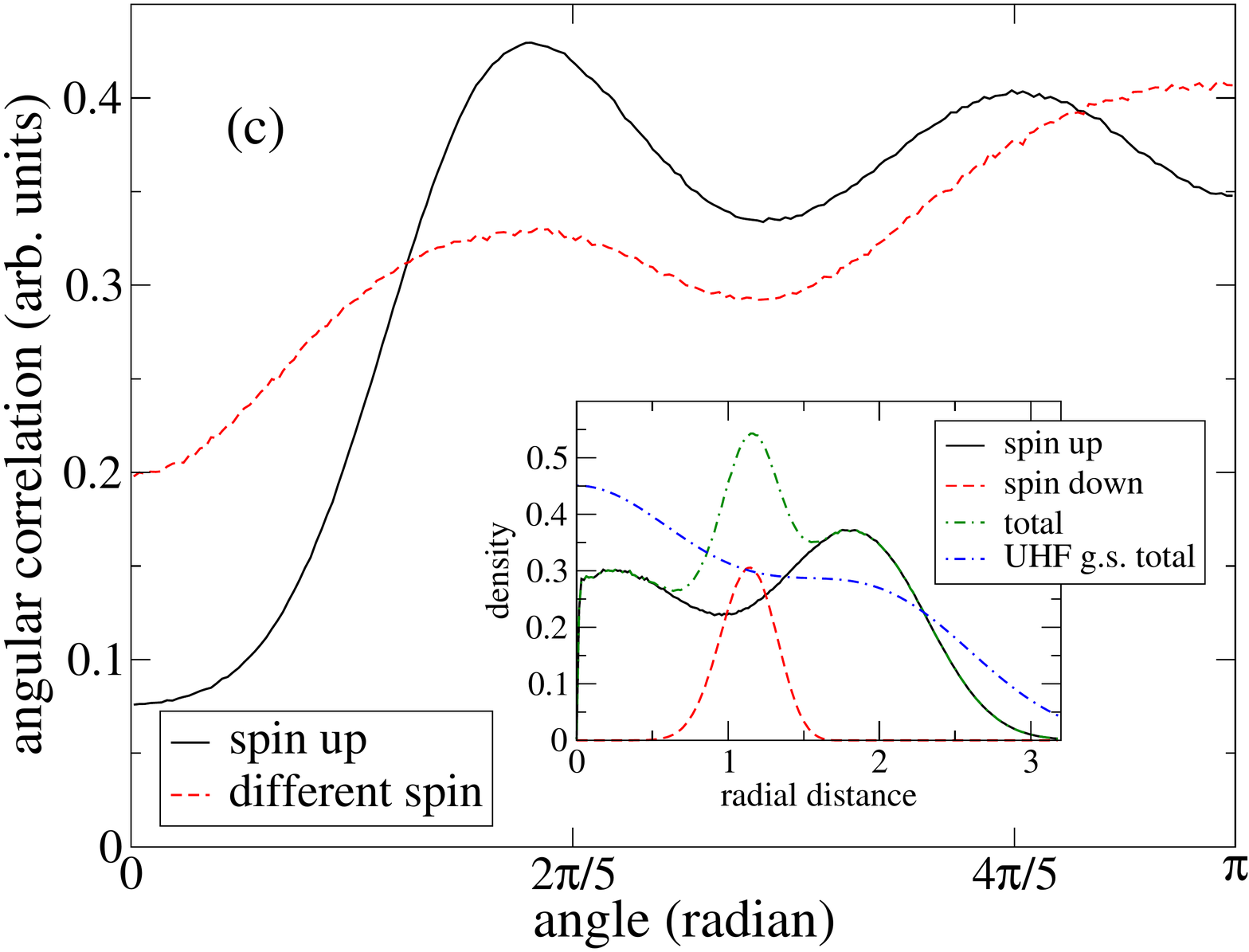}
\includegraphics[width=0.36\textwidth, keepaspectratio]{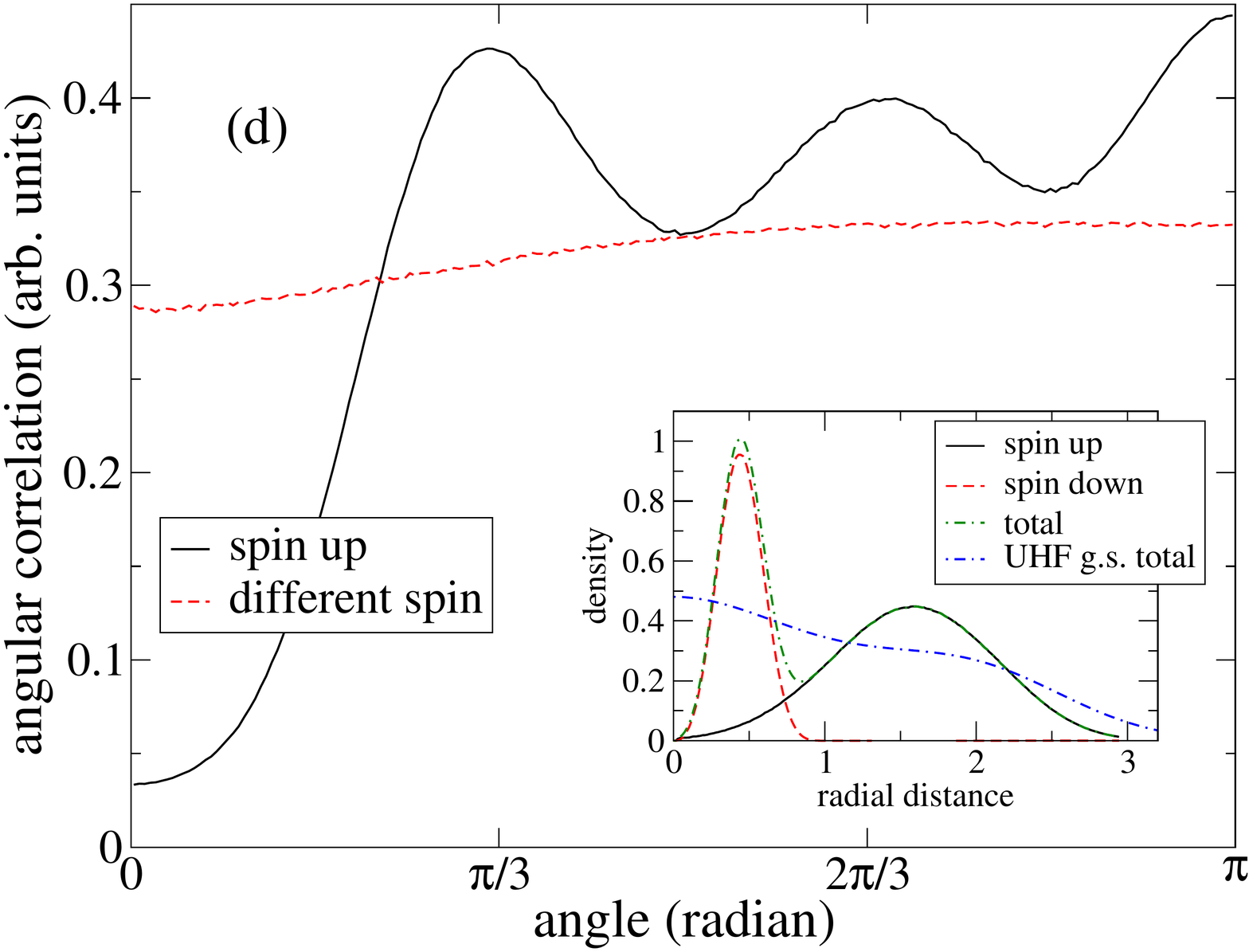}
\end{center}
\caption{\label{fig:msz2.5_7}
  The angular correlation functions and the radial particle density (inset)
  in the $S_z=\frac{5}{2}$ sector of quantum dot nitrogen ($N=7$).
  The parameters are (a) $\gamma=0.6$ and $\lambda=1.8$,
  (b) $\gamma=0.6$ and $\lambda=2.4$,
  (c) $\gamma=1.2$ and $\lambda=1.1$, and
  (d) $\gamma=1.4$ and $\lambda=0.8$.
  We use $\beta^\ast=2.4$, except for (d) where $\beta^\ast=3.2$. Label D in Fig.~\ref{fig:nitrogen}.
}
\end{figure*}

In this sector we seek for possible ordering in the rotationally invariant $(\frac{5}{2},C_\infty)$
phase, which is relevant in UHF in small intervals for $\gamma\ge0.6$.
We observe a remarkable phenomenon at $\gamma=0.6$, cf.\ Figs.~\ref{fig:msz2.5_7}(a) and \ref{fig:msz2.5_7}(b).
At the beginning of the $S_z=\frac{5}{2}$ interval, $\lambda=1.8$, we have a slightly distorted $(\frac{5}{2},C_s)$
structure with a spin-up electron at the center, and peaks at multiples of $\pi/3$
in the spin-up and the spin-independent angular correlation function.
On the other end of the interval, at $\lambda=2.4$, the center of the dot is depleted and
the angular correlation has peaks at multiples of $2\pi/7$; at $\theta=\pi$ it has a dip.
We denote this structure $(\frac{5}{2},C'_s)$, which never occurs in UHF.
In both cases the down-spin electron is located at a slightly smaller radius than the five or six up-spin electrons in the ring,
respectively.
At intermediate $\lambda$'s we see a gradual distortion of $(\frac{5}{2},C_s)$ to $(\frac{5}{2},C'_s)$,
the spin-up electrons in the ring gradually opening a slot for the central spin-up electron to occupy.
At $\gamma=0.8$ and $\lambda=1.8$ we see a similar intermediate state (not shown).
At $\gamma=1$ we find $(\frac{5}{2},C_s)$ in the entire $1.5\le\lambda\le2.1$ interval.
At a still stronger field, $\gamma=1.2$, we observe a further distortion of the $(\frac{5}{2},C_s)$-type
correlated structure: the spin-down electron pulls closer to the center,
and the five spin-up electrons in the ring assume a pentagonal coordination with
angular correlation peaks at multiples of $2\pi/5$, Fig.~\ref{fig:msz2.5_7}(c).
Finally, at $\gamma=1.4$ and $\lambda=0.8$ [Fig.~\ref{fig:msz2.5_7}(d)], the spin-down electron is located near the center,
surrounded by a hexagon of up spins, as in the $(\frac{5}{2},C_{6v})$ state;
but the central electron now occupies a small ring.

\subsection{Quantum dot oxygen, $N=8$}

\begin{figure*}
\begin{center}
  \includegraphics[width=\textwidth, keepaspectratio]{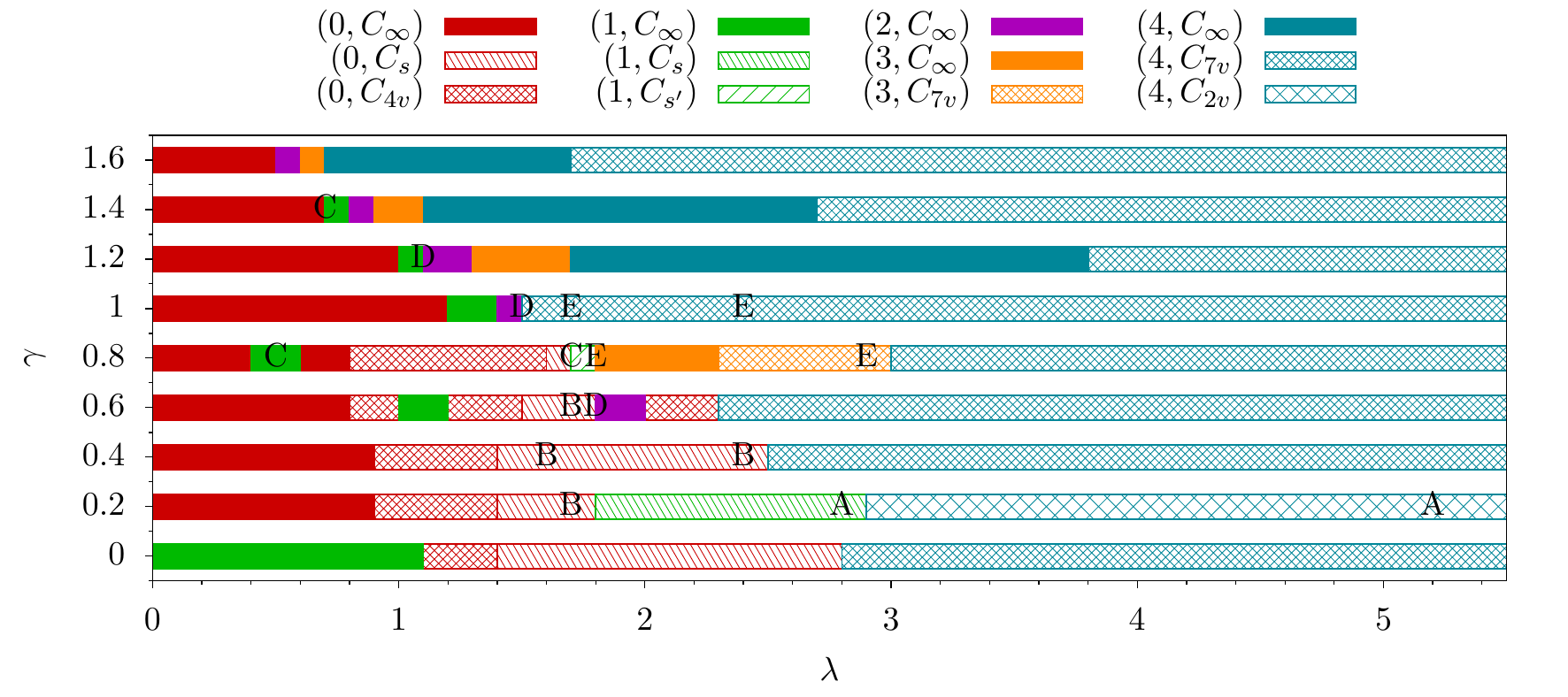}
   \def\svgwidth{\textwidth}
 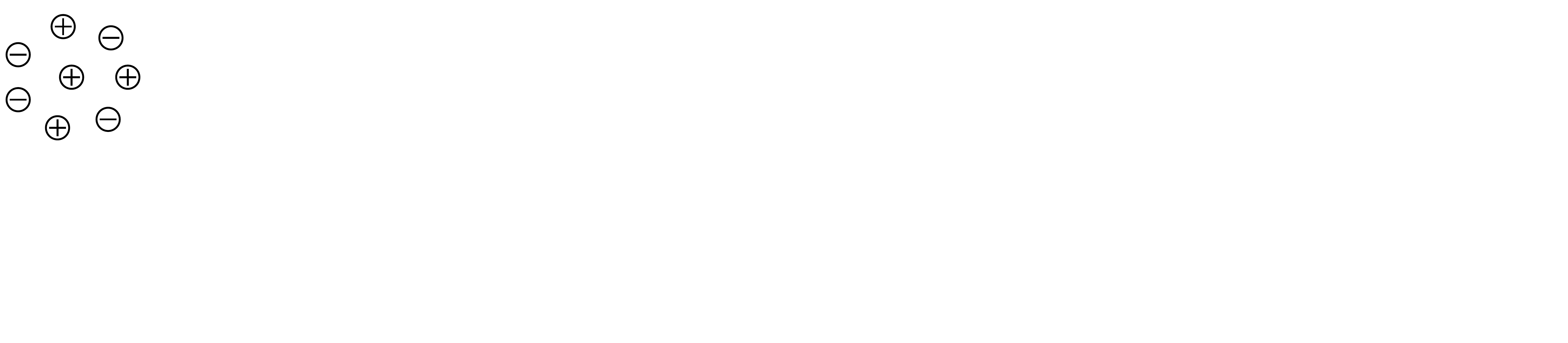
\end{center}
\caption{\label{fig:oxygen}
  (Top)
  Sections of the UHF phase diagram of quantum dot oxygen ($N=8$) at the magnetic field parameter $\gamma=\frac{\omega_c}{\omega}$
  fixed at multiples of 0.2.
  Labels A to E indicate PIMC parameters in Figs.~\ref{fig:mfp8}, \ref{fig:msz0_8}, \ref{fig:msz1_8},
  \ref{fig:msz2_8}~and~\ref{fig:msz3_8}, respectively.
  (Bottom)
  Sketch of the spin density peaks in the relevant symmetry-breaking states of quantum dot oxygen.
  Note that $(2,C_s)$ and  $(2,C_s')$ never occur as a UHF ground state, but such correlations do occur in PIMC.
  For $S_z=0$, equivalent structures are obtained by interchanging spin-up and spin-down.
}
\end{figure*}

Apart from the five rotationally invariant states $(S_z,C_\infty)$ (where $S_z=0,1,2,3,4$),
the relevant symmetry-breaking states are sketched in Fig.~\ref{fig:oxygen}.

The UHF phase diagram is shown in Fig.~\ref{fig:oxygen}.
For small $\gamma\le0.6$, spatial symmetry breaks first according to UHF, and it is followed by a sequence of spin flips.
Symmetry is apparently restored for short intervals, a mistake that PIMC immediately refutes (see below).
For $\gamma\ge1$, the rotationally invariant state first polarizes its spin,
and then undergoes a symmetry-breaking transition later according to UHF.
At the intermediate value $\gamma=0.8$, UHF predicts an intractably complicated sequence of transitions.
Further, it is surprising that the large-$\lambda$ limit at $\gamma=0.2$ is not a centered heptagonal structure, but a hexagonal ring with
two additional particle peaks inside, ($4,C_{2v}$).

\begin{figure*}[htbp]
  \begin{center}
    \includegraphics[width=.64\textwidth, keepaspectratio]{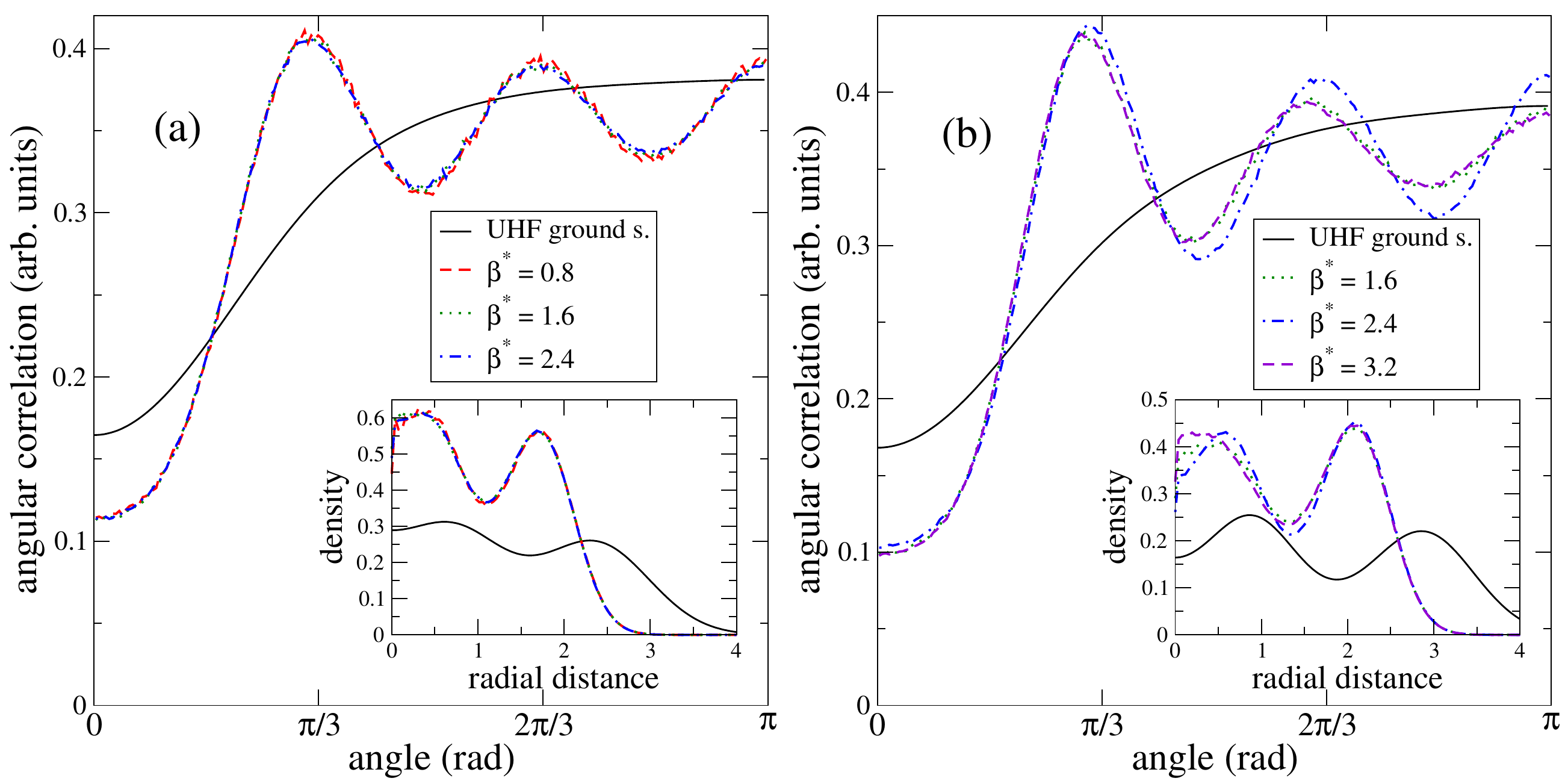}
\end{center}
\caption{\label{fig:mfp8}
  The angular correlation function $g(\theta)$ for fully spin-polarized quantum dot oxygen ($N=8$) for $\gamma=0.2$
  at (a) $\lambda=2.8$ and (b) $\lambda=5.2$;
  The insets show the radial density. Label A in Fig.~\ref{fig:oxygen}.
}
\end{figure*}


\subsubsection{The $S_z=4$ subspace of $N=8$}

We have decided to check the last peculiarity at $\gamma=0.2$ by PIMC.
As Fig.~\ref{fig:mfp8} shows, the sixfold ordering is always manifest in the correlations;
we have checked this in the range $2\le\lambda\le5.2$.
At $\lambda=5.2$, the strength of the correlation slightly depends on the temperature,
but it is unclear to us what this feature signifies.
The radial density shows a double peak-structure, as expected; the structure is always slightly more compact
according to PIMC than according to UHF, cf.\ the insets in Fig.~\ref{fig:mfp8}.
The corroboration of the $(4,C_{2v})$ correlations by PIMC is significant if we recall that
for $N=6$ our method does overrule the prediction of the suspicious symmetry by UHF, cf.\ Fig.~\ref{fig:mfp6}.

\begin{figure*}[htbp]
\begin{center}
\includegraphics[width=0.36\textwidth, keepaspectratio]{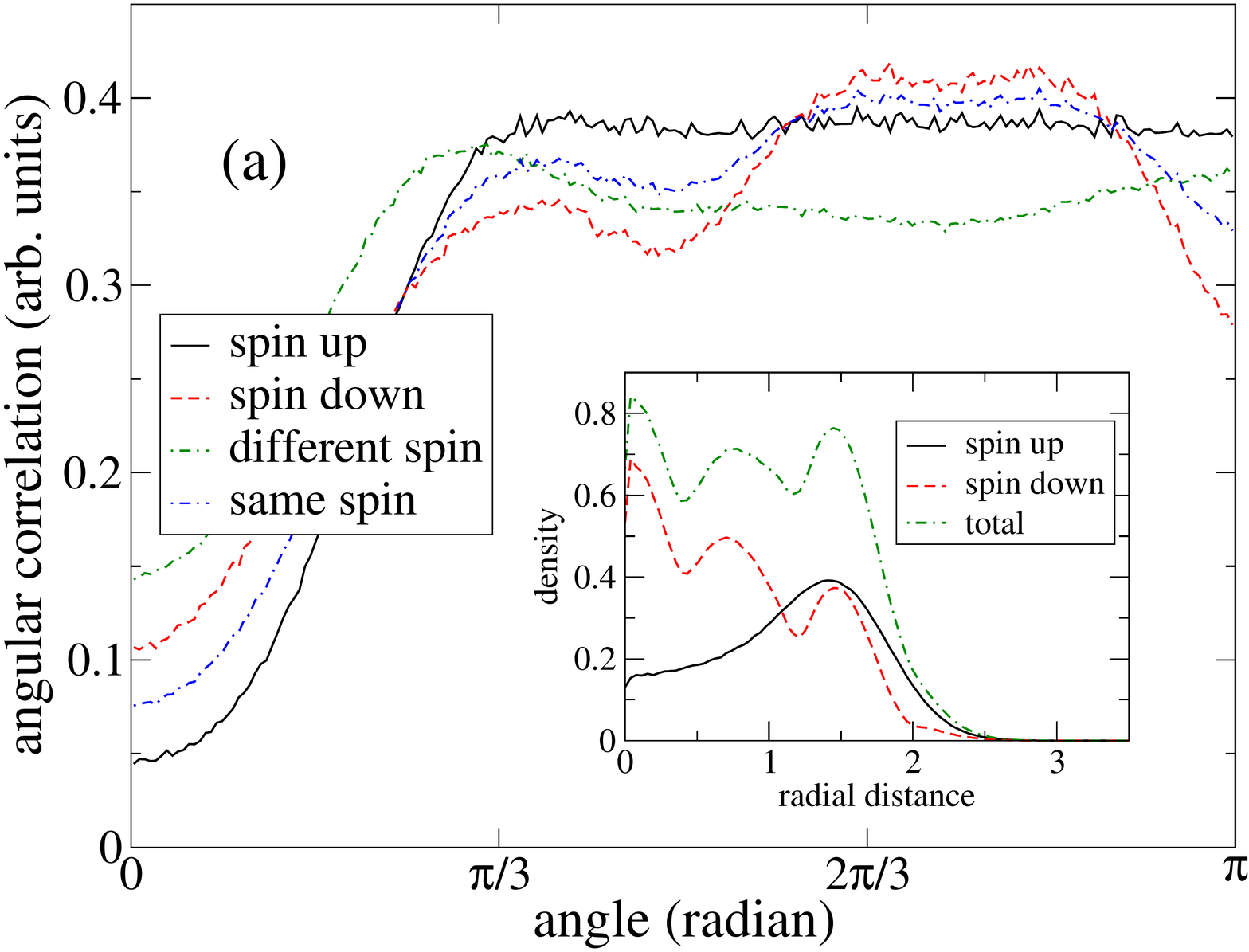}
\includegraphics[width=0.36\textwidth, keepaspectratio]{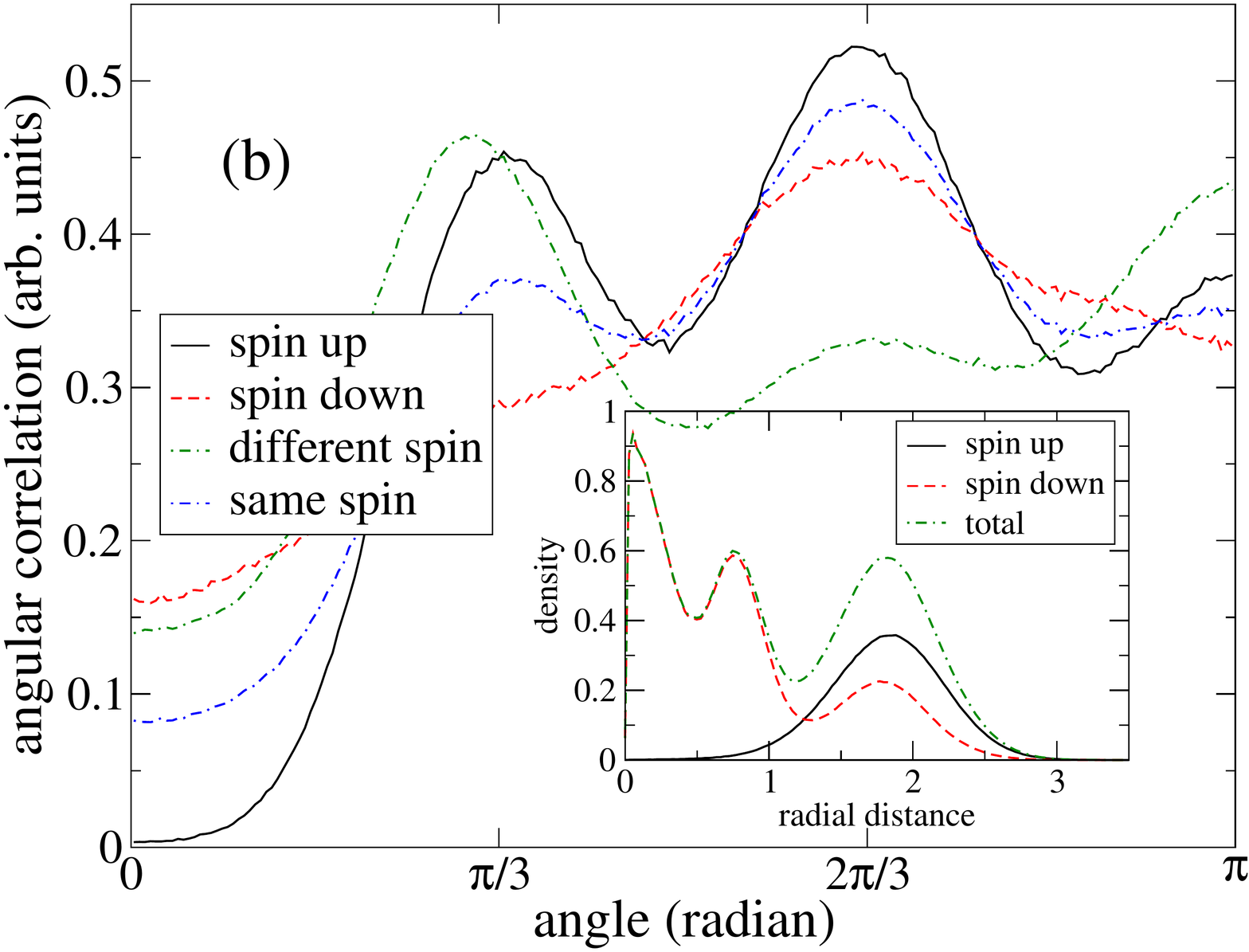}

\includegraphics[width=0.36\textwidth, keepaspectratio]{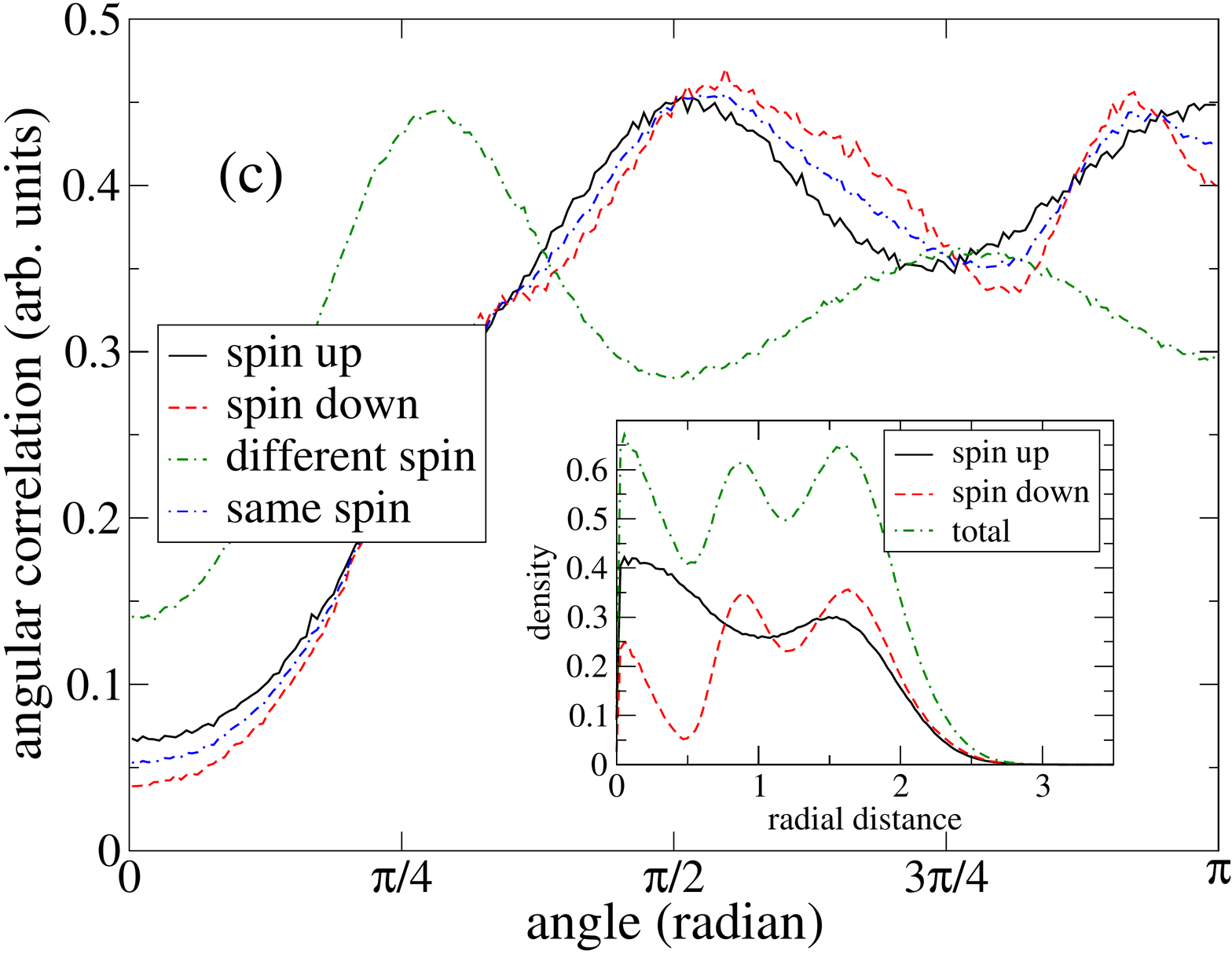}
\includegraphics[width=0.36\textwidth, keepaspectratio]{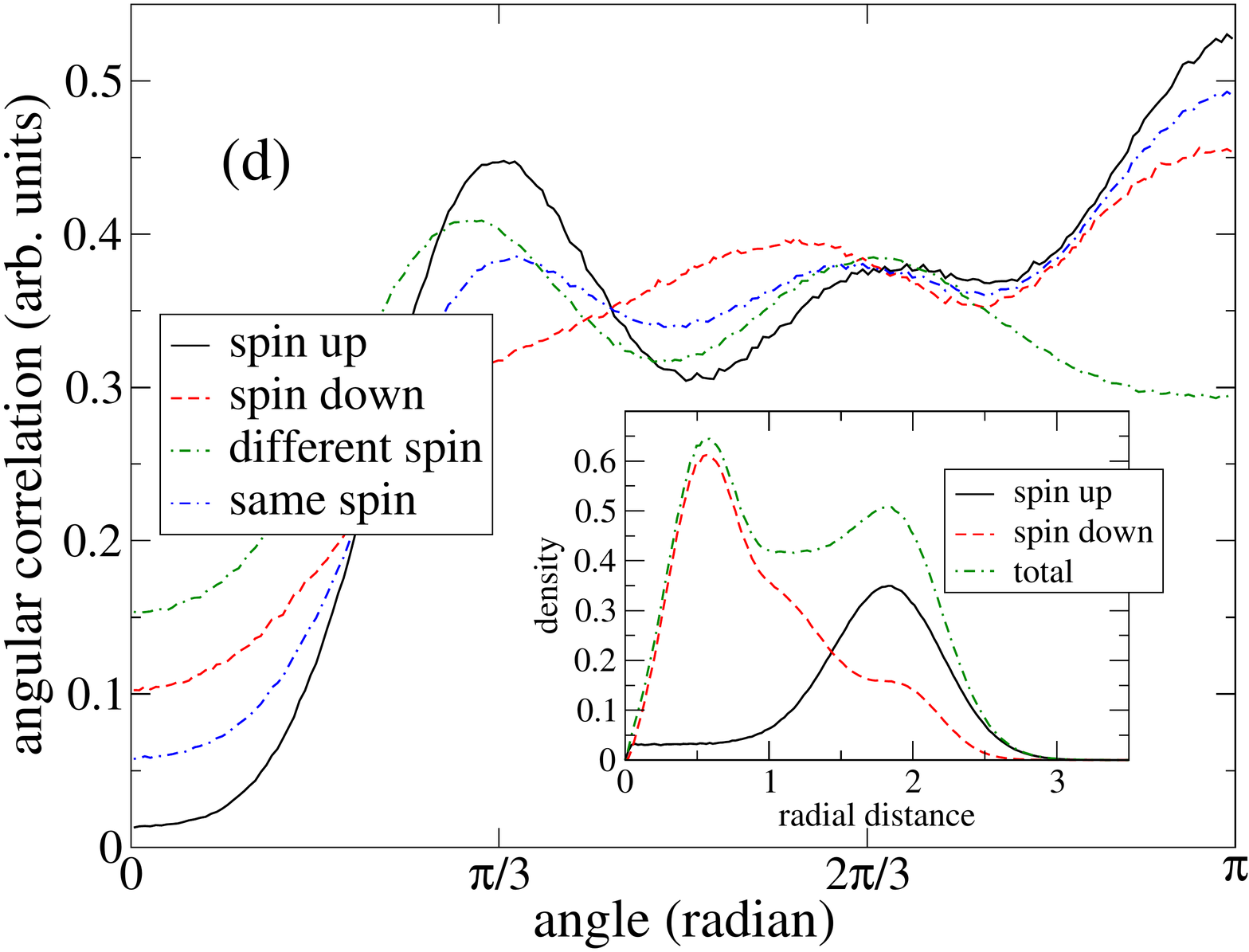}
\end{center}
\caption{\label{fig:msz0_8}
  The angular correlation functions and the radial density (insets) for quantum dot oxygen ($N=8$) of spin $S_z=0$
  (a) for $\gamma=0.2$ and $\lambda=1.7$,
  (b) for $\gamma=0.4$ and $\lambda=2.4$,
  (c) for $\gamma=0.4$ and $\lambda=1.6$, and
  (d) for $\gamma=0.6$ and $\lambda=1.7$.
  $\beta^\ast=2.4$ in all panels. Label B in Fig.~\ref{fig:oxygen}.
}
\end{figure*}


\subsubsection{The $S_z=0$ subspace of $N=8$}

In the $S_z=0$ sector, we find correlations by PIMC that are mostly independent of the UHF picture.
For the small field $\gamma=0.2$, UHF predicts a sequence of $(0,C_\infty)$, $(0,C_{4v})$, and $(0,C_s)$ states.
At the highest coupling of the unpolarized interval, $\lambda=1.7$,
we find that the spin-up electrons form a ring, while two spin-down electrons are located
around the same radius, one at the center, one at an intermediate radius, cf.\ Fig.~\ref{fig:msz0_8}(a).
The same-spin correlation has a sharp peak near $\theta=\pi/3$ and a broad one near $2\pi/3$,
suggesting a hexagonal ring.
The down-spin correlation function has a broad peak near $\theta=2\pi/3$, indicating that the down-spin
electrons in the ring are second neighbors.
Qualitatively, the structure is the same for smaller $\lambda$'s, even in the $(0,C_{4v})$ domain.
In the small coupling $(0,C_\infty)$ range the radial structure of spin-down electrons is
visible, as well as a Coulomb hole-type correlation in the different-spin ACF.
The high-coupling state at $\gamma=0.4$ is the same as at $\gamma=0.2$, as seen in Fig.~\ref{fig:msz0_8}(b),
but the small coupling behavior is different.
At $\lambda=0.7$, the $C_{4v}$ order appears, with a square of the spin-up electrons inside,
and four spin-down electrons rotated by $\pi/4$.
Two of the four spin-down electrons are roughly at the same radius as the spin-up ones,
two are farther off.
As $\lambda$ is increased, spin-down electrons approach the center,
one of them is inside at $\lambda=1.6$ (together with a spin-up electron) [Fig.~\ref{fig:msz0_8}(c)],
two of them inside at $\lambda=2.4$ (but the spin-up electron leaves).
The angles between the remaining six electrons get distorted gradually to make the above mentioned hexagon at $\lambda=2.4$.
At greater field, $\gamma=0.6$ and 0.8, the weak-coupling behavior is similar, but for high coupling
$\lambda=1.7$, the two spin-down electrons within the hexagon are equivalent, forming a small ring
with vanishing electron density at the center.
See cf.\ Fig.~\ref{fig:msz0_8}(d).


\subsubsection{The $S_z=1$ subspace of $N=8$}

We find puzzling results here.
For $\gamma=0.2$, UHF identifies a wide interval of the $(1,C_s)$ symmetry breaking state.
Using this for phase fixing, PIMC at $\lambda=1.8$, 2.3, and 2.8 does not fully confirm such an order.
For the smaller two couplings the expected peaks at multiples of $2\pi/7$
are not present in either the same spin, the different spin, the spin-up or the spin-down angular correlation function.
While a first peak can be crudely identified, the function flattens out for larger angles,
precluding any interpretation.
At $\lambda=2.8$, the upper limit of the $(1,C_s)$ interval in UHF, a second peak is located
near $\theta\approx4\pi/7$, but the strength of the peaks does not correspond to the count of
first, second, etc.\ neighbors in the ring, with the appropriate spin; it is also unclear
if there is a third peak near $\theta\approx6\pi/7$.
This is seemingly one of the few cases where PIMC finds a less ordered state than mean-field theory.
We leave the analysis of this regime to future studies.

\begin{figure*}[htbp]
\begin{center}
\includegraphics[width=0.32\textwidth, keepaspectratio]{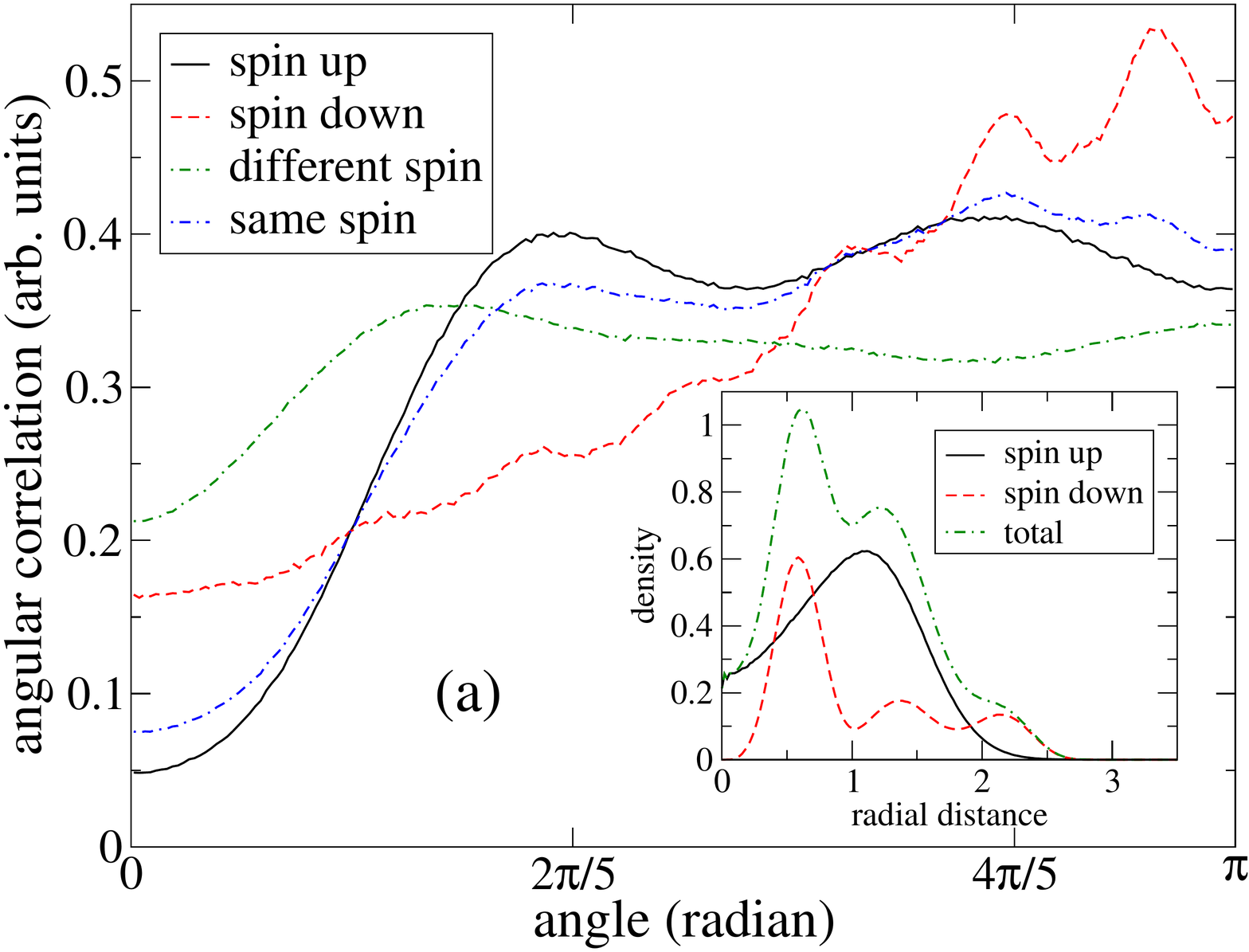}
\includegraphics[width=0.32\textwidth, keepaspectratio]{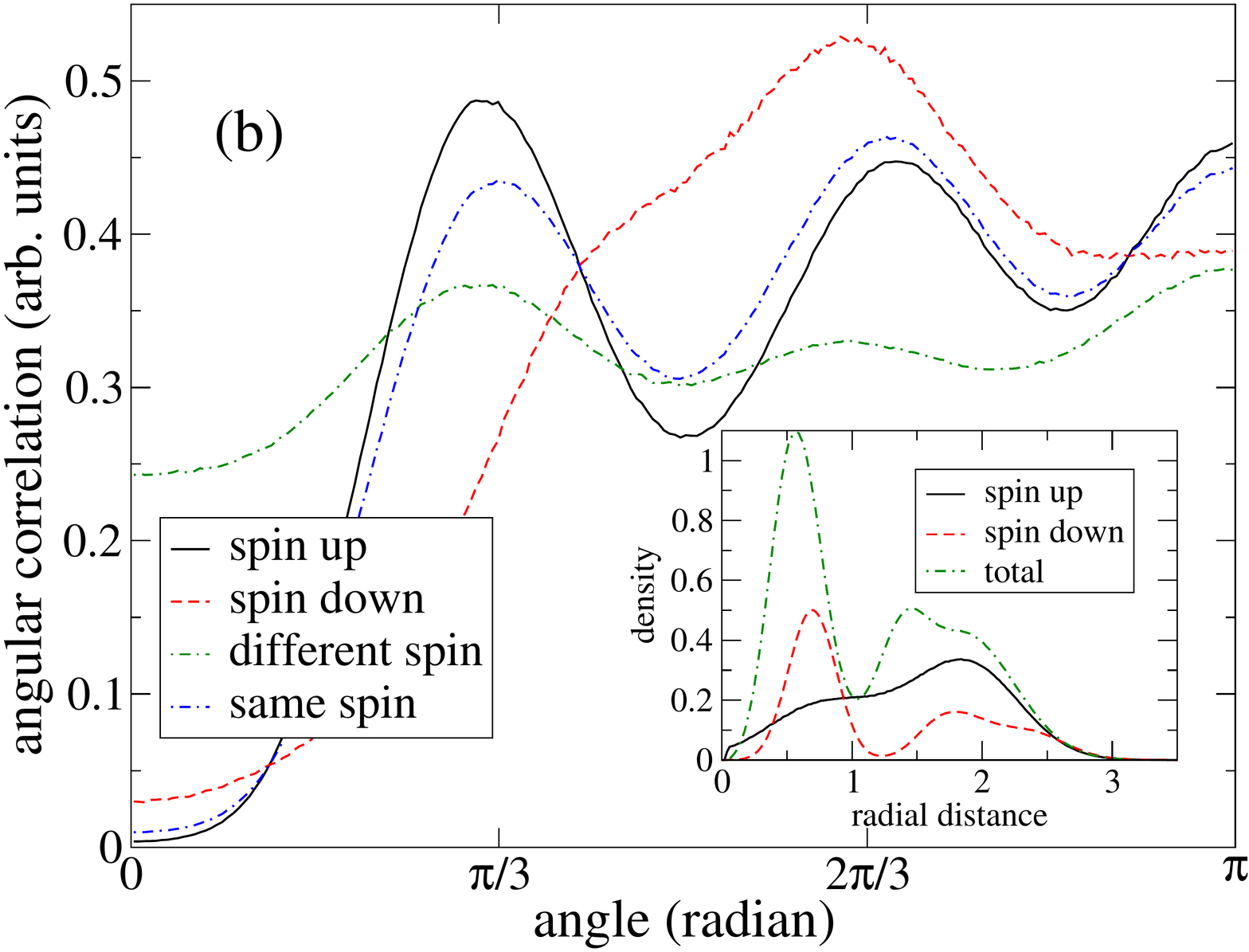}
\includegraphics[width=0.32\textwidth, keepaspectratio]{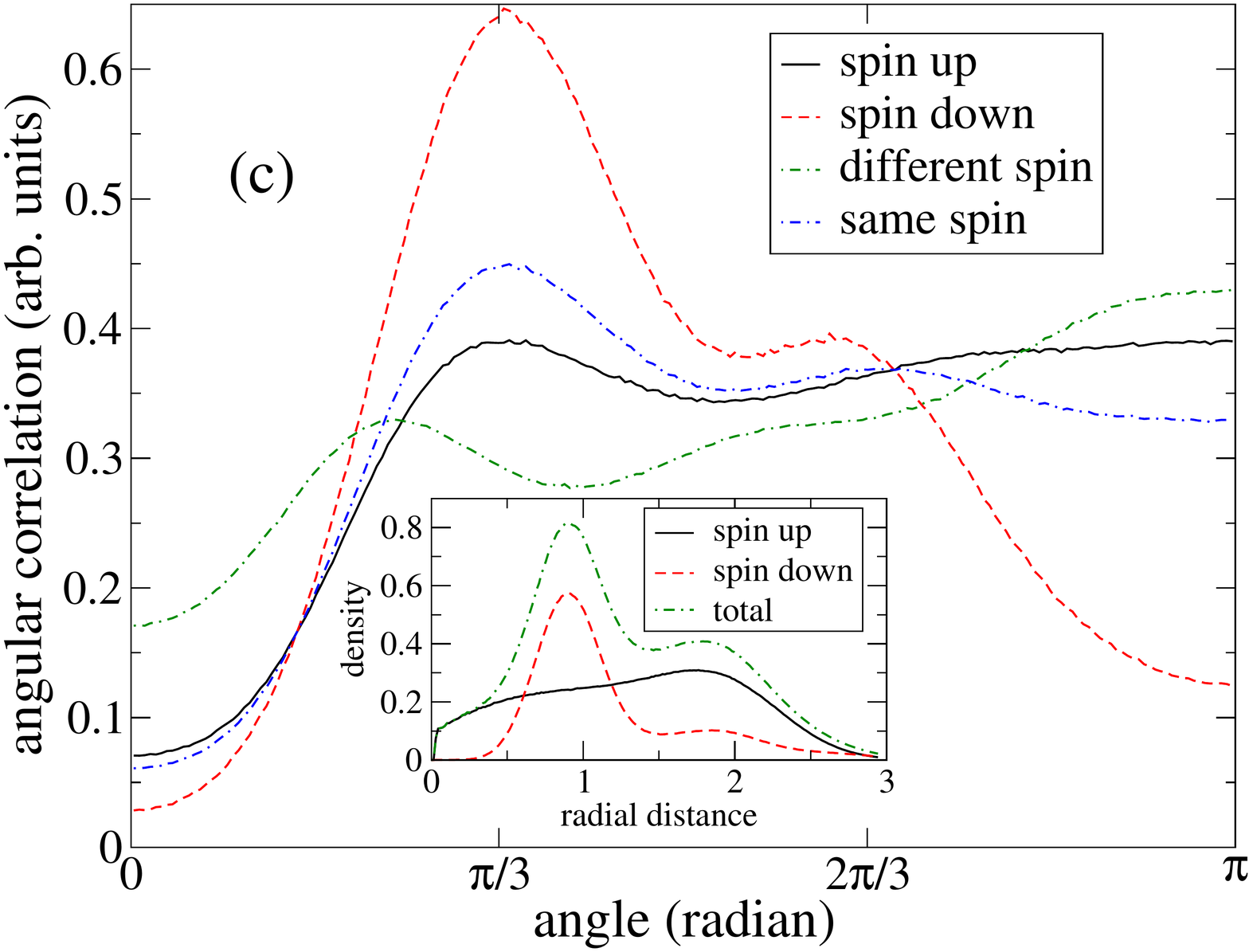}
\end{center}
\caption{\label{fig:msz1_8}
  The angular correlation functions and the radial density (insets) for quantum dot oxygen ($N=8$) of spin $S_z=1$
  (a) for $\gamma=0.8$ and $\lambda=0.5$,
  (b) for $\gamma=0.8$ and $\lambda=1.7$, and
  (c) for $\gamma=1.4$ and $\lambda=0.7$.
  $\beta^\ast=2.4$ in all panels. Label C in Fig.~\ref{fig:oxygen}.
}
\end{figure*}

We obtain clearer results for larger fields.
At $\gamma=0.8$ and $\lambda=0.5$, we can identify a pentagonal ring of spin-up electrons;
one of the down-spin electrons is near the origin, two are outside of the ring, typically on opposite sides ($\theta\approx\pi$),
see Fig.~\ref{fig:msz1_8}(a).
At $\lambda=1.7$, where UHF predicts $(1,C_s''')$, PIMC identifies a quite different molecular ordering:
the angular correlation function of spin-up electrons has peaks near multiples of $\pi/3$, and the same is true for the
different-spin correlation function.
There is a hexagonal ring of five spin-up and a spin-down electron.
The two remaining spin-down electrons are inside the ring; they are correlated with
each other and the electrons in the ring in a complex manner, cf. Fig.~\ref{fig:msz1_8}(b).
This structure also emerges at $\gamma=1.2$ and $\lambda=1$, where UHF predicts $(1,C_\infty)$ (not shown).
Yet another behavior is seen at $\gamma=1.4$ and $\lambda=0.7$: the spin-down ACF
lets us identify an isosceles triangle, but the majority electrons do not show azimuthal ordering;
the different spin correlation function only testifies a weak avoidance; cf. Fig.~\ref{fig:msz1_8}(c).
Finally, we have a puzzle at $\gamma=1$ and $\lambda=1.2$:
for $\beta^\ast\le 2.8$, we find a hexagonal ring as in Fig.~\ref{fig:msz1_8}(b),
but for $\beta^\ast=3.2$ the system switches to a pentagonal arrangement as in Fig.~\ref{fig:msz1_8}(a).
The change in the radial density is negligible during this angular rearrangement.


\subsubsection{The $S_z=2$ subspace of $N=8$}

\begin{figure*}[htbp]
\begin{center}
\includegraphics[width=0.32\textwidth, keepaspectratio]{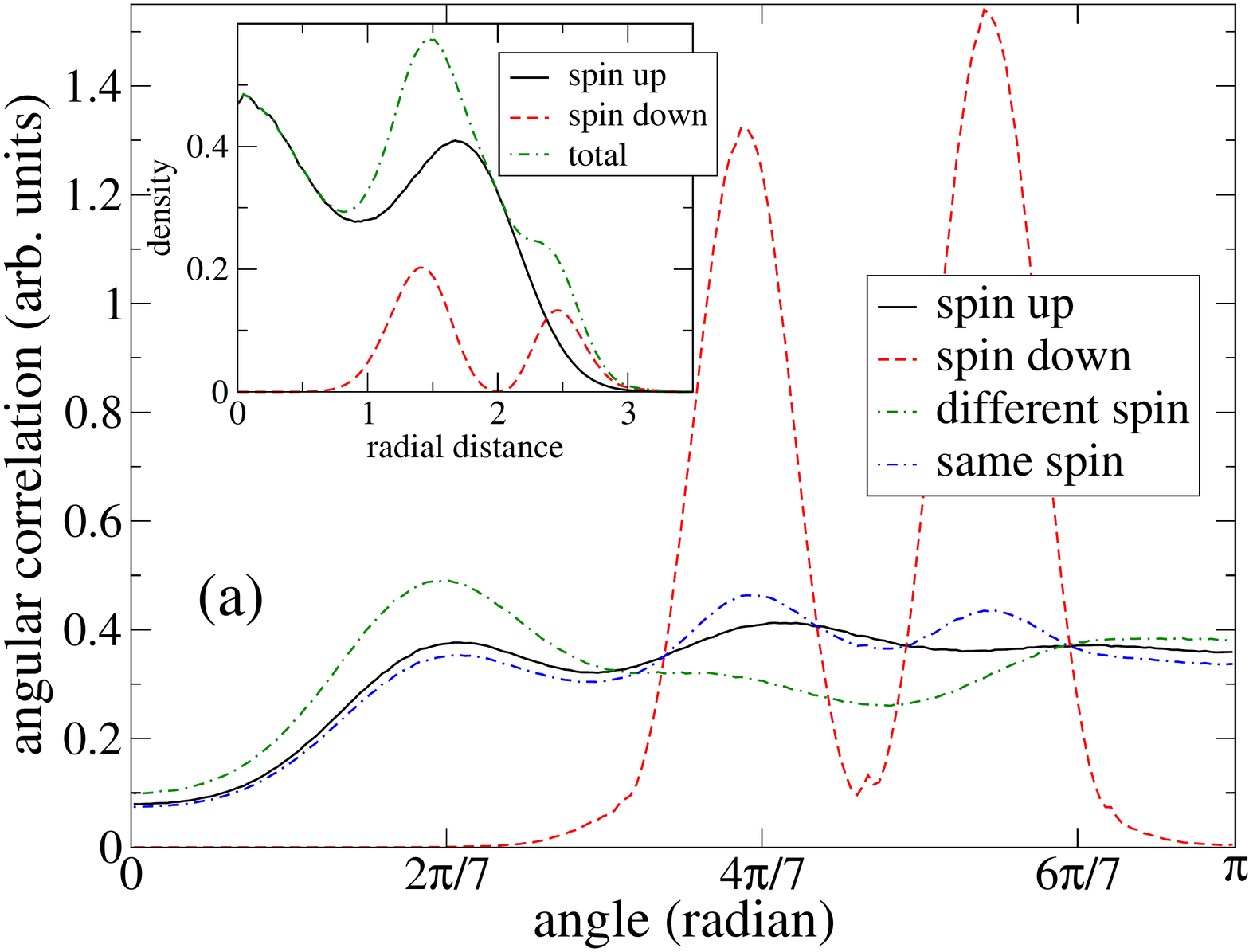}
\includegraphics[width=0.32\textwidth, keepaspectratio]{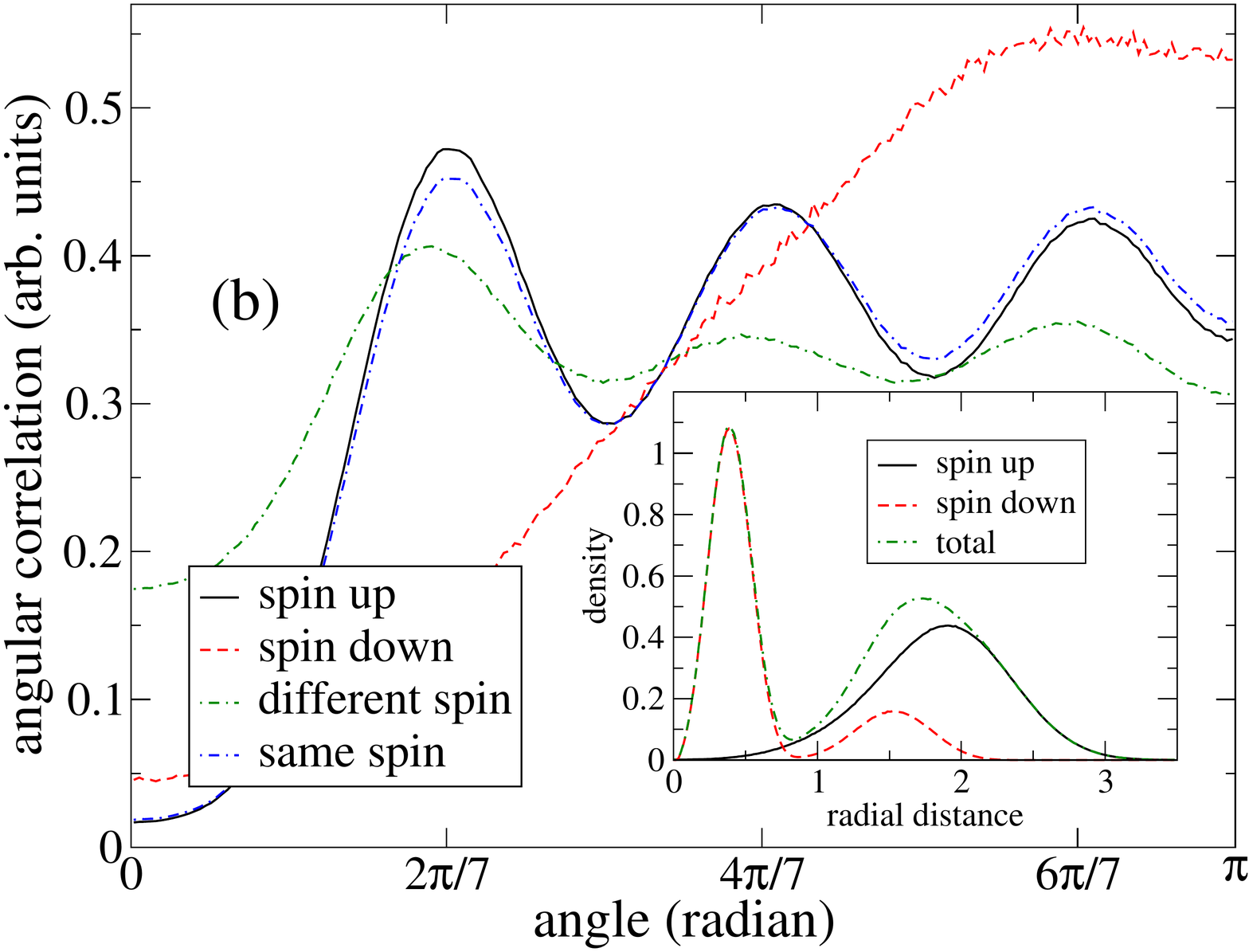}
\includegraphics[width=0.32\textwidth, keepaspectratio]{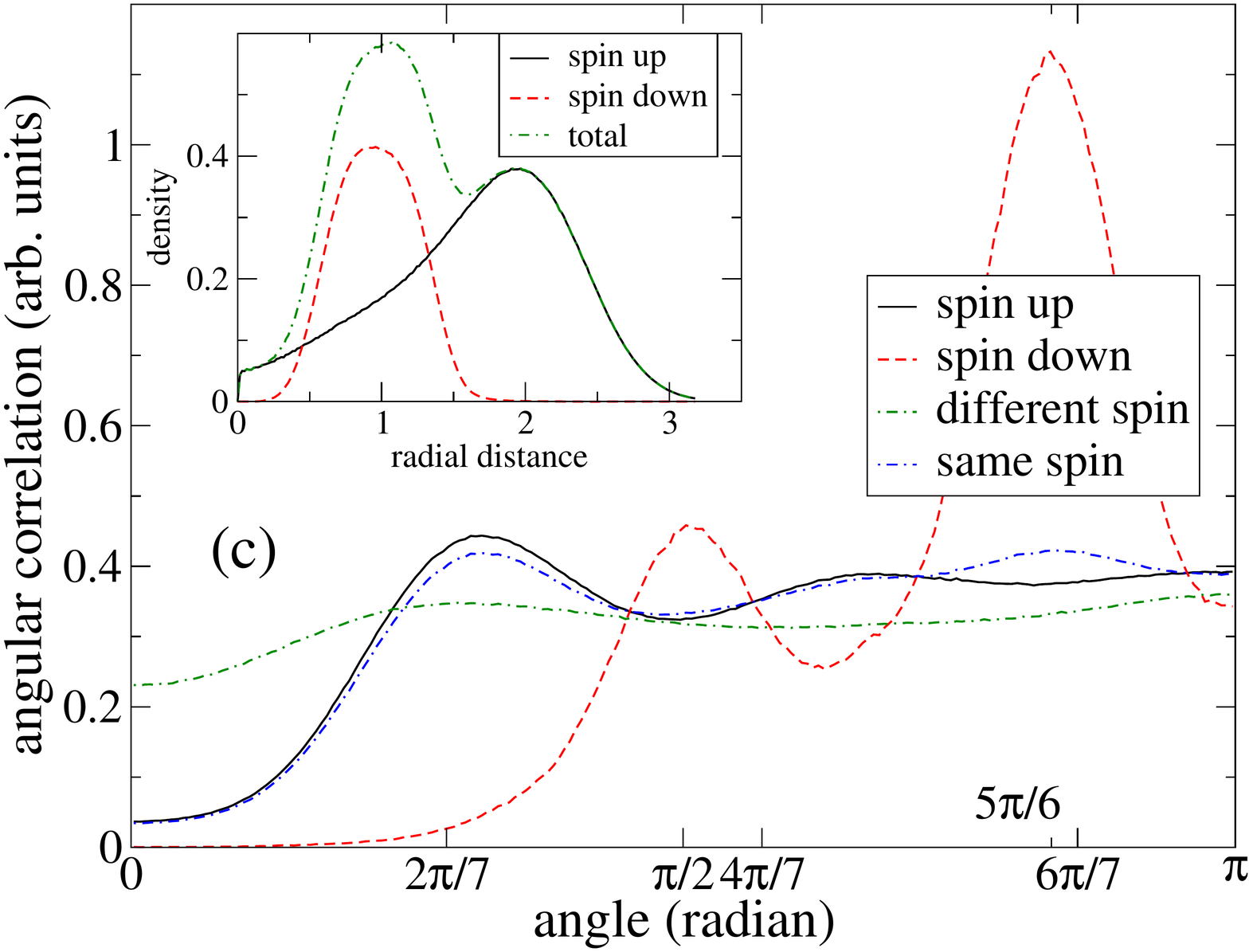}
\end{center}
\caption{\label{fig:msz2_8}
  The angular correlation functions and the radial density (insets) for quantum dot oxygen ($N=8$) of spin $S_z=2$
  (a) for $\gamma=0.6$ and $\lambda=1.8$ ($\beta^\ast=3.2$),
  (b) for $\gamma=1$ and $\lambda=1.5$ ($\beta^\ast=2.4$), and
  (c) for $\gamma=1.2$ and $\lambda=1.1$ ($\beta^\ast=3.2$). Label D in Fig.~\ref{fig:oxygen}.
}
\end{figure*}

While UHF identifies just narrow intervals of featureless $(2,C_\infty)$ states in the $S_z=2$ sector,
PIMC suggests a rich variety of correlated phases.
At $\gamma=0.6$ and $\lambda=1.8$, both the same-spin and the spin-up angular correlation function
shows peaks at multiples of $2\pi/7$, suggesting a seven-member ring around an electron near the origin;
c.f.\ Fig.~\ref{fig:msz2_8}(a).
The radial density confirms the existence of a spin-up ring,
but the spin-down electrons are located at slightly greater and slightly smaller radii, as shown in the inset.
The spin-down ACF has comparable peaks around $\theta\approx4\pi/7$ and $6\pi/7$,
which indicates that the location of spin-down electrons at these angles is close to degenerate.
Further, at $\gamma=1$ and $\lambda=1.5$ the radial density informs us that the central electron has spin down,
c.f.\ Fig.~\ref{fig:msz2_8}(b)
Each of $g_\text{same}(\theta)$, $g_{\uparrow\downarrow}(\theta)$ and $g_{\uparrow\uparrow}(\theta)$
has peaks at multiples of $2\pi/7$.
$g_{\downarrow\downarrow}(\theta)$, on the other hand, indicates that the central electron is slightly off the origin,
preferring locations opposite to that of the down-spin electron in the ring.
We denote this state by $(2,C_s)$.
A different picture emerges at $\gamma=1.2$, $\lambda=1.1$:
a hexagonal ring of up-spin electrons surrounds a smaller ring of two down-spin electrons,
c.f.\ Fig.~\ref{fig:msz2_8}(c).
The spin-down angular correlation has peaks near $\theta=\pi/2$ and $5\pi/6$,
suggesting that the spin-down electrons are located as in $(2,C_s')$ in the lower row of Fig.~\ref{fig:oxygen}.
The internal structure of the down-spin ring cannot be resolved with certainty;
the preferential angles depend on temperature and no clear tendency is discernible.
The same behavior is present at $\gamma=1.4$, $\lambda=0.8$, although the localization of the
down-spins is not resolvable (now shown).
At the strongest magnetic field we study, $\gamma=1.6$, $\lambda=0.5$,
all we can claim is that down-spin electrons are outside of a cloud of up-spin electrons.

\begin{figure*}[htbp]
  \begin{center}
    \includegraphics[width=0.32\textwidth, keepaspectratio]{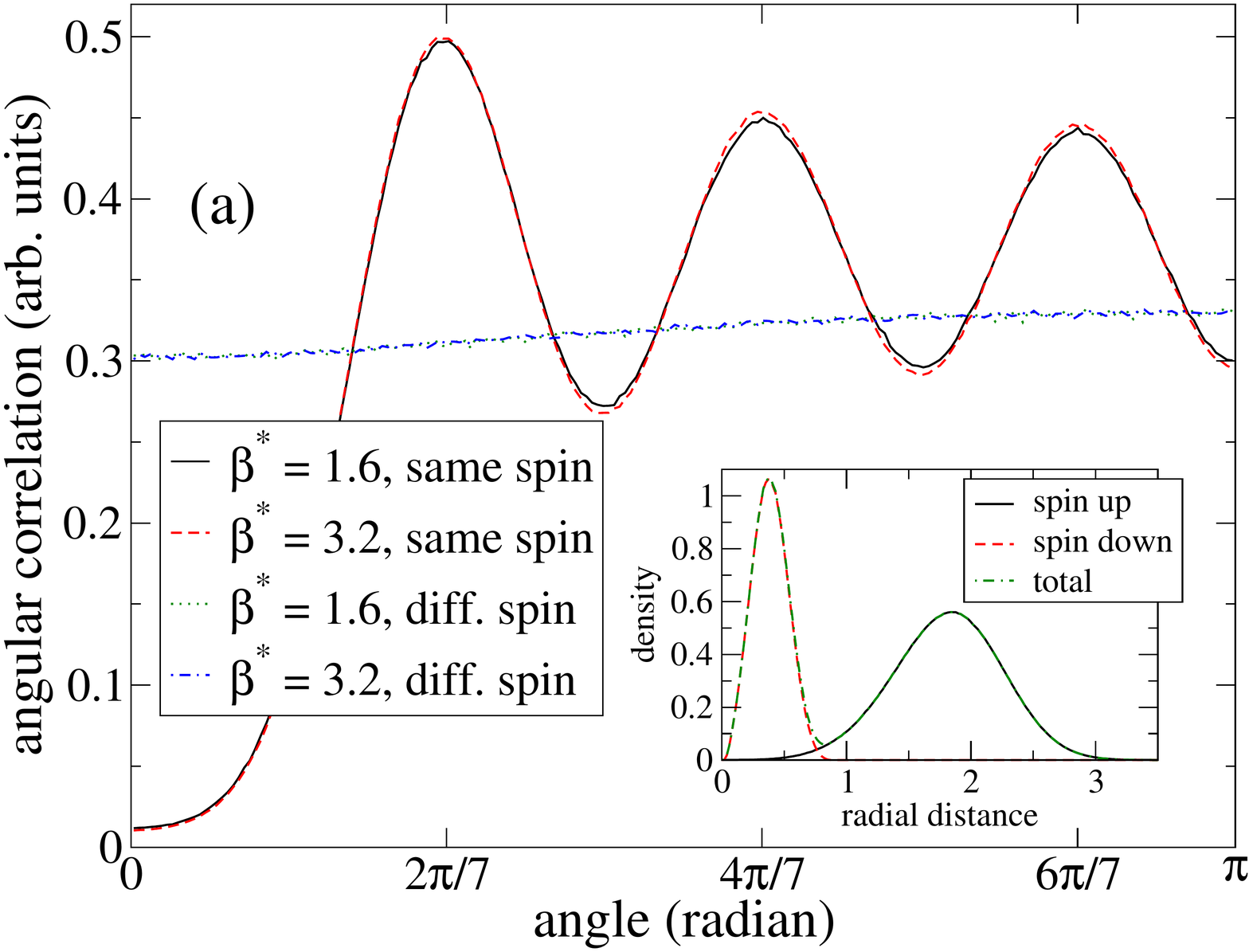}
    \includegraphics[width=0.32\textwidth, keepaspectratio]{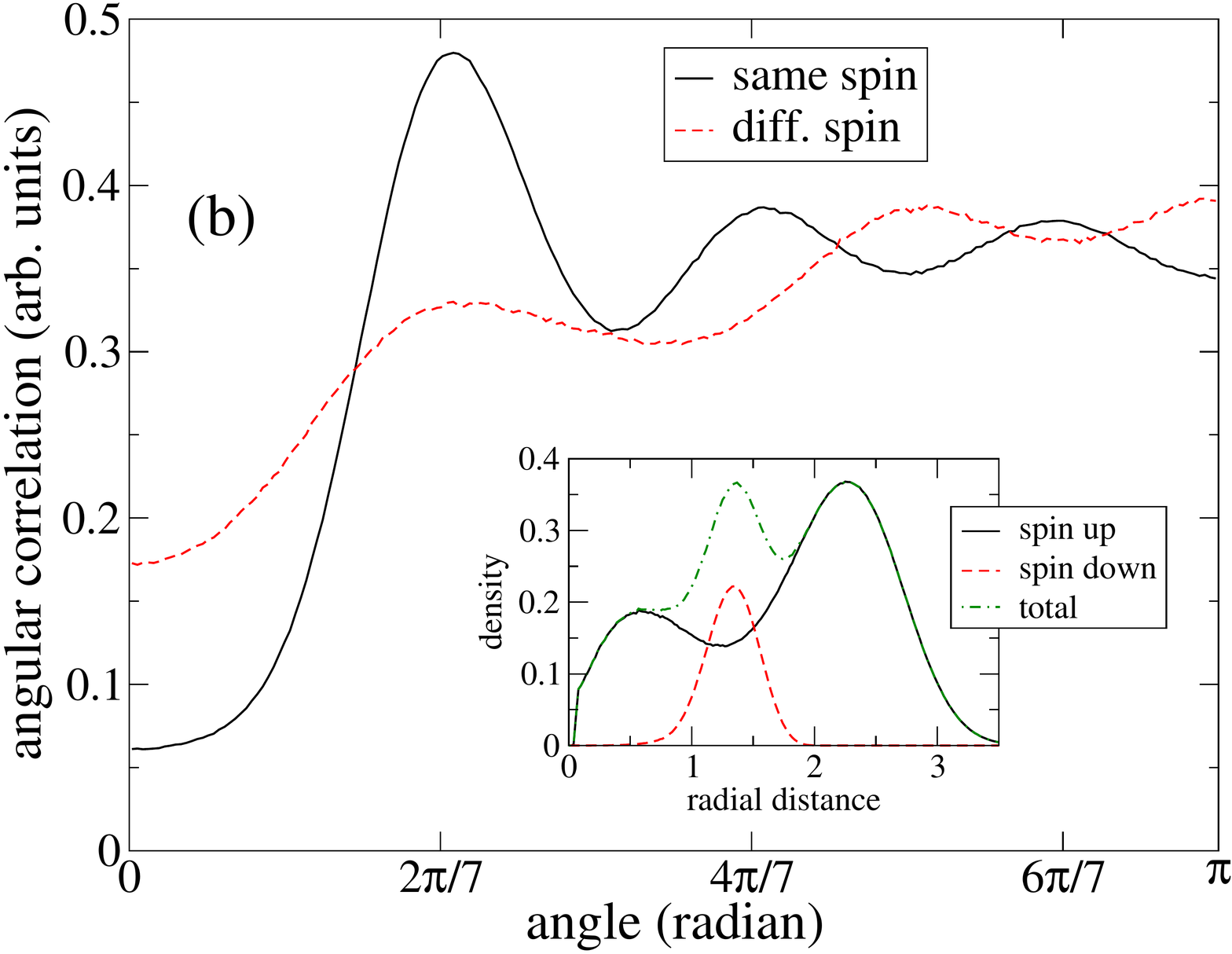}
    \includegraphics[width=0.27\textwidth, keepaspectratio]{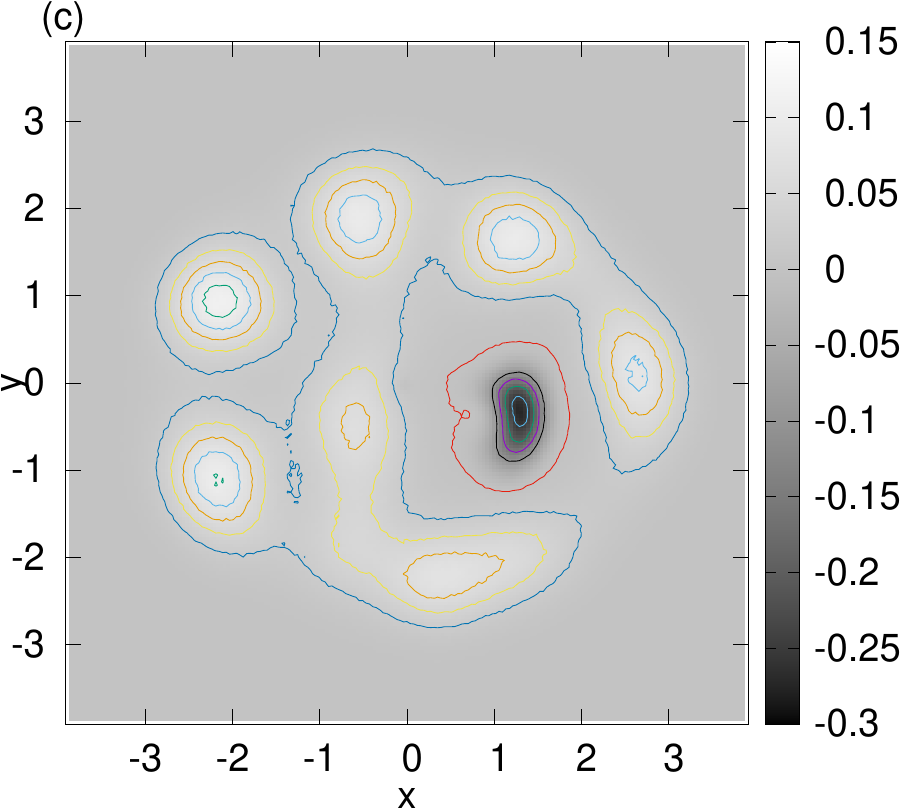}
    
    \includegraphics[width=0.32\textwidth, keepaspectratio]{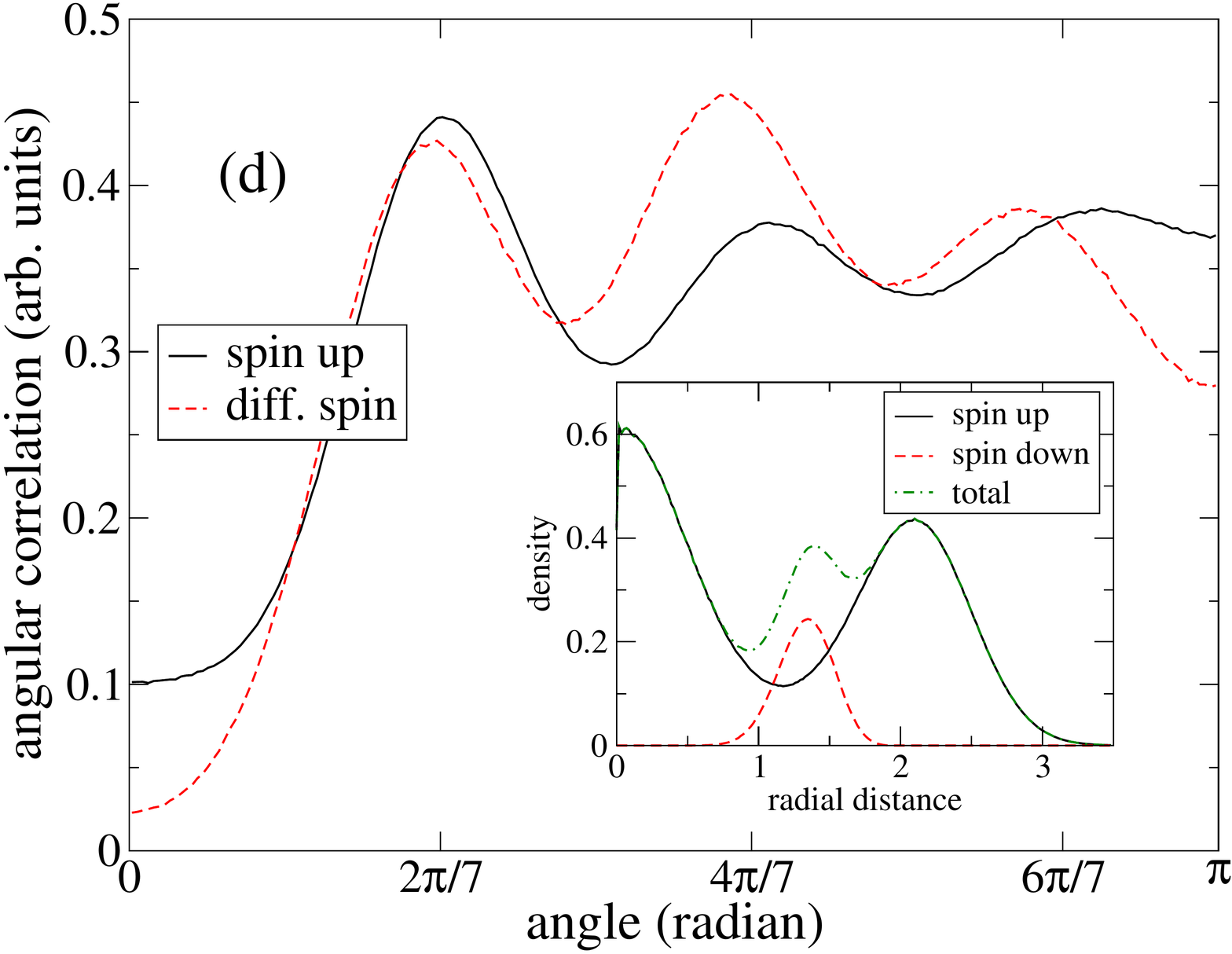}
    \includegraphics[width=0.32\textwidth, keepaspectratio]{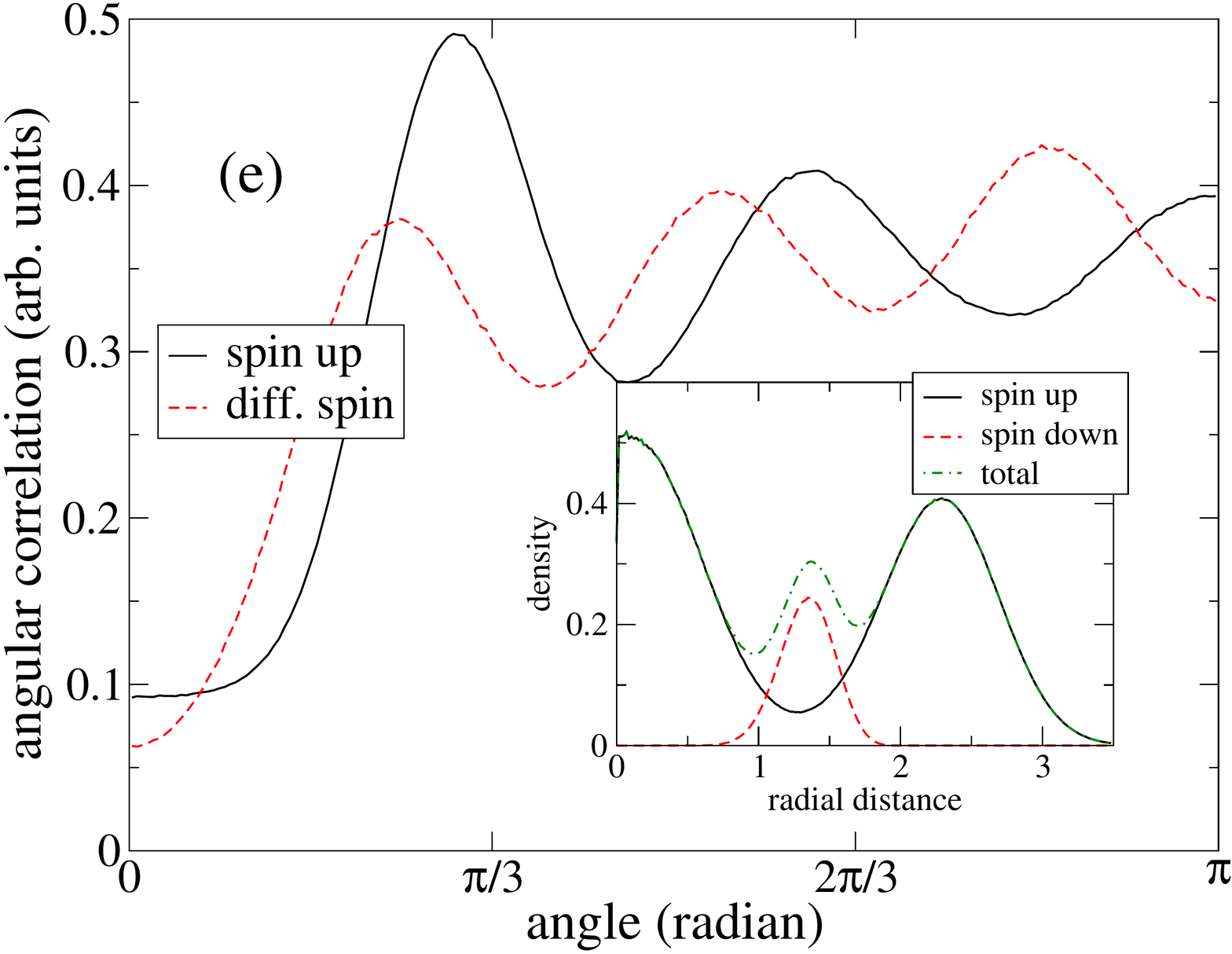}
    \includegraphics[width=0.27\textwidth, keepaspectratio]{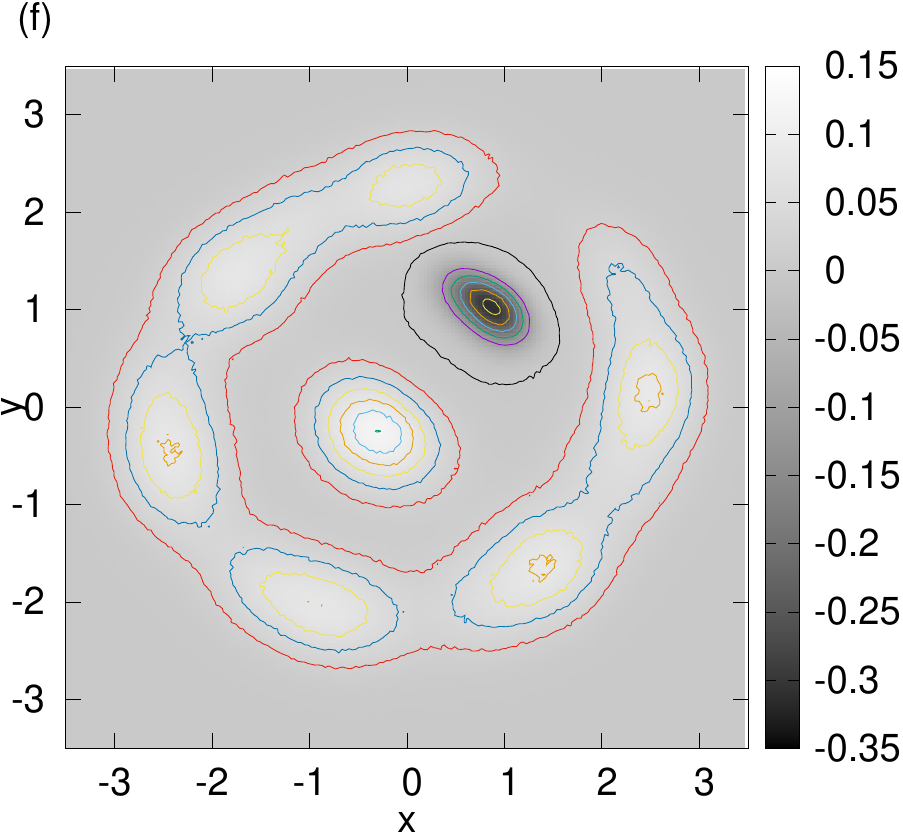}
\end{center}
\caption{\label{fig:msz3_8}
  First row: angular correlation functions and radial densities (insets) in the $S_z=3$ sector of quantum dot oxygen ($N=8$) at
  $\gamma=0.8$ and (a) $\lambda=1.8$, (b) $\lambda=2.9$.
  (c) shows a spin density histogram for (b) at $\beta^\ast=3.2$.
  Second row: the same for $\gamma=1$ and (d) $\lambda=1.7$, (e) $\lambda=2.4$.
  (f) shows a spin density histogram for (e) at $\beta^\ast=2.4$. Label E in Fig.~\ref{fig:oxygen}.
}
\end{figure*}


\subsubsection{The $S_z=3$ subspace of $N=8$}

At a moderate magnetic field $\gamma=0.8$, UHF predicts a transition from a featureless $(3,C_\infty)$ state
to a sevenfold symmetric $(3,C_{7v})$ ground state at $\lambda=2.3$.
By PIMC, however, we find that the heptagonal ring is already present at $\lambda=1.8$,
the lower limit of the $S_z=3$ range at $\gamma=0.8$, cf.\ Fig.~\ref{fig:msz3_8}(a).
The central spin-down electron shows hardly any angular correlation with the spin-up ones in the ring,
but it is displaced slightly from the center.
This behavior is already present at $\beta^\ast=1.6$ and hardly changes with decreasing temperature.
At $\lambda=2.3$, where the UHF ground state is $(3,C_{7v})$,
PIMC confirms the same state (not shown), but only at a low temperature $\beta^\ast=3.2$.
We see, however, a deviation from the $C_{7v}$ Wigner molecule picture at stronger couplings.
As Figs.~\ref{fig:msz3_8}(b) and \ref{fig:msz3_8}(c) demonstrate for $\lambda=2.9$, one spin-up electron enters the ring,
which starts to reorganize into a hexagonal structure.

Very interesting structures emerge at greater fields.
At $\gamma=1$, $\lambda=1.7$, we find a spin-up electron near the center, surrounded by a heptagon, cf.\ Fig.~\ref{fig:msz3_8}(d).
The spin-down electron is part of the latter but it is located at a slightly smaller distance from the center.
We denote this state $(3,C_s)$.
For a stronger coupling $\lambda=2.4$, the spin-down electron and the electron near the center approach each other,
while the ring assumes a hexagonal structure, as seen in Figs.~\ref{fig:msz3_8}(e) and \ref{fig:msz3_8}(f).
Similar order occurs at a higher field $\gamma=1.2$ and $\lambda=1.3$;
the spin-down electron moves further inward and it is less correlated with the electrons in the hexagon.
Further, at $\gamma=1.4$ and $\lambda=0.9$, the two electrons of different spins occupy two sites
symmetrically around the origin, surrounded by a hexagon.


\section{Conclusion}
\label{sec:conclusion}

We have applied the path-integral Monte Carlo method with phase fixing
to explore the correlated phases in few-electron quantum dots in a magnetic field.
We have found that the method is flexible enough to yield structures that strongly differ, even qualitatively,
from the mean-field state we use for phase fixing.
It is also applicable in moderate magnetic fields where the lowest Landau level approximation,
a frequently used simplifying assumption both in configuration interaction and in variational studies, does not hold.
Both temperature and correlation effects are fully taken into account; the most important
limitation of the method is that the simulation is performed at fixed $z$ component of the
total spin, and a comparison of different spin sectors is challenging.
Nevertheless, there are suggestions in the literature for obtaining free energies from PIMC simulations,
and this is a possible direction of further studies.

We have found that electrons in the quantum dot very often show correlations characteristic of Wigner molecules
at moderate couplings where the (unrestricted) Hartee-Fock method still predicts an unbroken symmetry.
Moreover, we have found cases where PIMC completely revises the symmetry-breaking structures
found in the mean-field approximation, as seen, for example, in Figs.~\ref{fig:mpp}, \ref{fig:mfp4b}(b),
\ref{fig:msz1_4}, and \ref{fig:mfp6}(a).
Further, PIMC identifies correlated structures that have no precedent in mean-field theory,
e.g., the $(\frac{3}{2},C_s)$, $(\frac{3}{2},C_{4v})$, and $(\frac{1}{2},C_s''')$ phases of quantum dot boron ($N=5$),
the $(2,C_s')$ state of quantum dot carbon ($N=6$), and the $(2,C_s')$ and $(2,C_s)$ states of quantum dot oxygen ($N=8$),
and a few others.
We have presented finite temperature results, but there are cases where the correlation functions
apparently no longer change with decreasing temperature, letting us infer with some degree of confidence the
ground-state properties of the system.

Experimental connections are no doubt remote at the moment.
This is mainly due to two reasons: the difficulty of accessing the moderate to strongly coupled range,
and the scarcity of experimental probes.
Using GaAs parameters, $\hbar\omega_0=3$ meV, and disregarding screening mechanisms,
the coupling constant in current experiments can be estimated as $\lambda<1.9$.
The strongly coupled regime might be approached by using a weaker confinement potential, but the larger
size of the dot also requires cleaner samples.
The coupling parameter can also be increased by using a semiconductor with greater effective mass,
e.g., Silicon or holes in GaAs.
Transport experiments detect the structural changes in the dot by the modulation of the addition energies,
a quantity we cannot obtain with certainty from a path-integral Monte Carlo calculation at the moment.
The development of estimators connected to optical properties is an option that we delegate to further studies.

\FloatBarrier


\appendix
\section{The derivation of the cumulant action in an external magnetic field}

Here we provide computationally tractable expressions for the cumulant action in Eqs.~(\ref{eq:cumulant1})-(\ref{eq:cumulant3}),
both the Coulomb interaction and the harmonic confinement potential.

For the harmonic confinement potential $V(R)=\frac{1}{2}m^\ast\omega_0^2\sum_{i=1}^Nr_i^2$ the
program sketched in Eqs.~(\ref{eq:cumulant1})-(\ref{eq:cumulant3}) can be performed analytically.
Focusing on a single particle $i$, a straightforward calculation yields
\begin{multline}
  U_\text{C}^\text{conf.}(r_{i,0},r_{i,1},\tau)=\frac{m^\ast\omega_0^2\ell_c^2}{4(1-u)}\left[
    \frac{2(r_{i,0}^2+r_{i,1}^2)}{\ell_c^2(1-u)} I_2\right.\\
    +\left.\left(4 + \frac{i}{2\ell_c^2}(x_0y_1-x_1y_0) + \frac{1+u}{1-u}\frac{2\mathbf r_{i,0}\cdot\mathbf r_{i,1}}{\ell_c^2}\right)I_1
    \right],
\end{multline}
where $u=e^{-\hbar\omega_c\tau}$, $\ell_c=\sqrt\frac{\hbar}{eB}$ is the magnetic length, and
\begin{equation}
  \begin{split}
  I_1 &= \tau(1+u) + \frac{2}{\hbar\omega_c}(u-1),\\
  I_2 &= \frac{1-u^2}{\hbar\omega_c} -2\tau u.
  \end{split}
\end{equation}

For the pair interaction, the cumulant action becomes a sum over all pairs:
\begin{equation}
  U_\text{C}(R_0,R_1,\tau)=\sum_{i<j}u_\text{C}^{i,j}
\end{equation}
where $u_\text{C}^{i,j}\equiv u_\text{C}(\mathbf r_{0,i},\mathbf r_{0,j},\mathbf r_{1,i},\mathbf r_{1,j},\tau)$ is
\begin{multline}
  u_\text{C}^{i,j}=
  \int_0^\tau dt\int d^2r_i d^2r_j\frac{e^2}{4\pi\epsilon|\mathbf r_i-\mathbf r_j|}\\
  \times\mu(\mathbf r_{i},\mathbf r_{j},t|
  \mathbf r_{0,i},\mathbf r_{0,j},\mathbf r_{1,i},\mathbf r_{1,j},\tau)
\end{multline}
with the complex weight
\begin{multline}
  \mu(\dots)=
  \frac{\rho_0(\mathbf r_{0,i},\mathbf r_i;t)\rho_0(\mathbf r_{i},\mathbf r_{1,i};\tau-t)}
       {\rho_0(\mathbf r_{0,i},\mathbf r_{1,i};\tau)}\\
  \times\frac{\rho_0(\mathbf r_{0,j},\mathbf r_j;t)\rho_0(\mathbf r_{j},\mathbf r_{1,j};\tau-t)}
           {\rho_0(\mathbf r_{0,j},\mathbf r_{1,j};\tau)}.
\end{multline}
Here, $\rho_0$ is taken from Eq.~(\ref{eq:openbc}).
Let us transform to center-of-mass and relative coordinates $\mathbf R=(\mathbf r_i+\mathbf r_j)/2$
and $\mathbf r=\mathbf r_i-\mathbf r_j$ for the initial, the final, and the running variables.
Further, let us insert $\int\frac{d^2k}{(2\pi)^2}e^{-i\mathbf k\cdot\mathbf r}V(k)$ for the pair interaction.
The complex weight separates as
\begin{equation}
  \mu(\dots)=\mu(\mathbf R,t|\mathbf R_{0},\mathbf R_{1},\tau)\mu(\mathbf r,t|\mathbf r_{0},\mathbf r_{1},\tau),
\end{equation}
and the center-of-mass integral $\int d^2R$ can be performed. Then
\begin{multline}
  u_\text{C}^{i,j}=
  \frac{1}{2}\int_0^\tau dt\int\frac{d^2k}{(2\pi)^2}V(k)\\
  \times\int d^2r\mu(\mathbf r,t|\mathbf r_{0},\mathbf r_{1},\tau)e^{-i\mathbf k\cdot\mathbf r}.
\end{multline}
The integral on the relative coordinate can also be performed.
Using the notation $u_0=e^{-\hbar\omega_c\tau}$, $u_1=e^{-\hbar\omega_c t}$, $u_2=e^{-\hbar\omega_c(\tau-t)}$, and
\begin{equation}
\begin{split}
  \sigma_t^2=\frac{(1-u_1)(1-u_2)}{(1-u_0)},\qquad
  \Delta\mathbf r=\mathbf r_1-\mathbf r_0,\\
  \mathbf r_t=\mathbf r_0 + \frac{(1-u_1)(1+u_2)}{2(1-u_0)}\Delta\mathbf r,
\end{split}
\end{equation}
we obtain
\begin{multline}
  u_\text{C}^{i,j}=
  \int_0^\tau dt\int\frac{d^2k}{(2\pi)^2}V(k)\\
  \times\exp\left(-\sigma_t^2\ell_c^2k^2
  -\frac{\sigma_t^2}{2}\mathbf{\hat z}\cdot\left(\mathbf k\times\Delta\mathbf r\right)-i\mathbf k\cdot\mathbf r_t\right).
\end{multline}
\begin{figure}[htbp]
\begin{center}
\includegraphics[width=0.7\columnwidth, keepaspectratio]{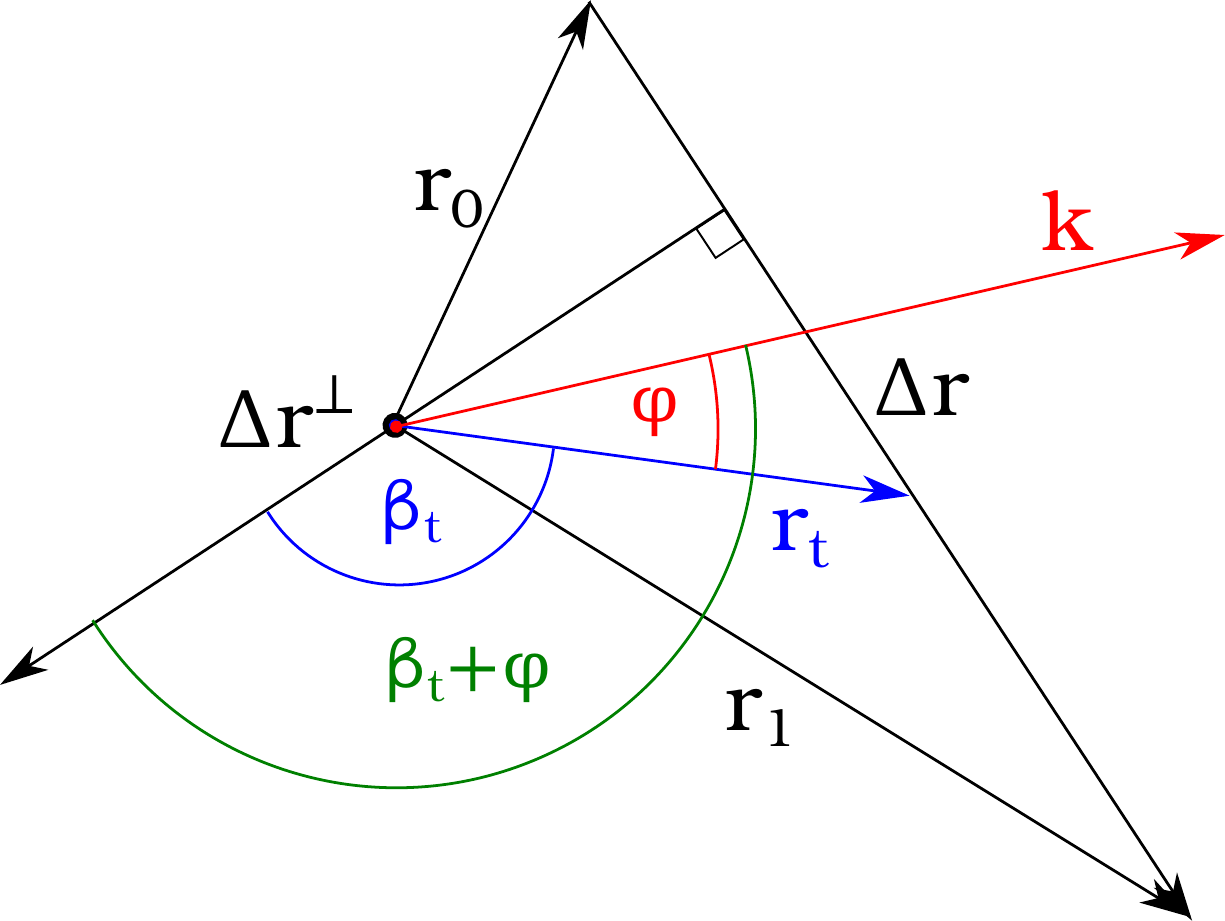}
\end{center}
\caption{\label{fig:angles}
  Angles defined for the integration of the cumulant action in Eq.~(\ref{eq:pairuc}).
  $\Delta\mathbf r^\perp$ is the clockwise $\pi/2$ rotation of $\Delta\mathbf r$, i.e., $(y_1-y_0,x_0-x_1)$.
}
\end{figure}
Let us change to polar coordinates as shown in Fig.~\ref{fig:angles}, i.e.,
measuring the angle of $\mathbf k$ from the direction specified
by $\mathbf r_t$. Then
\begin{multline}
  \label{eq:pairuc}
  u_\text{C}^{i,j}=
  \int_0^\tau dt\int_0^\infty\frac{kdk}{(2\pi)^2}V(k)e^{-\sigma_t^2\ell_c^2k^2}\\
  \times\int_0^{2\pi}d\phi\exp\left(
  -\frac{\sigma_t^2}{2}k\Delta r\cos(\beta_t+\phi)-ikr_t\cos(\phi)\right).
\end{multline}
In order to evaluate Eq.~(\ref{eq:pairuc}), we have to make another approximation.
Let us expand the exponent in the integrand to first order in $B$, i.e., $e^{f(B)}\approx e^{(f(B=0))}(1+Bf'(B=0))$.
Physically, this approximation can be interpreted in terms of a comparison of the de Broglie thermal
wave length $\ell_\tau=\sqrt{\hbar^2\tau/m}$ and the magnetic length:
\begin{equation}
  \label{eq:expansion}
\frac{\ell_\tau^2}{\ell_c^2}\ll 1,
\end{equation}
which becomes $\hbar\omega\tau\gamma\ll1$ in terms of our dimensionless parameters.
As we fix $\hbar\omega\tau=0.01$ and $\gamma\le1.6$, Eq.~(\ref{eq:expansion}) is fulfilled in our simulations.
Substitute $t=\lambda\tau$, $q=k\ell_\tau$, $\tilde r=r/\ell_\tau$, $\Delta\tilde r=\Delta r/\ell_\tau$,
and $\mathbf{\tilde r}_\lambda=\mathbf{\tilde r}_0+\lambda\Delta\mathbf{\tilde r}$, elementary algebra yields
\begin{multline}
  u_\text{C}^{i,j}=\tau\frac{e^2}{4\pi\epsilon\ell_\tau}\int_0^1d\lambda\int_0^\infty dq\int_0^{2\pi}\frac{d\phi}{2\pi}\\
  \times\exp\left(-\lambda(1-\lambda)q^2-iq\tilde r_\lambda\cos(\phi)\right) \\
  \times\left(
  1-\frac{\ell_\tau^2}{\ell_c^2}\frac{\lambda(1-\lambda)}{2}\Delta\tilde rq
  \cos(\beta_\lambda)\cos(\phi)
  \right).
\end{multline}
Performing the $q$ and the $\phi$ integrations, we obtain that the real and the imaginary parts are zeroth-order and first-order
in the magnetic field $B$, respectively.
Explicitly,
\begin{equation}
  \label{eq:ure}
  \text{Re}(u_\text{C}^{i,j})=\tau\frac{e^2}{4\pi\epsilon\ell_\tau}\frac{\sqrt\pi}{2}\int_0^1d\lambda
  \frac{\exp\left(-X\right)I_o\left(X\right)}{\sqrt{\lambda(1-\lambda)}},
\end{equation}
and
\begin{multline}
  \label{eq:uim}
  \text{Im}(u_\text{C}^{i,j})=\tau\frac{e^2}{4\pi\epsilon\ell_\tau}\frac{\sqrt\pi\Delta\tilde r}{16}\frac{\ell_\tau^2}{\ell_c^2}
  \int_0^1d\lambda
  \tilde r_\lambda\cos(\beta_\lambda)\\
  \times\frac{\exp\left(-X\right)}{\sqrt{\lambda(1-\lambda)}}
  \left( I_o\left(X\right) - I_1\left(X\right)
  \right),
\end{multline}
where $I_n$ are modified Bessel functions of the first kind, and $X=\frac{\tilde r_\lambda^2}{8\lambda(1-\lambda)}$.

Equations (\ref{eq:ure}) and (\ref{eq:uim}) still need to be evaluated numerically.
Introducing the variables
\begin{equation}
\begin{split}
  y=|\mathbf r_1| + |\mathbf r_0| - |\Delta\mathbf r| >0,\\
  s=|\Delta\mathbf r|,\qquad
  x=\left||\mathbf r_1| - |\mathbf r_0|\right|,
\end{split}
\end{equation}
we can factor out scales as
\begin{align}
  \text{Re}(u_\text{C}^{i,j})&=\tau\frac{e^2}{4\pi\epsilon\ell_\tau}\frac{\sqrt\pi}{2} T_0(y,s),\\
  \text{Im}(u_\text{C}^{i,j})&=\tau\frac{e^2}{4\pi\epsilon\ell_\tau}\frac{\sqrt\pi}{16}\frac{\ell_\tau^2}{\ell_c^2}
  \text{sgn}(x_0y_1 - x_1y_0)T_1(y,s,x).\nonumber
\end{align}
We then tabulate $T_0(y,s)$ and $T_1(y,s,x)$ in the interval $0\le y,s\le\tilde r_\text{max}$ and $0\le x\le\tilde r_\text{max}/2$.
We also tabulate first derivatives of $T_0$ and $T_1$, which are necessary for the thermodynamic estimator
$\frac{d u_\text{C}^{i,j}}{d\tau}$.
Note that $T_0$ has only two variables because it knows nothing about the magnetic field,
and for zero magnetic field the Runge-Lenz vector is also conserved.
We note the asymptotic behavior in the $y\to0$ and the $x\to s$ limits
\begin{equation}
\frac{\partial T_1}{\partial y}\propto \frac{x\sqrt{s-x}}{\sqrt y},\quad
\frac{\partial T_1}{\partial s}\propto \frac{\sqrt{xy}}{\sqrt{s-x}},\quad
\frac{\partial T_1}{\partial x}\propto \frac{\sqrt{xy}}{\sqrt{s-x}}.
\end{equation}
Because of these divergences, we extrapolate beyond the last grid point where $\frac{\partial T_1}{\partial y}$,
$\frac{\partial T_1}{\partial s}$ and $\frac{\partial T_1}{\partial x}$ is finite.
$T_0$ and its derivatives are regular everywhere.
We choose $\tilde r_\text{max}=6$ and use linear interpolation in the tabulated domain;
outside of this domain we use the semiclassical approximation to the action,
\begin{multline}
  u_\text{SC}^{i,j} = \tau\frac{e^2}{4\pi\epsilon |\Delta\mathbf r|}\ln\left(
  \frac{|\Delta\mathbf r||\mathbf r_1| + \Delta\mathbf r\cdot\mathbf r_1}
       {|\Delta\mathbf r||\mathbf r_0| + \Delta\mathbf r\cdot\mathbf r_0}
  \right).
\end{multline}


\begin{acknowledgments}
This research was supported by the National Research Development and Innovation Office of Hungary
within the Quantum Technology National Excellence Program  (Project No.\ 2017-1.2.1-NKP-2017-00001).
This work benefited from the TopMat program of the PSI2 project funded by the IDEX Paris-Saclay, ANR-11-IDEX-0003-02.
We are grateful to the HPC facility at the Budapest University of Technology and Economics.
T.\ H.\ G.\ acknowledges support from the ``Quantum Computing and Quantum Technologies'' PhD School of the University of Basel.
\end{acknowledgments}

\bibliography{bib/qmcbiblio}

\end{document}

%% file: final_fig/uhflll8.pdf_tex
\begingroup%
  \makeatletter%
  \providecommand\color[2][]{%
    \errmessage{(Inkscape) Color is used for the text in Inkscape, but the package 'color.sty' is not loaded}%
    \renewcommand\color[2][]{}%
  }%
  \providecommand\transparent[1]{%
    \errmessage{(Inkscape) Transparency is used (non-zero) for the text in Inkscape, but the package 'transparent.sty' is not loaded}%
    \renewcommand\transparent[1]{}%
  }%
  \providecommand\rotatebox[2]{#2}%
  \ifx\svgwidth\undefined%
    \setlength{\unitlength}{360bp}%
    \ifx\svgscale\undefined%
      \relax%
    \else%
      \setlength{\unitlength}{\unitlength * \real{\svgscale}}%
    \fi%
  \else%
    \setlength{\unitlength}{\svgwidth}%
  \fi%
  \global\let\svgwidth\undefined%
  \global\let\svgscale\undefined%
  \makeatother%
  \begin{picture}(1,0.7)%
    \put(0,0){\includegraphics[width=\unitlength,page=1]{uhflll8.pdf}}%
    \put(0.08561472,0.02440322){\color[rgb]{0,0,0}\makebox(0,0)[lb]{\smash{0}}}%
    \put(0,0){\includegraphics[width=\unitlength,page=2]{uhflll8.pdf}}%
    \put(0.17325361,0.02440322){\color[rgb]{0,0,0}\makebox(0,0)[lb]{\smash{1}}}%
    \put(0,0){\includegraphics[width=\unitlength,page=3]{uhflll8.pdf}}%
    \put(0.2607536,0.02440322){\color[rgb]{0,0,0}\makebox(0,0)[lb]{\smash{2}}}%
    \put(0,0){\includegraphics[width=\unitlength,page=4]{uhflll8.pdf}}%
    \put(0.34839166,0.02440322){\color[rgb]{0,0,0}\makebox(0,0)[lb]{\smash{3}}}%
    \put(0,0){\includegraphics[width=\unitlength,page=5]{uhflll8.pdf}}%
    \put(0.43589166,0.02440322){\color[rgb]{0,0,0}\makebox(0,0)[lb]{\smash{4}}}%
    \put(0,0){\includegraphics[width=\unitlength,page=6]{uhflll8.pdf}}%
    \put(0.52339165,0.02440322){\color[rgb]{0,0,0}\makebox(0,0)[lb]{\smash{5}}}%
    \put(0,0){\includegraphics[width=\unitlength,page=7]{uhflll8.pdf}}%
    \put(0.61089165,0.02440322){\color[rgb]{0,0,0}\makebox(0,0)[lb]{\smash{6}}}%
    \put(0,0){\includegraphics[width=\unitlength,page=8]{uhflll8.pdf}}%
    \put(0.69853054,0.02440322){\color[rgb]{0,0,0}\makebox(0,0)[lb]{\smash{7}}}%
    \put(0,0){\includegraphics[width=\unitlength,page=9]{uhflll8.pdf}}%
    \put(0.78603054,0.02440322){\color[rgb]{0,0,0}\makebox(0,0)[lb]{\smash{8}}}%
    \put(0,0){\includegraphics[width=\unitlength,page=10]{uhflll8.pdf}}%
    \put(0.87366942,0.02440322){\color[rgb]{0,0,0}\makebox(0,0)[lb]{\smash{9}}}%
    \put(0.47550277,-0.00476344){\color[rgb]{0,0,0}\makebox(0,0)[lb]{\smash{$\lambda$}}}%
    \put(0,0){\includegraphics[width=\unitlength,page=11]{uhflll8.pdf}}%
    \put(0.05634611,0.05218111){\color[rgb]{0,0,0}\makebox(0,0)[lb]{\smash{0}}}%
    \put(0,0){\includegraphics[width=\unitlength,page=12]{uhflll8.pdf}}%
    \put(0.03086422,0.12037555){\color[rgb]{0,0,0}\makebox(0,0)[lb]{\smash{0.2}}}%
    \put(0,0){\includegraphics[width=\unitlength,page=13]{uhflll8.pdf}}%
    \put(0.03086422,0.18870888){\color[rgb]{0,0,0}\makebox(0,0)[lb]{\smash{0.4}}}%
    \put(0,0){\includegraphics[width=\unitlength,page=14]{uhflll8.pdf}}%
    \put(0.03086422,0.25690333){\color[rgb]{0,0,0}\makebox(0,0)[lb]{\smash{0.6}}}%
    \put(0,0){\includegraphics[width=\unitlength,page=15]{uhflll8.pdf}}%
    \put(0.03086422,0.32509721){\color[rgb]{0,0,0}\makebox(0,0)[lb]{\smash{0.8}}}%
    \put(0,0){\includegraphics[width=\unitlength,page=16]{uhflll8.pdf}}%
    \put(0.05634611,0.39329166){\color[rgb]{0,0,0}\makebox(0,0)[lb]{\smash{1}}}%
    \put(0,0){\includegraphics[width=\unitlength,page=17]{uhflll8.pdf}}%
    \put(0.03086422,0.4614861){\color[rgb]{0,0,0}\makebox(0,0)[lb]{\smash{1.2}}}%
    \put(0,0){\includegraphics[width=\unitlength,page=18]{uhflll8.pdf}}%
    \put(0.03086422,0.52981943){\color[rgb]{0,0,0}\makebox(0,0)[lb]{\smash{1.4}}}%
    \put(0,0){\includegraphics[width=\unitlength,page=19]{uhflll8.pdf}}%
    \put(0.03086422,0.59801387){\color[rgb]{0,0,0}\makebox(0,0)[lb]{\smash{1.6}}}%
    \put(0.00601333,0.329){\color[rgb]{0,0,0}\rotatebox{90}{\makebox(0,0)[lb]{\smash{$\gamma$}}}}%
    \put(0,0){\includegraphics[width=\unitlength,page=20]{uhflll8.pdf}}%
    \put(0.95401664,0.05587444){\color[rgb]{0,0,0}\makebox(0,0)[lb]{\smash{0.2}}}%
    \put(0,0){\includegraphics[width=\unitlength,page=21]{uhflll8.pdf}}%
    \put(0.95401664,0.12406889){\color[rgb]{0,0,0}\makebox(0,0)[lb]{\smash{0.3}}}%
    \put(0,0){\includegraphics[width=\unitlength,page=22]{uhflll8.pdf}}%
    \put(0.95401664,0.19226333){\color[rgb]{0,0,0}\makebox(0,0)[lb]{\smash{0.4}}}%
    \put(0,0){\includegraphics[width=\unitlength,page=23]{uhflll8.pdf}}%
    \put(0.95401664,0.26045777){\color[rgb]{0,0,0}\makebox(0,0)[lb]{\smash{0.5}}}%
    \put(0,0){\includegraphics[width=\unitlength,page=24]{uhflll8.pdf}}%
    \put(0.95401664,0.32879166){\color[rgb]{0,0,0}\makebox(0,0)[lb]{\smash{0.6}}}%
    \put(0,0){\includegraphics[width=\unitlength,page=25]{uhflll8.pdf}}%
    \put(0.95401664,0.3969861){\color[rgb]{0,0,0}\makebox(0,0)[lb]{\smash{0.7}}}%
    \put(0,0){\includegraphics[width=\unitlength,page=26]{uhflll8.pdf}}%
    \put(0.95401664,0.46518054){\color[rgb]{0,0,0}\makebox(0,0)[lb]{\smash{0.8}}}%
    \put(0,0){\includegraphics[width=\unitlength,page=27]{uhflll8.pdf}}%
    \put(0.95401664,0.53337499){\color[rgb]{0,0,0}\makebox(0,0)[lb]{\smash{0.9}}}%
    \put(0,0){\includegraphics[width=\unitlength,page=28]{uhflll8.pdf}}%
    \put(0.95401664,0.60156943){\color[rgb]{0,0,0}\makebox(0,0)[lb]{\smash{1}}}%
    \put(0,0){\includegraphics[width=\unitlength,page=29]{uhflll8.pdf}}%
  \end{picture}%
\endgroup%

%% file: final_fig/four.pdf_tex
\begingroup%
  \makeatletter%
  \providecommand\color[2][]{%
    \errmessage{(Inkscape) Color is used for the text in Inkscape, but the package 'color.sty' is not loaded}%
    \renewcommand\color[2][]{}%
  }%
  \providecommand\transparent[1]{%
    \errmessage{(Inkscape) Transparency is used (non-zero) for the text in Inkscape, but the package 'transparent.sty' is not loaded}%
    \renewcommand\transparent[1]{}%
  }%
  \providecommand\rotatebox[2]{#2}%
  \ifx\svgwidth\undefined%
    \setlength{\unitlength}{566.72169122bp}%
    \ifx\svgscale\undefined%
      \relax%
    \else%
      \setlength{\unitlength}{\unitlength * \real{\svgscale}}%
    \fi%
  \else%
    \setlength{\unitlength}{\svgwidth}%
  \fi%
  \global\let\svgwidth\undefined%
  \global\let\svgscale\undefined%
  \makeatother%
  \begin{picture}(1,0.29497606)%
    \put(0,0){\includegraphics[width=\unitlength,page=1]{four.pdf}}%
    \put(0.0323442,0.01755093){\color[rgb]{0,0,0}\makebox(0,0)[lb]{\smash{$(0,C_{2v})$}}}%
    \put(0,0){\includegraphics[width=\unitlength,page=2]{four.pdf}}%
    \put(0.82260756,0.02255232){\color[rgb]{0,0,0}\makebox(0,0)[lb]{\smash{$(2,C_{4v})$}}}%
    \put(0,0){\includegraphics[width=\unitlength,page=3]{four.pdf}}%
    \put(0.56752116,0.01755093){\color[rgb]{0,0,0}\makebox(0,0)[lb]{\smash{$(1,C_s)$}}}%
    \put(0,0){\includegraphics[width=\unitlength,page=4]{four.pdf}}%
    \put(0.30743584,0.01755093){\color[rgb]{0,0,0}\makebox(0,0)[lb]{\smash{$(0,C_s)$}}}%
  \end{picture}%
\endgroup%

%% file: final_fig/five.pdf_tex
\begingroup%
  \makeatletter%
  \providecommand\color[2][]{%
    \errmessage{(Inkscape) Color is used for the text in Inkscape, but the package 'color.sty' is not loaded}%
    \renewcommand\color[2][]{}%
  }%
  \providecommand\transparent[1]{%
    \errmessage{(Inkscape) Transparency is used (non-zero) for the text in Inkscape, but the package 'transparent.sty' is not loaded}%
    \renewcommand\transparent[1]{}%
  }%
  \providecommand\rotatebox[2]{#2}%
  \ifx\svgwidth\undefined%
    \setlength{\unitlength}{994.73934624bp}%
    \ifx\svgscale\undefined%
      \relax%
    \else%
      \setlength{\unitlength}{\unitlength * \real{\svgscale}}%
    \fi%
  \else%
    \setlength{\unitlength}{\svgwidth}%
  \fi%
  \global\let\svgwidth\undefined%
  \global\let\svgscale\undefined%
  \makeatother%
  \begin{picture}(1,0.46373027)%
    \put(0,0){\includegraphics[width=\unitlength,page=1]{five.pdf}}%
    \put(0.02412661,0.25925144){\color[rgb]{0,0,0}\makebox(0,0)[lb]{\smash{$(1/2,C_s)$}}}%
    \put(0,0){\includegraphics[width=\unitlength,page=2]{five.pdf}}%
    \put(0.2349924,0.25925144){\color[rgb]{0,0,0}\makebox(0,0)[lb]{\smash{$(1/2,C_s')$}}}%
    \put(0,0){\includegraphics[width=\unitlength,page=3]{five.pdf}}%
    \put(0.4430087,0.25925144){\color[rgb]{0,0,0}\makebox(0,0)[lb]{\smash{$(1/2,C_s'')$}}}%
    \put(0,0){\includegraphics[width=\unitlength,page=4]{five.pdf}}%
    \put(0.84346334,0.25925144){\color[rgb]{0,0,0}\makebox(0,0)[lb]{\smash{$(1/2,C_s''')$}}}%
    \put(0,0){\includegraphics[width=\unitlength,page=5]{five.pdf}}%
    \put(0.77640557,0.0099991){\color[rgb]{0,0,0}\makebox(0,0)[lb]{\smash{$(5/2,C_{4v})$}}}%
    \put(0,0){\includegraphics[width=\unitlength,page=6]{five.pdf}}%
    \put(0.57123826,0.0099991){\color[rgb]{0,0,0}\makebox(0,0)[lb]{\smash{$(5/2,C_{5v})$}}}%
    \put(0,0){\includegraphics[width=\unitlength,page=7]{five.pdf}}%
    \put(0.36892095,0.0099991){\color[rgb]{0,0,0}\makebox(0,0)[lb]{\smash{$(3/2,C_s)$}}}%
    \put(0,0){\includegraphics[width=\unitlength,page=8]{five.pdf}}%
    \put(0.15687887,0.0099991){\color[rgb]{0,0,0}\makebox(0,0)[lb]{\smash{$(3/2,C_{4v})$}}}%
    \put(0.6433301,0.25925144){\color[rgb]{0,0,0}\makebox(0,0)[lb]{\smash{$(1/2,C_{2v})$}}}%
  \end{picture}%
\endgroup%

%% file: final_fig/six.pdf_tex
\begingroup%
  \makeatletter%
  \providecommand\color[2][]{%
    \errmessage{(Inkscape) Color is used for the text in Inkscape, but the package 'color.sty' is not loaded}%
    \renewcommand\color[2][]{}%
  }%
  \providecommand\transparent[1]{%
    \errmessage{(Inkscape) Transparency is used (non-zero) for the text in Inkscape, but the package 'transparent.sty' is not loaded}%
    \renewcommand\transparent[1]{}%
  }%
  \providecommand\rotatebox[2]{#2}%
  \ifx\svgwidth\undefined%
    \setlength{\unitlength}{1689.6438046bp}%
    \ifx\svgscale\undefined%
      \relax%
    \else%
      \setlength{\unitlength}{\unitlength * \real{\svgscale}}%
    \fi%
  \else%
    \setlength{\unitlength}{\svgwidth}%
  \fi%
  \global\let\svgwidth\undefined%
  \global\let\svgscale\undefined%
  \makeatother%
  \begin{picture}(1,0.21063378)%
    \put(0,0){\includegraphics[width=\unitlength,page=1]{six.pdf}}%
    \put(0.10143946,0.15952293){\color[rgb]{0,0,0}\makebox(0,0)[lb]{\smash{$(0,C_s)$}}}%
    \put(0,0){\includegraphics[width=\unitlength,page=2]{six.pdf}}%
    \put(0.29939597,0.16958836){\color[rgb]{0,0,0}\makebox(0,0)[lb]{\smash{$(0,C_{3v})$}}}%
    \put(0,0){\includegraphics[width=\unitlength,page=3]{six.pdf}}%
    \put(0.75905734,0.15784506){\color[rgb]{0,0,0}\makebox(0,0)[lb]{\smash{$(2,C_s)$}}}%
    \put(0,0){\includegraphics[width=\unitlength,page=4]{six.pdf}}%
    \put(0.95030474,0.16120021){\color[rgb]{0,0,0}\makebox(0,0)[lb]{\smash{$(2,C_s')$}}}%
    \put(0,0){\includegraphics[width=\unitlength,page=5]{six.pdf}}%
    \put(0.59297678,0.15616719){\color[rgb]{0,0,0}\makebox(0,0)[lb]{\smash{$(1,C_s')$}}}%
    \put(0,0){\includegraphics[width=\unitlength,page=6]{six.pdf}}%
    \put(0.20377286,0.04712338){\color[rgb]{0,0,0}\makebox(0,0)[lb]{\smash{$(2,C_{5v})$}}}%
    \put(0,0){\includegraphics[width=\unitlength,page=7]{six.pdf}}%
    \put(0.4117954,0.04544587){\color[rgb]{0,0,0}\makebox(0,0)[lb]{\smash{$(3,C_{5v})$}}}%
    \put(0,0){\includegraphics[width=\unitlength,page=8]{six.pdf}}%
    \put(0.63491875,0.02028175){\color[rgb]{0,0,0}\makebox(0,0)[lb]{\smash{$(3,C_{6v})$}}}%
    \put(0.71297154,0.02028175){\color[rgb]{0,0,0}\makebox(0,0)[lb]{\smash{$(3,C_{3v})$}}}%
  \end{picture}%
\endgroup%

%% file: final_fig/seven.pdf_tex
\begingroup%
  \makeatletter%
  \providecommand\color[2][]{%
    \errmessage{(Inkscape) Color is used for the text in Inkscape, but the package 'color.sty' is not loaded}%
    \renewcommand\color[2][]{}%
  }%
  \providecommand\transparent[1]{%
    \errmessage{(Inkscape) Transparency is used (non-zero) for the text in Inkscape, but the package 'transparent.sty' is not loaded}%
    \renewcommand\transparent[1]{}%
  }%
  \providecommand\rotatebox[2]{#2}%
  \ifx\svgwidth\undefined%
    \setlength{\unitlength}{1408.58247532bp}%
    \ifx\svgscale\undefined%
      \relax%
    \else%
      \setlength{\unitlength}{\unitlength * \real{\svgscale}}%
    \fi%
  \else%
    \setlength{\unitlength}{\svgwidth}%
  \fi%
  \global\let\svgwidth\undefined%
  \global\let\svgscale\undefined%
  \makeatother%
  \begin{picture}(1,0.13880256)%
    \put(0,0){\includegraphics[width=\unitlength,page=1]{seven.pdf}}%
    \put(0.01502572,0.01309829){\color[rgb]{0,0,0}\makebox(0,0)[lb]{\smash{$(1/2,C_{2v})$}}}%
    \put(0,0){\includegraphics[width=\unitlength,page=2]{seven.pdf}}%
    \put(0.1418031,0.00706135){\color[rgb]{0,0,0}\makebox(0,0)[lb]{\smash{$(1/2,C_s)$}}}%
    \put(0,0){\includegraphics[width=\unitlength,page=3]{seven.pdf}}%
    \put(0.40743255,0.01309829){\color[rgb]{0,0,0}\makebox(0,0)[lb]{\smash{$(3/2,C_s)$}}}%
    \put(0,0){\includegraphics[width=\unitlength,page=4]{seven.pdf}}%
    \put(0.28266669,0.01108577){\color[rgb]{0,0,0}\makebox(0,0)[lb]{\smash{$(1/2,C_s')$}}}%
    \put(0,0){\includegraphics[width=\unitlength,page=5]{seven.pdf}}%
    \put(0.91253324,0.01511052){\color[rgb]{0,0,0}\makebox(0,0)[lb]{\smash{$(7/2,C_{6v})$}}}%
    \put(0,0){\includegraphics[width=\unitlength,page=6]{seven.pdf}}%
    \put(0.52616042,0.01309829){\color[rgb]{0,0,0}\makebox(0,0)[lb]{\smash{$(3/2,C_{2v})$}}}%
    \put(0,0){\includegraphics[width=\unitlength,page=7]{seven.pdf}}%
    \put(0.66098745,0.01108577){\color[rgb]{0,0,0}\makebox(0,0)[lb]{\smash{$(5/2,C_{6v})$}}}%
    \put(0,0){\includegraphics[width=\unitlength,page=8]{seven.pdf}}%
    \put(0.79178774,0.01309829){\color[rgb]{0,0,0}\makebox(0,0)[lb]{\smash{$(5/2,C_s)$}}}%
  \end{picture}%
\endgroup%

%% file: final_fig/eight.pdf_tex
\begingroup%
  \makeatletter%
  \providecommand\color[2][]{%
    \errmessage{(Inkscape) Color is used for the text in Inkscape, but the package 'color.sty' is not loaded}%
    \renewcommand\color[2][]{}%
  }%
  \providecommand\transparent[1]{%
    \errmessage{(Inkscape) Transparency is used (non-zero) for the text in Inkscape, but the package 'transparent.sty' is not loaded}%
    \renewcommand\transparent[1]{}%
  }%
  \providecommand\rotatebox[2]{#2}%
  \ifx\svgwidth\undefined%
    \setlength{\unitlength}{1580.90735728bp}%
    \ifx\svgscale\undefined%
      \relax%
    \else%
      \setlength{\unitlength}{\unitlength * \real{\svgscale}}%
    \fi%
  \else%
    \setlength{\unitlength}{\svgwidth}%
  \fi%
  \global\let\svgwidth\undefined%
  \global\let\svgscale\undefined%
  \makeatother%
  \begin{picture}(1,0.21862572)%
    \put(0,0){\includegraphics[width=\unitlength,page=1]{eight.pdf}}%
    \put(0.10187918,0.16220613){\color[rgb]{0,0,0}\makebox(0,0)[lb]{\smash{$(0,C_s)$}}}%
    \put(0,0){\includegraphics[width=\unitlength,page=2]{eight.pdf}}%
    \put(0.4365332,0.16220613){\color[rgb]{0,0,0}\makebox(0,0)[lb]{\smash{$(0,C_{4v})$}}}%
    \put(0,0){\includegraphics[width=\unitlength,page=3]{eight.pdf}}%
    \put(0.26598677,0.16220613){\color[rgb]{0,0,0}\makebox(0,0)[lb]{\smash{$(0,C_{s}')$}}}%
    \put(0,0){\includegraphics[width=\unitlength,page=4]{eight.pdf}}%
    \put(0.61403948,0.16220613){\color[rgb]{0,0,0}\makebox(0,0)[lb]{\smash{$(1,C_s)$}}}%
    \put(0,0){\includegraphics[width=\unitlength,page=5]{eight.pdf}}%
    \put(0.76464729,0.16220613){\color[rgb]{0,0,0}\makebox(0,0)[lb]{\smash{$(1,C_s'')$}}}%
    \put(0,0){\includegraphics[width=\unitlength,page=6]{eight.pdf}}%
    \put(0.93677629,0.16220613){\color[rgb]{0,0,0}\makebox(0,0)[lb]{\smash{$(1,C_s''')$}}}%
    \put(0,0){\includegraphics[width=\unitlength,page=7]{eight.pdf}}%
    \put(0.17296368,0.04028274){\color[rgb]{0,0,0}\makebox(0,0)[lb]{\smash{$(2,C_s)$}}}%
    \put(0,0){\includegraphics[width=\unitlength,page=8]{eight.pdf}}%
    \put(0.70010208,0.04028274){\color[rgb]{0,0,0}\makebox(0,0)[lb]{\smash{$(4,C_{7v})$}}}%
    \put(0,0){\includegraphics[width=\unitlength,page=9]{eight.pdf}}%
    \put(0.52797562,0.04207563){\color[rgb]{0,0,0}\makebox(0,0)[lb]{\smash{$(4,C_{2v})$}}}%
    \put(0,0){\includegraphics[width=\unitlength,page=10]{eight.pdf}}%
    \put(0.87760774,0.04028274){\color[rgb]{0,0,0}\makebox(0,0)[lb]{\smash{$(3,C_{7v})$}}}%
    \put(0.17334389,0.04924761){\color[rgb]{0,0,0}\makebox(0,0)[lb]{\smash{}}}%
    \put(0.35437432,0.05063345){\color[rgb]{0,0,0}\makebox(0,0)[b]{\smash{}}}%
    \put(0.36517512,0.04028274){\color[rgb]{0,0,0}\makebox(0,0)[lb]{\smash{$(2,C_s')$}}}%
  \end{picture}%
\endgroup%